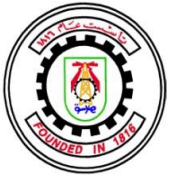 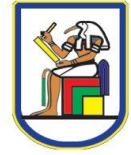

# INFRARED SOLAR ENERGY HARVESTING USING NANO-RECTENNAS

By

Islam Esmat Mohamed Mohamed Hashem Sayed

A Thesis Submitted to the
Faculty of Engineering at Cairo University
in Partial Fulfillment of the
Requirements for the Degree of
MASTER OF SCIENCE
in
Engineering Physics

FACULTY OF ENGINEERING, CAIRO UNIVERSITY
GIZA, EGYPT
2013

# INFRARED SOLAR ENERGY HARVESTING USING NANO-RECTENNAS

By

Islam Esmat Mohamed Mohamed Hashem Sayed

A Thesis Submitted to the
Faculty of Engineering at Cairo University
in Partial Fulfillment of the
Requirements for the Degree of
MASTER OF SCIENCE
in
Engineering Physics

Under the Supervision of

| | |
|---|---|
| Prof. Dr. Nadia H. Rafat | Prof. Dr. Ezzeldin A. Soliman |
| Professor | Professor |
| Department of Engineering Math. and Physics, | Department of Physics, School of Sciences and Engineering, |
| Faculty of Engineering, Cairo University | The American University in Cairo |

FACULTY OF ENGINEERING, CAIRO UNIVERSITY
GIZA, EGYPT
2013

# INFRARED SOLAR ENERGY HARVESTING USING NANO-RECTENNAS

By

Islam Esmat Mohamed Mohamed Hashem Sayed

A Thesis Submitted to the
Faculty of Engineering at Cairo University
in Partial Fulfillment of the
Requirements for the Degree of
MASTER OF SCIENCE
in
Engineering Physics

Approved by the
Examining Committee

____________________________
Prof. Dr. Nadia Hussein Rafat, Thesis Advisor

____________________________
Prof. Dr. Alaa Korany Abdelmageed, Internal Examiner

____________________________
Prof. Dr. Mohammed Ahmed El Shaer, External Examiner (Dept. of Engineering Mathematics and Physics, Zagazig University)

FACULTY OF ENGINEERING, CAIRO UNIVERSITY
GIZA, EGYPT
2013

**Engineer's Name:** Islam Esmat Mohamed Mohamed Hashem Sayed
**Date of Birth:** 04/10/1988
**Nationality:** Egyptian
**E-mail:** eemm_hashem@ieee.org
**Phone:** …………………..
**Address:** Engineering Physics Dept., 12613 Cairo University

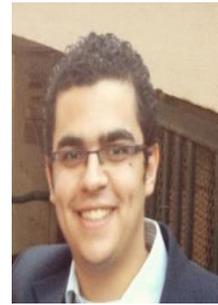

**Registration Date:** 01/03/2011
**Awarding Date:** …./…./……..
**Degree:** Master of Science
**Department:** Department of Engineering Mathematics and Physics
**Supervisors:**

Prof. Dr. Nadia H. Rafat
Prof. Dr. Ezzeldin A. Soliman

**Examiners:**

Prof. Dr. Mohammed Ahmed El Shaer (External examiner)
Prof. Dr. Alaa Korany Abdelmageed  (Internal examiner)
Prof. Dr. Nadia Hussein Rafat   (Thesis advisor)

**Title of Thesis:**

Infrared Solar Energy Harvesting using Nano-Rectennas

**Summary:**

In this thesis, rectifying antennas or rectennas are implemented to harvest the infrared solar energy. The nano-rectifier is modeled using the transfer matrix method based on Airy Function and the Non-Equilibrium Green's Function. The figures of merit of the nano-rectifier are analyzed. Two structures for the nano-rectennas are proposed, where the nanodipole is either fed vertically or laterally by plasmonic metal insulator metal transmission lines. In each structure, the nantenna is successfully matched to its associated plasmonic line. The overall efficiency of the proposed travelling wave rectennas are calculated and compared.

# Acknowledgments

First and foremost I would like to express my indebtedness and gratefulness to my academic advisors: Prof. Nadia H. Rafat and Prof. Dr. Ezzeldin A. Soliman. It has been a pleasure to work with them and learn from such extraordinary advisors. They have always made themselves available for help and advice with their boundless enthusiasm and positive thinking.

I am indebted to Dr. Nadia for the *countless hours* that she spent with me on discussing problems and making suggestions, and for many many discussions that made my research possible. In the past two and half years, she taught me invaluable skills such as communicating ideas effectively and logically, critical thinking and documenting and reporting.

I am also indebted to Dr. Ezzeldin for giving me an *invaluable opportunity* to work on challenging and extremely interesting project over the past two and half years. His insight, enthusiasm and passion to scientific research have great impacts on my study and research.

Also, I would like to express my special appreciations to Prof. Dr. Sherif Sedky for introducing me to Dr. Ezzeldin.

Special thanks to Prof. Ahmed Alaa Abouelsaood, Prof. Salah El Nahwy, Prof. Nadia Rafat, Prof. Alaa K. Abdelmageed, Dr. Yasser El-Batawy, and Dr. Essam El Karamany for teaching such a great graduate courses at Cairo University. I benefited from the breadth of the courses that gave me comprehensive exposure to the core areas of Engineering Physics.

I would like to express my gratitude to the American University in Cairo (AUC) for making the research facilities at Youssef Jameel Science and Technology Research Center (YJ-STRC) available for my research.

I would like also to express my appreciations to Ms. Mai Sallam, a Ph.D. student working under the supervision of Dr. Ezzeldin, for giving me a numerous advices on the antenna design using CST Microwave Studio. Also, I would like to thank Mr. Robert Rizk, an M.Sc. student working under the supervision of Dr. Ezzeldin, for the useful discussion we had on the characteristic impedance calculation using COMSOL.

Special thanks to Omar Habbak, Kareem Adel, Yasser, Moanes, Elsisi, Omar Mosaad, MAS Zaghloul, Rehab Kotb, Hazem Aly, Tarek Ameen and Hesham Yamani for helping me to get a monthly license for Lumerical FDTD Solver.

Thanks to my colleagues studying abroad: Ahmed Alaa, Mohamed Fatouh, Amr Abdulzahir, Alhassan Fattin and Shehab Yomn for helping to get the JAP research papers.

Special thanks to my colleagues at Cairo University; MAS Zaghloul, Tarek Ameen, Ahmed Abdo, Eslam Sayed and Amr Mahmoud for their continuous support and the great times we had together throughout my work as a teaching assistant at the Department of Engineering Mathematics and Physics at Cairo University. They made the graduate school a very enjoyable experience.

Lastly, I would like to thank my parents. It is a dream that come true to achieve this degree, which I share with them. I thank my sister Marwa and brother Amr, and their families for the constant love and support, with a high five to Adam, Zane, and Aya for enlivening my spare time.

Islam E. Hashem
Cairo, June 2013II

*to my family and friends*

# Table of Contents

# List of Publications

**Published Papers:**

[1] I. E. Hashem, N. H. Rafat, and E. A. Soliman, "Theoretical Study of Metal-Insulator-Metal Tunneling Diode Figures of Merit," *Quantum Electronics, IEEE Journal of,* vol. 49, pp. 72-79, 2013.

[2] I. E. Hashem, N. H. Rafat, and E. A. Soliman, "Characterization of MIM diodes based on Nb/ $Nb_2O_5$," in *Nanoelectronics Conference (INEC), 2013 IEEE 5th International*, 2013, pp. 61-64.

[3] I. E. Hashem, N. H. Rafat, and E. A. Soliman, "Dipole Nantennas Terminated by Travelling Wave Rectifiers for Ambient Thermal Energy Harvesting," *Nanotechnology, IEEE Transaction on,* vol. PP, 2014.

[4] I. E. Hashem, N. H. Rafat, and E. A. Soliman, "Nanocrescent Antenna for Optical Communication System" in *2014 IEEE Symposium on Electromagnetic Compatibility, Raleigh, NC*.

[5] I. E. Hashem, N. H. Rafat, and E. A. Soliman, "Analysis of Metal Insulator Metal Plasmonic Transmission Lines for Improved Rectenna's Coupling Efficiency" *2015 9th European Conference on Antennas and Propagation (EUCAP), Lisbon, Portugal*.

[6] I. E. Hashem, N. H. Rafat, and E. A. Soliman, "Harvesting Thermal Infrared Emission using Nanodipole Terminated by Traveling Wave Rectifier," *2015 9th European Conference on Antennas and Propagation (EUCAP), Lisbon, Portugal*.

# List of Symbols

| Symbol | Description |
|---|---|
| $\lambda$ | Wavelength |
| $\lambda_g$ | Guided wavelength |
| $\phi$ | Work function |
| $\chi$ | Electron-affinity |
| $\boldsymbol{\varepsilon}$ | Permittivity |
| $E_f$ | Fermi level |
| $V_o$ | Amplitude of infrared radiation |
| $V_b$ | Bias voltage |
| $d$ | Insulator thickness |
| $U$ | Barrier height of the MIM junction |
| $e$ | Charge of an electron |
| $J$ | Tunneling current density |
| $m$ | Effective mass |
| $f$ | Fermi-Dirac distributions |
| $\psi$ | Wave-function |
| $\hbar$ | Reduced Plank's constant |
| $T(E_x)$ | Tunneling probability |
| $Ai$ | Airy function of the first kind |
| $Bi$ | Airy function of the second kind |
| $k$ | Electron's wave vector |
| $H$ | Hamiltonian |
| $\Gamma_{L,R}$ | Electron's escape rate |
| $G$ | Green's function |
| Nb | Niobium |
| $Nb_2O_5$ | Niobium dioxide |
| $\phi_{R-L}$ | Barrier asymmetry |
| $\alpha$ | Attenuation constant |
| $\alpha^{-1}$ | Plasmon decay length (Propagation length) |
| $L_{prop.}$ | Propagation length |
| Au | Gold |
| Ag | Silver |
| $\varepsilon_\infty$ | Contribution of bound electrons to dielectric constant |
| $f_p$ | Plasmon frequency |
| $\Gamma$ | Damping collision frequency |
| $\delta$ | Skin depth |
| $k_0$ | Free space permittivity |
| $c$ | Speed of wave propagation in free space |
| $v$ | Speed of wave propagation in a medium |
| $\underline{E}$ | Electric field |
| $\underline{J}$ | Conduction current density |
| $n$ | Refractive index |
| $n_{eff}$ | Effective refractive index |
| $\omega$ | Angular frequency |
| $Z_c$ | Characteristic impedance |
| $Z_A$ | Antenna impedance |
| $R_D$ | Diode Semiclassical resistance |
| $C_D$ | Diode capacitance |
| $A$ | Overlapping area |
| $Z_D$ | Diode impedance |
| $\eta$ | Overall system efficiency |
| $\eta_c$ | Coupling efficiency |
| $\eta_D$ | Diode responsivity |
| $\eta_{ant.}$ | Antenna efficiency |
| $\eta_{rad.}$ | Radiation efficiency |
| $P(\lambda, T)$ | Plank's irradiance for Blackbody's radiation |
| $\beta$ | Responsivity |

# List of Acronyms

| | |
|---|---|
| LWIR | Long Wavelength Infrared |
| THz | Terahertz |
| IR | Infrared |
| Rectenna | Rectifying antenna |
| MIM | Metal Insulator Metal |
| MIIM | Metal Insulator Insulator Metal |
| WKB | Wentzel-Kramers-Brillouin |
| TMM | Transfer Matrix Method |
| AFTMM | Transfer Matrix Method based on Airy's Function |
| NEGF | Non-Equilibrium Green's Function |
| FWHM | Full-Width at Half Maximum |
| ALD | Atomic Layer Deposition |
| EBL | Electron Beam Lithography |
| AC | Alternating Current |
| DC | Direct Current |
| FEM | Finite Element Method |
| FDTD | Finite Difference Time Domain |
| TW | Traveling Wave |
| TM | Transverse magnetic |
| VCS | Vertical Coupled Strips |
| LCS | Lateral Coupled Strips |
| NS | Nano-Strip |

# List of Figures

# Abstract

Research in clean renewable sources of energy attracts considerable interest recently. The Sun represents a huge source of energy that can solve all human-kind energy problems if utilized efficiently. The Sun emits its energy in the visible range centered at 0.5 $\mu$m. 50% of the Sun's energy arrives and absorbed by the Earth and remitted in the form of radiation in the Long Wavelength Infrared (LWIR) centered at 10 μm (30 THz). At the LWIR range, the rectifying antenna (rectenna) is one of the promising candidates that might be used as a reliable source for providing DC power. The rectenna is a nantenna connected to a rectifier. The Metal Insulator Metal (MIM) diode is the only available rectifier that can operate at 30 THz.

In this research, rectennas formed from nano-dipole antennas terminated by plasmonic transmission lines formed from two strips coupled either vertically or laterally, are developed for infrared solar energy harvesting. The length of the transmission lines used are made longer than the plasmon decay length, which facilitates the matching with the nantennas.

For the rectifier, an analytical expression using the Transfer Matrix Method (TMM) based on the Airy function is derived to model the tunneling current throughout the MIM structure and is implemented using Matlab. In addition, the current is also modeled using the Non-Equilibrium Green's Function (NEGF) for the sake of validating the results. Both the TMM and the NEGF results match with each other and with published experimental works. The figures of merit of the rectifier are analyzed, and the effect of the metals and insulator choices on these merits is investigated. Adding another insulator layer to enhance the rectifier's characteristics is also considered.

The transmission lines used as both rectifiers and matched antenna terminations are analyzed. Full design curves are prepared for these transmission lines. These curves show the variation of the propagation characteristics with geometrical dimensions and material parameters. These characteristics include the effective refractive index, attenuation constant, plasmon decay length, characteristic impedance, and the dispersion of the fundamental modes along the lines. Two well-known full-wave electromagnetic simulators are used to perform this study, namely COMSOL and Lumerical. Very good agreement between the two solvers is noticed.

The nantenna simulation is performed using CST and cross-validated using COMSOL. A systematic design approach is presented in this research that shows how different components can be integrated with each other with maximum radiation receiving efficiency and maximum coupling efficiency at 30 THz.

The overall rectenna system efficiency is calculated. Due to the extremely small spacing between its two strips, the nanodipole terminated by vertical coupled strips plasmonic transmission line offers significantly higher efficiency than that terminated by a lateral coupled strips line. However, the overall efficiency of the vertical coupled strips rectenna is still very low due to the extremely low efficiency of the MIM rectifier.

A novel nanocrescent antenna operating at 1.55 µm is introduced in this research. The proposed nantenna enjoys remarkable wideband behavior, characterized by a FWHM of 61% compared with only 35% for conventional nantennas. This makes it a very good candidate for optical communication systems that require intensive date streaming. The new antenna is studied thoroughly via both Lumerical and CST and a full parametric study, including both the geometrical and material parameters, is performed. The design approach used to engineer the spectral response of the proposed nanocrescent nantenna is outlined.

# Chapter 1 Introduction

## 1.1. Rectennas

Recent advances in the fabrication technology permits the fabrication of nanoscale devices with high precision. A few years ago, the technology to fabricate nanophotonic devices such as the nantennas was not available. Hence, there were no antennas operating at the infrared range of frequencies. Now, it becomes possible to fabricate such nano-scale devices with reasonable accuracy due to the development of fabrication tools like electron beam lithography and Atomic Layer Deposition (ALD). Hence, the idea proposed by R. L. Bailey in 1972 to harvest solar energy using rectennas (rectifying antennas) becomes applicable [1, 2].

The spectrum of the Sun, that reaches the Earth can be divided into three main parts [3][4]: Ultraviolet (9%), Visible (39%), and Infrared (52%), where the bracketed quantities represent the approximate percentage of each frequency range of the power density received. Almost half of the energy reaches the boundary of the atmosphere is either absorbed or reflected back and the remaining half hits the Earth and re-radiated in the Long Wavelength Infrared Range (LWIR), $7\mu m < \lambda < 14\ \mu m$, with peak at 10 µm (30 THz), as it is clear from Fig. 1.1.

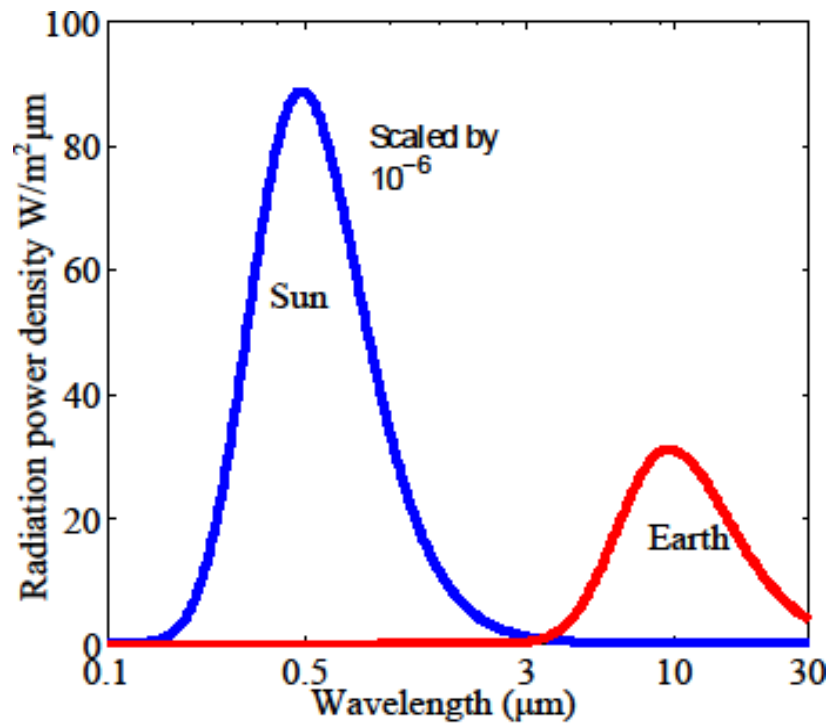

**Figure 1.1: Radiation power density of the Sun and the Earth as calculated by the Plank's blackbody radiation.**

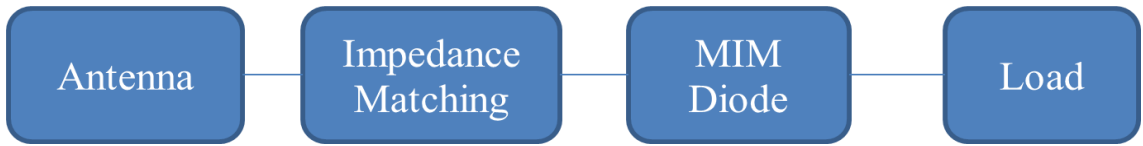

**Figure 1.2: Rectenna system block diagram.**

The *visible* radiation from the *Sun* is unidirectional and is currently harvested via solar cells. The *infrared* radiation from the *Earth* is isotropic and the harvesting of this radiation is in the curiosity phase. The main technique used to harvest this energy is via the rectenna. The rectenna, or rectifying antenna, is made of a nano-antenna coupled to a rectifier, as shown in Fig. 1.2. A rectifier is essential, as the nantenna harvests the solar incident radiation and converts it into a THz Alternating Current (AC) voltage difference at its gap. The rectifier converts this AC voltage into Direct Current (DC) voltage difference. At such high frequency of operation, 30 THz that corresponds to 10 µm wavelength, conventional semiconducting diodes fail to operate due to the fact that the time needed to build charge along the depletion region is no longer negligible with respect to the periodic time of the waveform. The successful alternative is the Metal-Insulator-Metal (MIM) tunneling diode. Impedance matching should be satisfied between the nantenna and rectifier in order to enhance the coupling efficiency between them. The net DC harvested power is either applied directly across a load resistance or stored in batteries for later use.

The infrared rectenna has the potential to be used in photovoltaic to harvest the solar radiation at that frequency range, which is not collected by semiconductor PV solar cells for a few reasons:

(1) The absence of bandgap implies that the efficiency limitation of the semiconductor solar cells is not present.
(2) The infrared rectenna can operate at day and night.
(3) The infrared rectenna works at any weather condition.
(4) The infrared rectenna is less sensitive to the orientation as the infrared radiation is isotropic.

## 1.2. Nanoantennas (Nantennas)

Research in nano-antennas, or simply nantennas, is still in the exploration phase. The frequency scaling law states that the increase of the operation frequency is accompanied by decrease in the antenna dimensions [5]. This law is applicable along the entire frequency range from the extremely low frequency, passing through the microwave, millimeter wave, Terahertz, infrared, up to the visible light. Nantennas as optical transceivers have a number of advantages, such as: small size, fast response, high directivity, doubling the information capacity via sensitivity to polarization, and broadband behavior. These advantages over semiconductor based transceivers promote the use of nantennas in several applications, such as: energy harvesting [6-11], infrared detectors [12-15], multispectral imaging [16], near field optical microscopy [17-19], chemical and biological sensing [20], single molecule spectroscopy [21], cancer tumors treatment.

A number of nantenna configurations have been proposed in the literature recently, such as monopoles [9], dipoles [22-27], bowties [22-24, 28], fractal bowties [29, 30], spirals [8, 16, 31, 32], Yagi-Uda [33-36], nano-ring [37], nanowire [38, 39], and nanocrescent [5, 40-42].

The metal type plays a major role in determining the efficiency and losses of the nantenna. Metals at microwave frequencies possess constant very large imaginary permittivity, which prohibits the presence of electromagnetic fields within metals. At infrared frequencies, the permittivity becomes complex and frequency dependent. Fig. 1.3 shows the real and imaginary parts of the permittivity of gold, silver, aluminum, as provided by Palik [43] and copper as provided by [44]. The strong dependency of the permittivity function on the wavelength can be easily noticed in the figure. Moreover, the real and imaginary parts are in the same order, which means that the metal volume at such high wave frequencies is somewhere between perfect metal and perfect dielectric. In other words, its volume contains electromagnetic field like the dielectric, with losses like the conductor. Hence, fields and currents can exist inside the metal volume in the infrared frequencies. Consequently, the selection of the metal type used for realizing nantennas is much more critical than the case of microwave antennas.

## 1.3. Rectifier

The operating frequency of the rectenna determines the suitable rectifier to be used. At the gigahertz frequency range, semiconductor rectifiers can be used. Beyond the gigahertz range, the transit time of the electrons through the semiconductor junctions hinders their operation at the terahertz range. A way to solve that problem was to use the Schottky diodes as that can be used at the terahertz and far-infrared frequency range [45-48]. However, beyond 12 THz the Metal-Insulator-Metal (MIM) tunnel diode is the most convenient rectifier [49]. The use of metal/insulator extended the frequency of operation of the diode as the transport of charges through the junction is based on the tunneling, which has a time constant of around 1 $fs$.

The metal insulator metal diode is made of two metal electrodes with few nanometers insulator in between. The band diagram of the MIM is presented in Fig. 1.4 at thermal equilibrium. Two dissimilar metals of work functions $\phi_L$ and $\phi_R$ are used here. The work function is defined for the metals as the minimum energy required to liberate an electron or the energy required to excite the electron from the Fermi level to the vacuum level [50]. The Fermi level is aligned horizontally here in Fig. 1.4 as there is no applied bias voltage. The barrier height is defined as the difference between the metal's work function and the electron affinity of the insulator, $\chi$. It is worth noting that the use of dissimilar metal electrodes results in built in field, which is defined as the ratio of the difference between the work function of the two metal electrodes to the insulator thickness. The use of dissimilar metal electrodes enhances the diodes figures of merit, as it will be illustrated in Chapter 2.

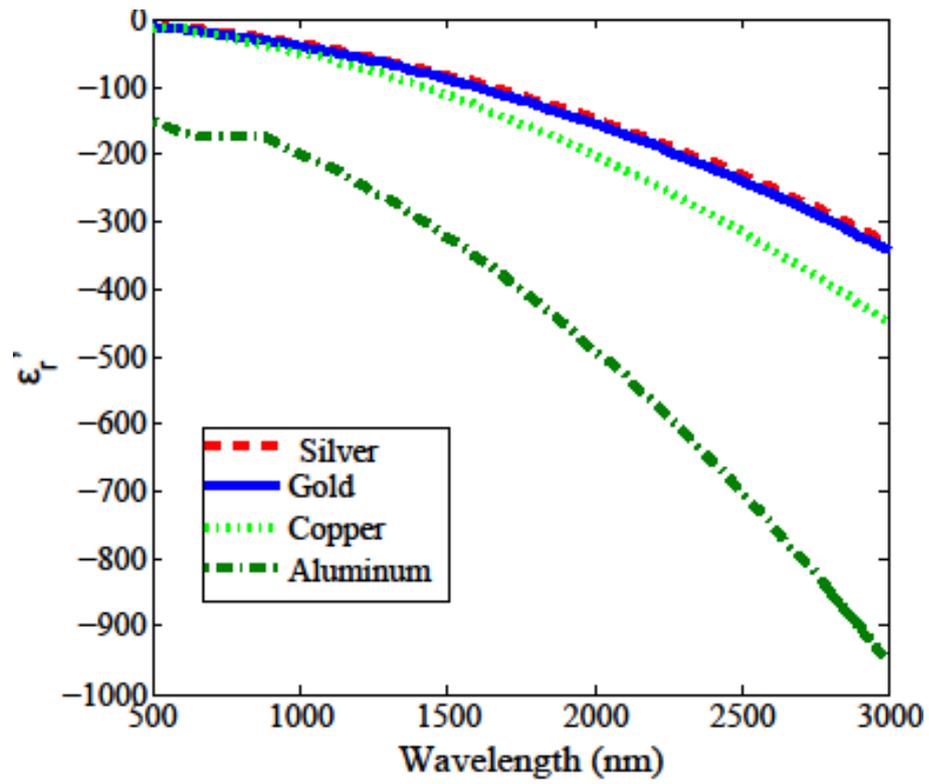

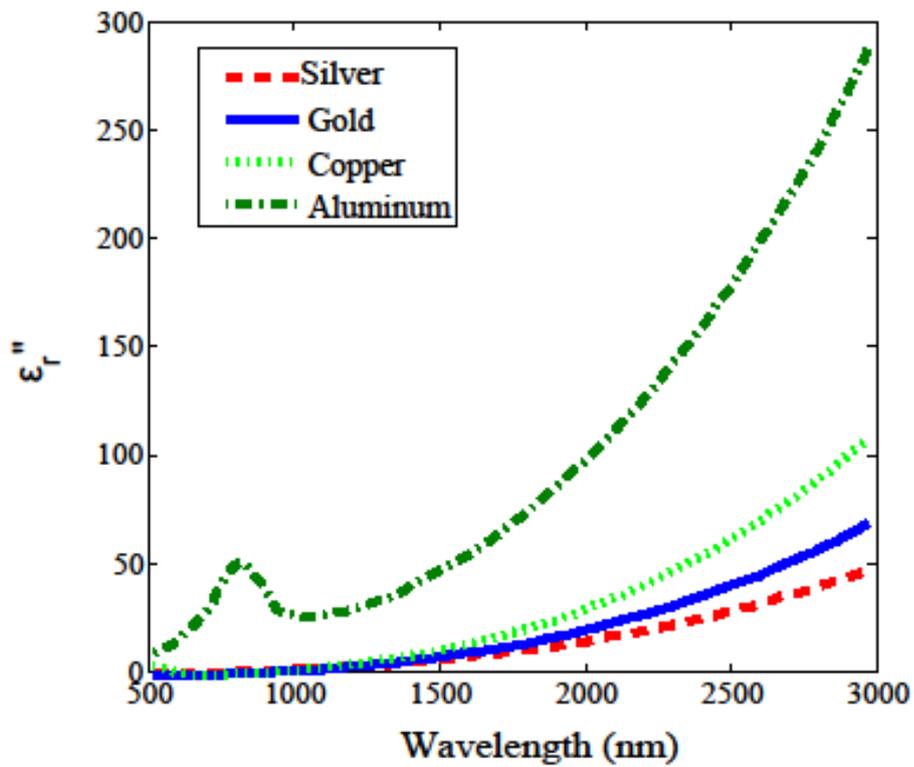

**Figure 1.3:** Permittivity of the gold, silver, copper, and aluminum, versus the wavelength for various metals. a) Real part, $\varepsilon'_r$, b) Imaginary part, $\varepsilon''_r$.

In case of applying negative DC bias voltage, $V_b$, on the right metal electrode, as shown in Fig. 1.5, the energy states of the right metal electrode shift up. The field across the insulator changes by the amount of the added bias voltage divided by the insulator thickness. By keeping the insulator thickness less than 5 nm, the electrons in one metal electrode can tunnel across the barrier to the other side. For the applied bias at Fig. 1.5, that definitely gives rise to the electrons on the right side to move to the unoccupied states in the left metal electrode. Details on the calculation of the tunneling probability and the current voltage characteristics are presented in Chapter 2.

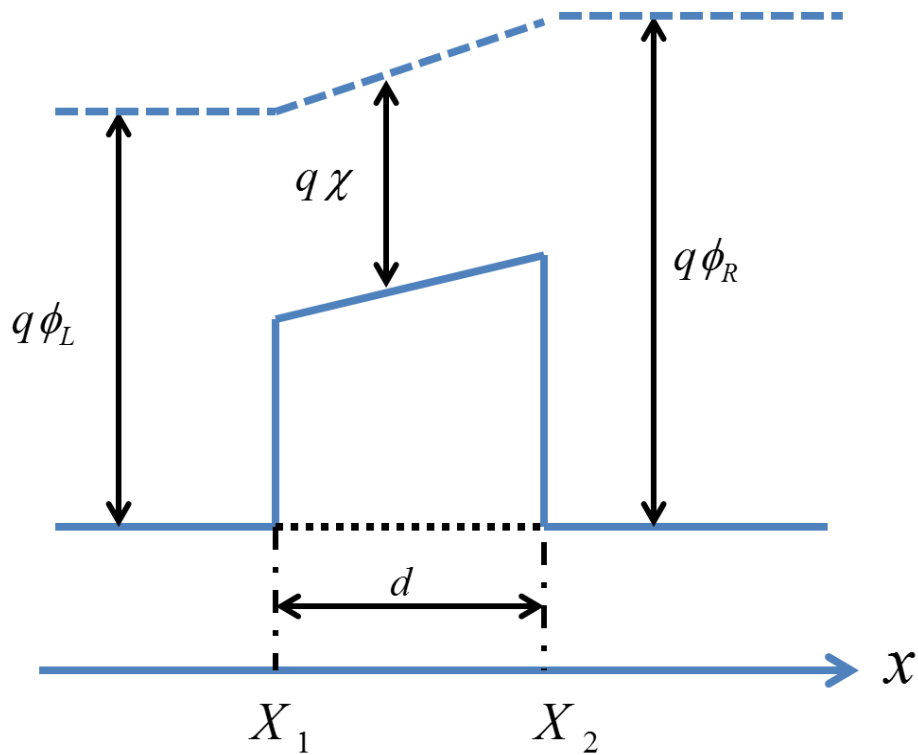

**Figure 1.4: Energy band diagram of the MIM at equilibrium.**

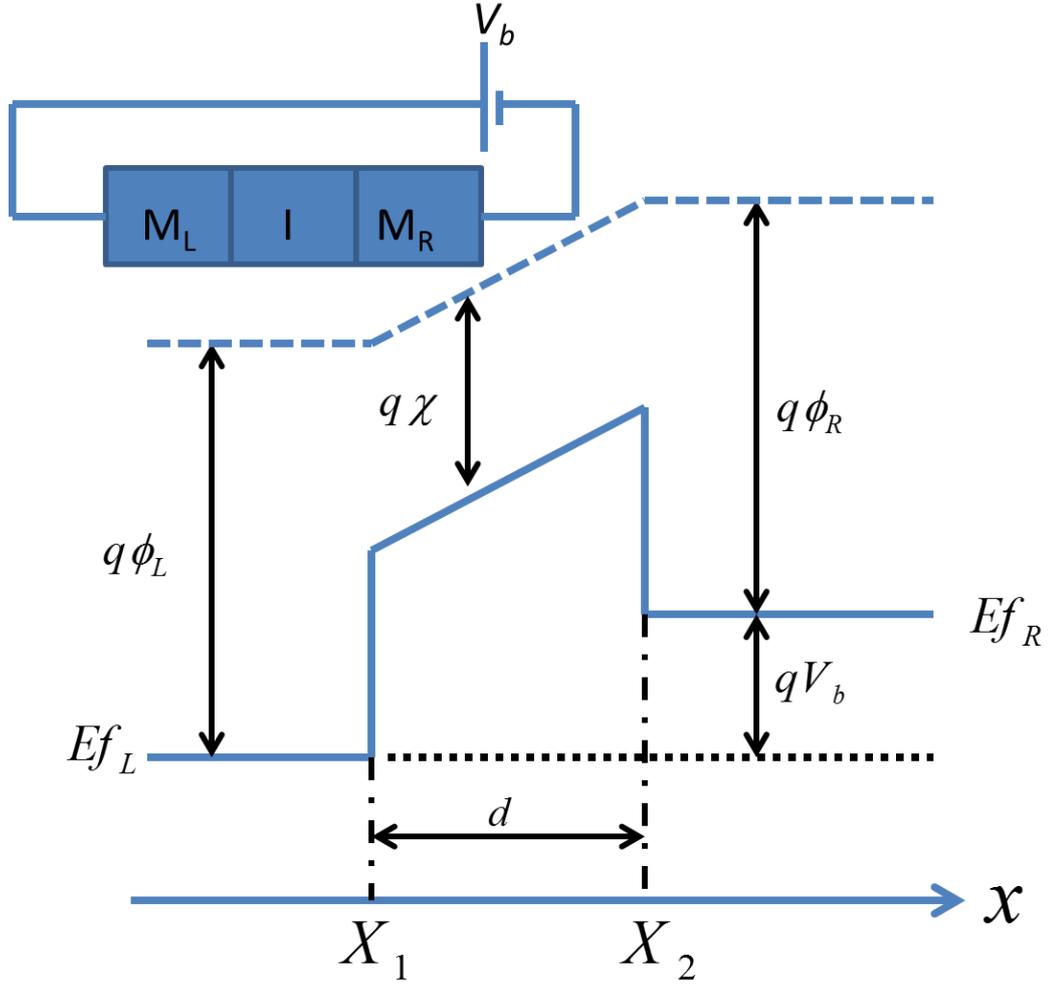

**Figure 1.5: Energy band diagram of the MIM in case of applied bias voltage, $V_b$.**

As mentioned earlier, the nantenna harvests the solar radiation and converts it into an AC voltage difference. The total applied voltage on the diode, $V(t) = V_b + V_0 \cos(\omega t)$, where $V_0$ is the amplitude of the induced voltage at the nantenna gap and $\omega$ is the angular frequency of the incident radiation [13]. Fig. 1.6 shows the variation of the band diagram due to such DC bias and applied infrared radiation. When the induced AC voltage has the same polarity as the applied DC bias, the separation between the Fermi levels of the two metal electrodes is increased, as shown in Fig. 1.6. The probability of an electron to tunnel through the potential barrier from right to left increases. On the other hand, when the induced infrared voltage is of opposite polarity to the DC applied bias, the Fermi levels come closer together and the tunneling probability of an electron through the potential barrier decreases. Hence, for each half cycle, there is a different probability of the electron's tunneling. This *non-linear* behavior of the MIM diode makes it act as a rectifier.

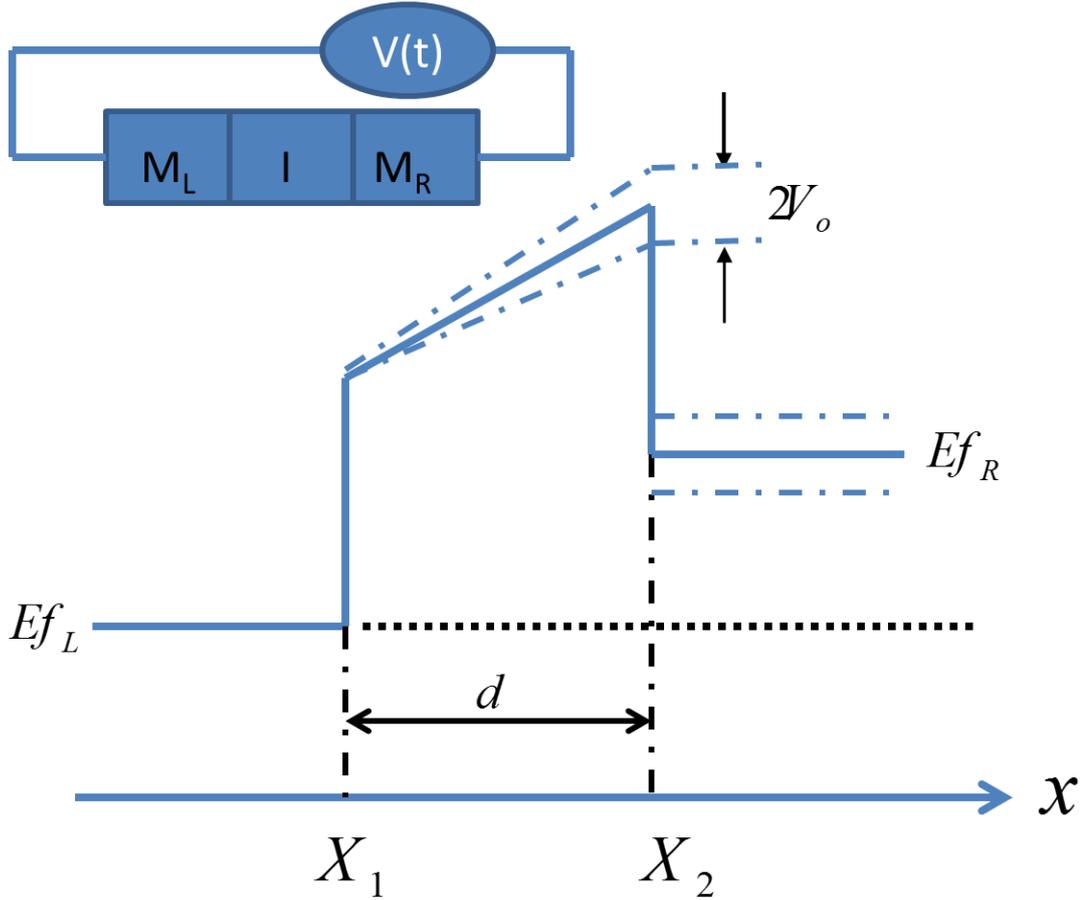

**Figure 1.6: Energy band diagram of the MIM in Energy band diagram of MIM in case of incident infrared radiation in addition to the applied bias voltage.**

## 1.4. Traveling wave MIM diode

Though the use of nantennas definitely increases the efficiency of harvesting solar power in its AC form, the overall system efficiency does not exceed 0.1%. This is mainly due to the extremely low MIM rectifier efficiency. In addition, the coupling efficiency between the nantenna and rectifier has some effect on the overall efficiency of the system. The impedance of the nantenna is less than 1 kΩ, while the MIM resistance is in the order of MΩ - GΩ. Hence, the quality of matching between them is very bad. A step towards solving this problem is proposed by Berland et. al., by using plasma oxidation in order to lower the diode impedance [51]. However, that could not lower the diode impedance very much and the coupling efficiency is still very low.

In 2006, Estes and Moddel [10] proposed an alternative solution for the impedance matching problem. The matching can be achieved by using the traveling wave metal insulator metal structure, shown in Fig. 1.7. This structure serves as an electrically long *transmission line* and a *rectifier* at the same time. The electromagnetic waves, captured by the nantenna, propagate through the MIM structure. Due to the relatively long length and high attenuation of this transmission line, almost no reflected wave will be generated at termination whatever the type of termination is. This allows replacing the MIM line with its characteristic impedance, which is in the same order of the nantenna input impedance. This makes it possible to achieve perfect coupling from nantenna to the rectifier, as illustrated in Chapter 4.

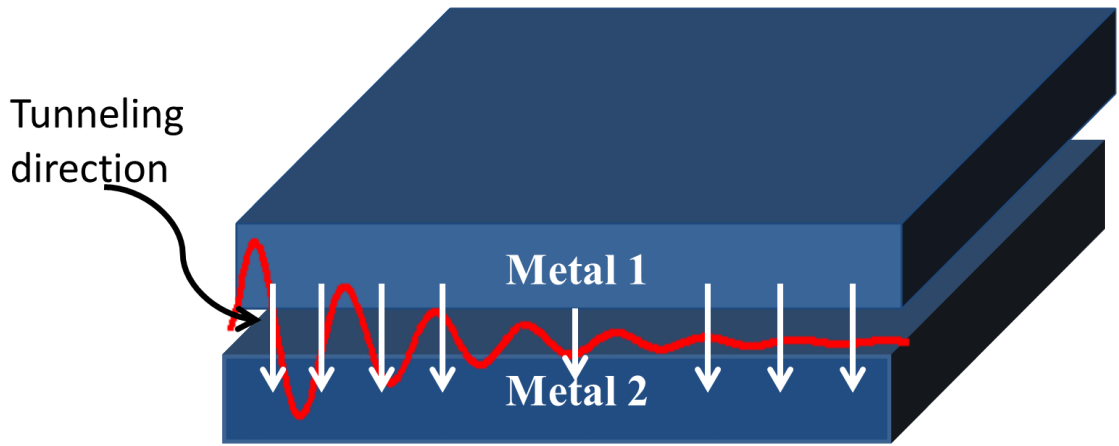

**Figure 1.7: Isometric view of the 3D traveling wave metal-insulator-metal plasmonic transmission line.**

## 1.5. Thesis objectives and contribution

A number of objectives have been indentified for this research. These objectives are listed below:

First, modeling the current throughout the MIM structure and providing a reliable model that can be used for design purposes. The current is modeled using the Transfer Matrix Method and the results are validated with the Non Equilibrium Green's Functions approach. Using the developed models, the figures of merits of the MIM diode are investigated. Parametric study had been performed, which facilitates the selection of metal types and insulators for best diode's resistance, responsivity, non-linearity, and asymmetry.

Second, studying traveling wave plasmonic transmission lines with different topologies. The full set of characteristics of each waveguide, has been obtained using well-known commercial software packages. The characteristics include: effective refractive index, attenuation constant, propagation length, and characteristic impedance. The simulations are carried out using the FEM solver and validated with FDTD solver. Full parametric study and design curves are prepared, which represent great aid to the design process of a complete rectenna based on the travelling wave rectifier concept.

Third, the integration of the proposed travelling wave rectifier with dipole nantenna has been performed successfully. Two types of plasmonic transmission lines are used in the integration that leads to two versions of the proposed dipole rectennas. A systematic approach for achieving best coupling between system' components is proposed. The two rectenna versions are thoroughly compared and the overall system efficiency for both of them are predicted.

## 1.6. Organization of the thesis

In chapter 2, the models used to calculate the current through the MIM structures are presented. The tunneling current calculation is based on the Transfer Matrix Method based on the Airy Function (AF-TMM). Also, the current calculation is validated using the Non-Equilibrium Green's Function (NEGF). In addition, the figures of merit of the devices, such as: responsivity, resistance, nonlinearity and asymmetry, are studied in details in this chapter.

In chapter 3, different traveling wave metal insulator metal plasmonic transmission lines are studied. These transmission lines are: vertical coupled strips, lateral coupled strips, and nanostrip waveguides. These transmission lines are fully characterized and compared. The study performed in this chapter has led to design curves that can be found attractive for travelling wave rectenna designers. Each of the three proposed plasmonic transmission lines in this chapter is integrated later to a nantenna.

In chapter 4, nanodipoles fed by vertical and lateral coupled strips are investigated. A full parametric study for the dipole nantenna is presented and discussed. A systematic approach to match the components of the system is adopted in this chapter. This lead to a complete system suitable for harvesting infrared solar energy. The figures of merit of the overall system, such as: nantenna harvesting efficiency, matching efficiency, rectifier's responsivity and the overall system efficiency, are defined and obtained.

Chapter 5 presents a nanocrescent antenna that can be used for optical communication systems. This nantenna is characterized by two resonance frequencies. Thus it can cover a wider spectrum than nano-dipole antennas. The study shows the effect of various geometrical parameters on the electric field intensity enhancement at the nantenna gap field point. The proposed nanocrescent nantenna can be terminated by nanostrip waveguide and can operate at the LWIR for infrared solar energy harvesting.

The thesis is concluded in chapter 6.

# Chapter 2 Analysis of Metal-Insulator-Metal Tunneling Diodes

## 2.1. Introduction

The study of tunneling Metal-Insulator-Metal (MIM) diodes performance represents an important topic for the aim of the development of rectennas for energy harvesting and infrared detection applications. Although the interest in MIM diodes started about 50 years ago [52-55], they attracted the attention again in the last few years due to these new applications, namely; energy harvesting [56-59] and infrared/terahertz detection [12, 13, 60]. The diode characteristics depend greatly on the choice of metals, insulator materials, and insulator thicknesses. The thickness of the insulator should be kept below 4 nm to keep the tunneling current as the main transport in a MIM diode [55]. For a low barrier height and greater than 4 nm insulator thickness, the thermionic transport dominates [4]. The figures of merit of these MIM diodes are: high degree of asymmetry, responsivity, and nonlinearity in addition to low diode resistance. The asymmetry is achieved when the two metallic layers are made of different materials; otherwise, the diode is symmetric. At zero bias, high responsivity is achieved by using high barrier asymmetric diodes. High nonlinearity is achieved by using high barrier heights with thick insulator layer. Low resistance can be reached by using low barrier heights in both sides with thin insulator layer. Unfortunately, the required figures of merit cannot be achieved in a single MIM diode. Nonetheless, they can be optimized by adding another insulator layer, Metal Insulator Insulator Metal (MIIM). MIIM contains a low barrier insulator and a high barrier one.

Different analytical formulas for the tunneling transmission probability through MIM diodes were developed based on WKB approximation [53-55, 61, 62]. However, the WKB does not take into consideration the wave function reflections at the interface between different layers. Therefore, other models are required to simulate the tunneling probability. Non Equilibrium Green Function (NEGF) numerical method is used by different authors to calculate the tunneling transmission probability [63-65]. It is an accurate numerical method but it needs long time of calculations on a PC in comparison to analytical models. Transfer matrix method (TMM) is a well known method used to analyze the propagation, reflection and transmission of waves through different layers of materials. The calculations using this method have been done before assuming constant potential in each layer or sub layers [66, 67]. Such constant potential leads to a simple formulation of the resultant wave equations when solving Schrödinger equation. Expressing the electron wave function as a plane wave is no longer valid in case of a constant electric field (constant gradient of potential), where the importance of the use of Airy functions [68] here glows up. Some previous works used this approach for calculating the wave function for a quantum well [69], resonant tunneling diodes [70] and waveguides [71, 72].

In this Chapter, a detailed quantum mechanical modeling of the tunneling current through MIM diodes is presented. An analytical expression for the tunneling transmission probability is derived using the Airy Functions based Transfer Matrix Method (AF-TMM) for any number of insulator layers. The simulation time is short in comparison to other numerical simulation times as it is based on analytical expressions rather than numerical methods. This is a great advantage to our simulator as it can be used as a tool to expect, before fabrication, the behavior of any MIM diode. We compare the AF-TMM results with that of the Non Equilibrium Green's Function (NEGF), and the well known WKB method. Comparisons between the predictions of our model with antenna-MIM diode responses published in previous works are also done to check its validity. To our knowledge, this is the first work, theoretically investigating the MIM diode performance due to the insulator thickness change, and the use of different metals at the left and right electrodes.

This chapter is organized as follows: in Section 2.2, the AF-TMM and the NEGF approaches are described in detail. The governing equations and their numerical implementation of these two approaches are outlined. The material parameters used in the simulation are summarized and the simulation results are presented in Section 2.3. The two theoretical approaches are compared, and their advantages and disadvantages are discussed. The comparison with experimental results also takes place in this section. The effect of work function differences and insulator thicknesses on the diode performance are discussed in Section 2.4. Also the effect of the bias voltage on the peak value of the diode responsivity and nonlinearity is discussed in this section. The effect of using two insulator layers on enhancing the rectifier performance is discussed in Section 2.5. Finally, the conclusion and summary of this chapter is presented in Section 2.6.

## 2.2. Theoretical models

### 2.2.1. Tunneling current model

The MIM diode structure consists of insulator layers sandwiched between two metals. The shape of the tunneling barrier is determined by the work function difference between these two metals, the applied bias voltage, and the electron affinities and thicknesses of the insulators. The 1D schematic band diagram of the device under applied voltage bias, $V_b$, is shown in Fig. 2.1. Where, $V_j$ is the potential difference across the $j^{\text{th}}$ insulator layer, and is given by:

$$V_j = \frac{d_j / \varepsilon_j}{\sum_{k=1}^{N} d_k / \varepsilon_k} (V_b + \phi_R - \phi_L) \tag{2.1}$$

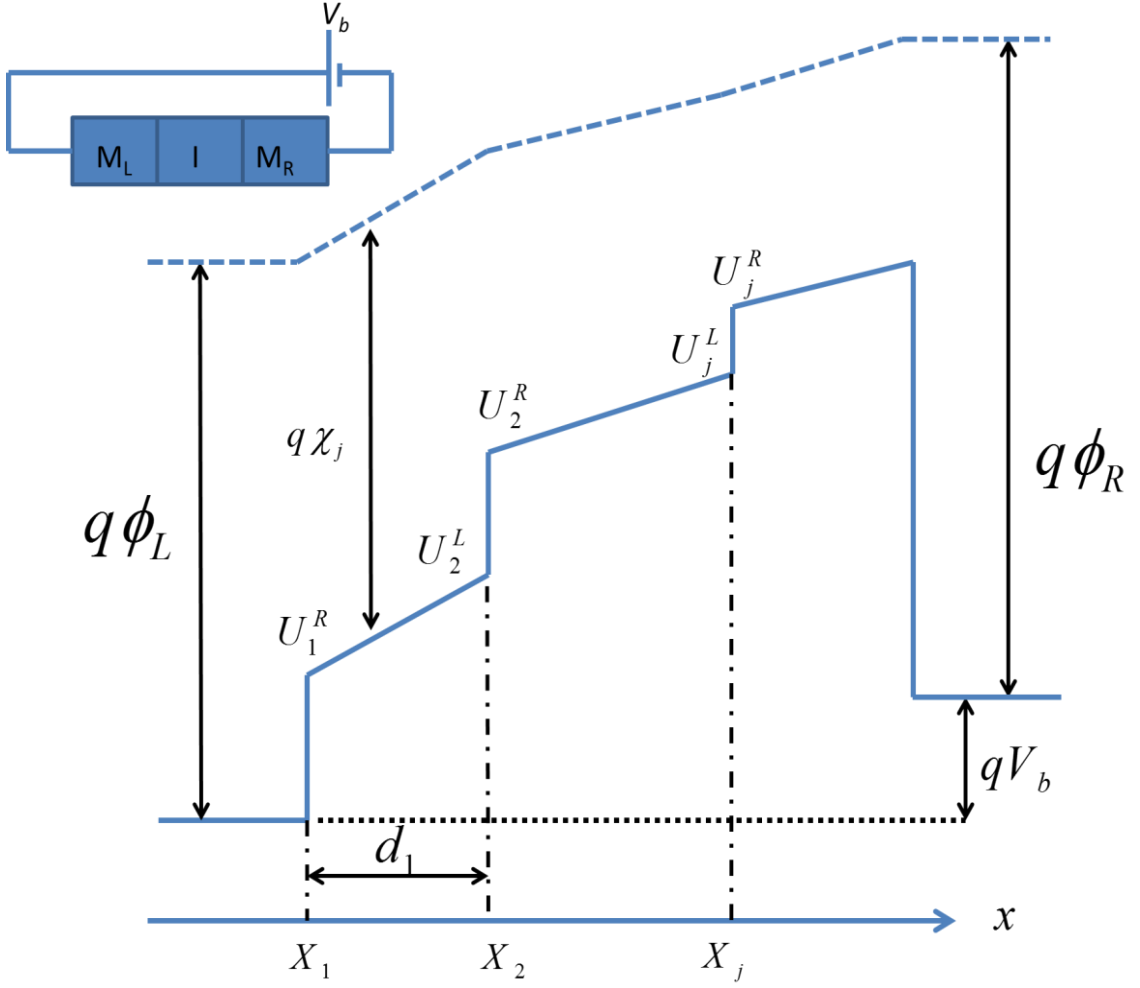

**Figure 2.1: The potential of a stack of N insulator materials under applied bias, $V_b$. Each insulator layer is characterized by a barrier height ($U_j$), a thickness ($d_j$), a dielectric constant ($\varepsilon_j$), and an effective mass ($m_j$).**

where $d_j$ and $\varepsilon_j$ are the thickness and dielectric constant of the $j^{th}$ layer respectively, $q\phi_L$ and $q\phi_R$ are the work functions of the left and right metal electrodes respectively.

With the assumption that the electron tunneling energy and the momentum transverse component are conserved, the tunnel current density is given by [55]:

$$J = J_{Forward} - J_{Backward} \qquad (2.2)$$

where

$$J_{Forward} = \frac{4\pi m_L^2 q}{h^3} \int_0^{E_m} T(E_x) \int_0^\infty f_L(E) dE dE_x \qquad (2.3a)$$

$$J_{Backward} = \frac{4\pi m_R^2 q}{h^3} \int_0^{E_m} T(E_x) \int_0^\infty f_R(E - qV_b) dE dE_x \quad (2.3b)$$

where $E$, $E_x$ and $T(E_x)$ are the total energy, the transverse energy of a tunneling electron, and the transmission probability respectively. $m_L$ and $m_R$ are the transverse electron effective masses in the left and right electrodes, respectively. $f_L$ and $f_R$ are the Fermi-Dirac distributions of the left and right metal electrodes.

For an efficient rectifier, the diode responsivity (curvature coefficient), $\frac{1}{2}\left(\frac{d^2I}{dV^2}/\frac{dI}{dV}\right)$, the nonlinearity, $\left(\frac{dI}{dV}/\frac{I}{V}\right)$, and the asymmetry $(I_+/I_-)$ in the current voltage characteristics of the MIM diode to be increased, while the diode resistance $\left(\frac{dI}{dV}\right)^{-1}$ should be decreased. $I_+$ and $I_-$ are the tunnel current under positive and negative applied bias voltage, respectively.

The 1D time-independent single-particle Schrödinger equation is given by:

$$-\frac{\hbar^2}{2}\frac{d}{dx}\left[\frac{1}{m}\frac{d\psi(x)}{dx}\right] + U(x)\psi(x) = E_x \psi(x) \quad (2.4)$$

where $\hbar$ is the reduced Plank constant, $\psi(x)$ is the electron wave-function, $m$ is the effective mass and $U(x)$ is the potential energy.

### 2.2.2. AF-TMM equations

In general, the MIM diode is assumed to contain $N$ insulator layers as shown in Fig. 2.1. The transmission probability $T(E_x)$ is calculated by assuming a one-band parabolic dispersion model for all the insulator layers. The solutions of the Schrödinger equation within the left and right metal regions are:

$$\psi_L(x) = A_L e^{ik_L x} + B_L e^{-ik_L x}, \quad x < X_1 \quad (2.5)$$

$$\psi_R(x) = A_R e^{ik_R x} + B_R e^{-ik_R x}, \quad x > X_{N+1} \quad (2.6)$$

where $\psi_L, \psi_R$ are the wave functions attributed to the left and right electrodes respectively, $k_{L,R}$ are the electron wave vectors of the plane wave functions in the left and right electrodes, respectively, and are defined as follows:

$$k_L = \sqrt{2m_L E_x}/\hbar \quad (2.7a)$$

$$k_R = \sqrt{2m_R(E_x - eV_b)}/\hbar \quad (2.7b)$$

The potential profile contains $N+1$ spaced grid points. By applying the normalization factor at each interface using the variable $u_j$ as follows:

$$u_j = s_j \left( \frac{2m_j}{\hbar^2} |F_j| \right)^{\frac{1}{3}} \left( \frac{U_j^L - E_x}{F_j} + (x - X_j) \right) \tag{2.8}$$

where $F_j = \dfrac{U_{j+1}^L - U_j^R}{X_{j+1} - X_j}$, and $s_j = \text{sgn}(F_j)$ (2.9)

Eq. (2.4) turns to $\dfrac{d^2 \psi_j}{du_j^2} - u_j \psi_j = 0$ (2.10)

The electron wave function through the insulator can be expressed as a linear combination of Airy functions, *Ai* and *Bi* [68]:

$$\psi_j(x) = A_j Ai(u_j) + B_j Bi(u_j), \quad X_j < x < X_{j+1} \tag{2.11}$$

The matching boundary conditions necessities the continuity of the wave function $\psi(x)$ as well as its effective mass multiplied spatial derivative (i.e. $\dfrac{1}{m}\dfrac{d\psi(x)}{dx}$) at the interfaces between any two different materials. This would result in relating the individual wave functions with each other at the interfaces as follows:

$$\begin{pmatrix} A_L \\ B_L \end{pmatrix} = \begin{pmatrix} T_{11} & T_{12} \\ T_{21} & T_{22} \end{pmatrix} \begin{pmatrix} A_R \\ B_R \end{pmatrix} = T \begin{pmatrix} A_R \\ B_R \end{pmatrix} \tag{2.12}$$

where

$$T = \frac{1}{2} \pi^N \begin{pmatrix} 1 & \dfrac{m_L}{ik_L} \\ 1 & \dfrac{m_L}{ik_L} \end{pmatrix} \Pi_{j=1}^N \{M_j N_j\} \begin{pmatrix} 1 & 1 \\ \dfrac{ik_R}{m_R} & -\dfrac{ik_R}{m_R} \end{pmatrix} \tag{2.13}$$

where $M_j$ and $N_j$ are the matrices attributed to the $j^{th}$ layer and are given by:

$$M_j = \begin{pmatrix} Ai(u_j^-) & Bi(u_j^-) \\ \dfrac{u_j'}{m_j} Ai'(u_j^-) & \dfrac{u_j'}{m_j} Bi'(u_j^-) \end{pmatrix}, N_j = \begin{pmatrix} Bi'(u_j^+) & -\dfrac{m_j}{u_j'} Bi(u_j^+) \\ -Ai'(u_j^+) & \dfrac{m_j}{u_j'} Ai(u_j^+) \end{pmatrix} \tag{2.14}$$

where $u' = s_j \left( \frac{2m_j}{\hbar^2} |F_j| \right)^{\frac{1}{3}}$, and the "−" and "+" symbols are attributed to the left and right sides of a specific insulator layer where $u_j^- = u_j(x = X_j)$, $u_j^+ = u(x = X_{j+1})$.

To find simple expressions for the probability amplitude, initial conditions are applied. Suppose the electron is incident from the left with $A_L = 1$, and there is no reflected traveling wave in metal 2, i.e. $B_R = 0$. It follows that $A_R = 1/T_{11}$. So, the tunneling transmission probability $T(E_x)$ can be computed as follows:

$$T(E_x) = \frac{m_L k_R}{m_R k_L} \frac{1}{|T_{11}|^2} \tag{2.15}$$

### 2.2.3. NEGF equations

Assuming the insulator layers are divided into $M$ grid points with uniform spacing, $a$. Finite difference discretization on the 1D grid is applied to Schrödinger equation eqn. (2.4) at each node $i$ as follows [63]:

$$-t_{i-1}\psi_{i-1} + (t_{i-1} + t_i + U_i)\psi_i - t_i \psi_{i+1} = E_x \psi_i \tag{2.16}$$

where $t_i \equiv \hbar^2 / 2m_i a^2$ represents the interaction between the nearest neighbor grid points $i$ and $i+1$, $U_i \equiv U(x_i)$, and $m_i$ is the electron effective mass between the nodes $i$ and $i+1$. The coupling of the potential barrier to the left and right metal electrodes is taken into consideration by rewriting eqn. (2.4) for $i=1$ and $i=M$ with open boundary conditions expressed at Metal-1/Insulator and Insulator/Metal-2 interfaces. So, Schrödinger equation now takes the following form [63]:

$$[E_x I - H - \Sigma_L - \Sigma_R]\psi = S \tag{2.17}$$

where $H$ is the $M$ x $M$ Hamiltonian matrix of the insulator potential, $I$ is the $M$ x $M$ identity matrix, $\psi$ is the wave-function $M$ x 1 vector and $S$ is $M$ x 1 vector. $\Sigma_L$ and $\Sigma_R$ are the $M$ x $M$ self energies of the left and right contacts respectively. $H$ can be rewritten under a tri-diagonal form:

$$H = \begin{bmatrix} t_1 + t_L + U_1 & -t_1 & 0 & \cdots & 0 \\ -t_1 & t_2 + t_1 + U_2 & -t_2 & \ddots & \vdots \\ 0 & -t_{i-1} & t_i + t_{i-1} + U_i & -t_i & \\ \vdots & \ddots & -t_i & \ddots & \ddots \\ & & & & \ddots \\ 0 & & 0 & \ddots & \end{bmatrix}$$

(2.18)

$\Sigma_L$ and $\Sigma_R$ are given by:

$$\Sigma_L = \begin{bmatrix} -t_L e^{ik_L a} & 0 & \cdots & 0 \\ 0 & 0 & & \vdots \\ \vdots & & \ddots & \\ 0 & \cdots & & 0 \end{bmatrix}, \Sigma_R = \begin{bmatrix} 0 & \cdots & & 0 \\ & \ddots & & \vdots \\ \vdots & & 0 & \\ 0 & & 0 & -t_R e^{ik_R a} \end{bmatrix}$$

(2.19)

The solution of eqn. (2.4) is simply given by $\psi = G(E_x)S$ where $G(E_x)$ is $M$ x $M$ Green's function:

$$G(E_x) = [E_x I - H - \Sigma_L - \Sigma_R]^{-1} \quad (2.20)$$

The escape rate of the electron from a particular state to the left or right metal contact are taken into account by defining the two quantities; $\Gamma_L$ and $\Gamma_R$:

$$\Gamma_{L,R} = i(\Sigma_{L,R} - \Sigma_{L,R}^+) \quad (2.21)$$

Finally, the tunneling transmission probability $T(E_x)$ can be computed as follows [63]:

$$T(E_x) = Trace(\Gamma_R G \Gamma_L G^+) \quad (2.22)$$

### 2.2.4. WKB approximation

Assuming a one band parabolic dispersion relation for the tunneling electrons, the transmission probability through $N$ insulator layers using the WKB approximation is given by [62]:

$$T(E_x) = \exp\left\{-\frac{2}{\hbar} \sum_{j=1}^{N} \int_{X_j}^{X_{j+1}} \sqrt{2m_j [E_x - U_j]} dx\right\} \quad (2.23)$$

where the integration limits existing in the tunneling transmission formula, $X_j$ and $X_{j+1}$, are the classical turning points of a certain insulator layer.

## 2.3. Models comparison

In this simulation study, the electron transport through MIM diodes of different Metal/Insulator structures is simulated. In the following, the AF-TMM, NEGF formalisms and WKB approximation are compared in terms of electron transmission probability and tunneling current through various MIM structures. In order to compare the simulation time of the two simulators, three parameters are defined: $t_{total}$, total simulation time, $t_E$, average simulation time per energy, and $t_{bias}$, average simulation time per bias point. These simulations were carried out with Intel Core i7 processor, 6MB Cache, and 8GB RAM. Also it is worth noting here that the parameter spacing, a, for the NEGF calculation is assumed equal to the hundredth of the insulator layer thickness. This was found adequate for reasonable simulation time.

The effective mass is assumed equal to the free mass of the electron throughout the MIM structure. This assumption was done as there is no consistent data available for the effective mass calculation for the studied metals and insulators with this thin thickness. Also we checked that the effective mass variation will not alter the conclusion we will derive later.

### 2.3.1. Transmission probability calculation

First, Nb is used as the left and right metal electrodes, which, from the experimental point of view, can quickly form a native oxide layer ($Nb_2O_5$) without forming an unintended interfacial layer. Two structures are simulated:

(1) A single 2 nm $Nb_2O_5$ layer, and

(2) A double layer stack made of 2 nm $Nb_2O_5$ and 1 nm $Ta_2O_5$.

Fig. 2.2 shows the tunneling transmission probability $T(E_x)$, versus electron energy calculated using the AF-TMM, the NEGF formalism, and the WKB approximation for Nb/$Nb_2O_5$/Nb MIM diode. The work function of Nb is 4.3 eV [73], and the electron affinity and dielectric constant of $Nb_2O_5$ are 4 V and 41, respectively [74]. The structure is simulated under applied bias, $V_b = 0.1$ V. A complete match between the AF-TMM $T(E_x)$ and the NEGF $T(E_x)$ for all ranges of energies is clear. That sheds light on the importance of our AF-TMM analytical equations and short simulation time with respect to the long time accurate NEGF numerical simulator.

The transmission probabilities clarifies that the WKB results differ significantly from that of AF-TMM and NEGF. Actually, the WKB approach does not take into account the reflections of the wave function at each Metal/ Insulator or Insulator/ Insulator potential discontinuity. Disregarding these discontinuities leads to the overestimation of the tunneling probability. However, these abrupt potential discontinuities are taken into consideration by the AFTMM and NEGF.

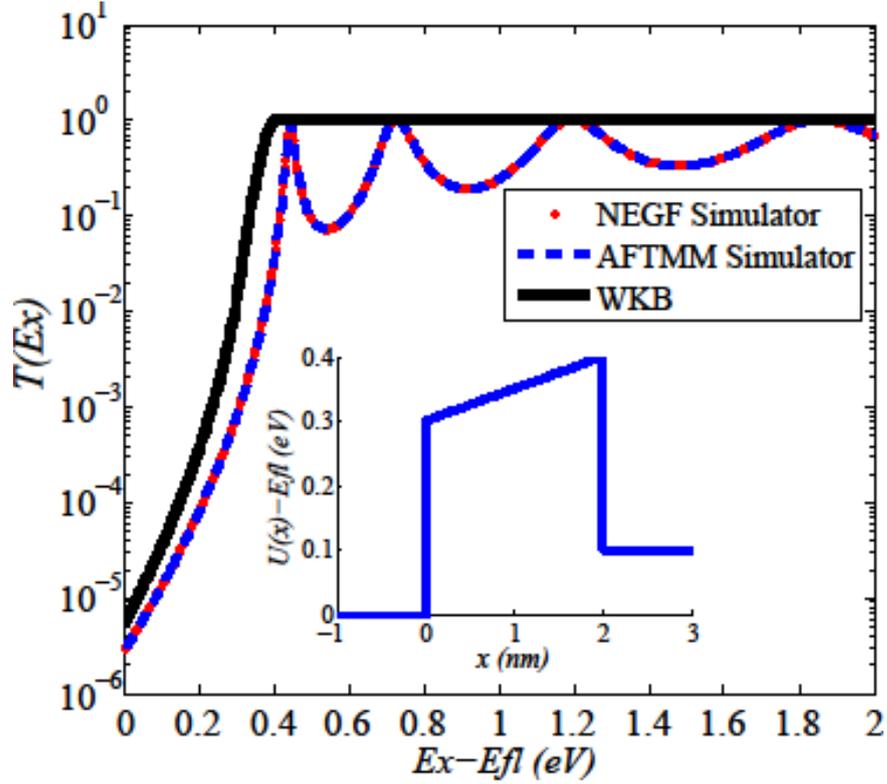

**Figure 2.2:** The transmission probability $T(E_x)$ versus the electron transmission energy calculated using AF-TMM, NEGF, and WKB at $V_b = 0.1$ V for Nb/Nb$_2$O$_5$/Nb MIM diode. The inset shows the energy band diagram of the device.

We can also observe the differences between (AF-TMM, NEGF) and WKB calculations when the electron energy is above the conduction band offset, where the WKB approach gives a tunneling transmission equal to 1, while the AF-TMM and NEGF quantum mechanical simulations give a transmission probability lower than 1 due to the reflection of the wave-functions due to the potential barrier discontinuities. Also, the simulation times, ($t_{total}$, $t_E$) for the single MIM are (30 seconds, 0.025 seconds) using the AFTMM, while they are (6 minutes, 0.35 second) in case of using the NEGF.

In addition, Fig. 2.3 shows $T(E_x)$ of the simulated Metal-Insulator-Insulator-Metal (MIIM) of Nb/Nb$_2$O$_5$(2 nm)-Ta$_2$O$_5$(1 nm)/Nb at 0.1 V applied voltage. The electron affinity and dielectric constant of Ta$_2$O$_5$ are 3.2 eV [75] and 22 [76], respectively. A complete matching between AF-TMM and NEGF results takes place like that of the previous MIM. Moreover, the times ($t_{total}$, $t_E$) for this MIIM are (50 seconds, 0.04 seconds) using the AFTMM and are (15 minutes, 1 second) using the NEGF. It is worth noting here, that the added insulating layer will result in a higher diode figures of merit as it will be illustrated in Section 2.5.

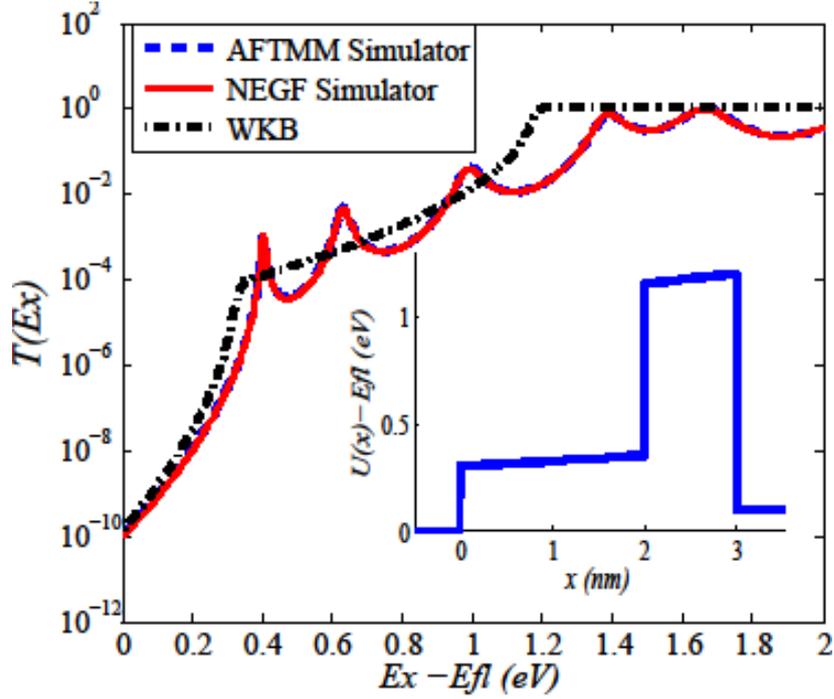

**Figure 2.3:** The transmission probability $T(E_x)$ versus the electron transmission energy calculated using AF-TMM, NEGF, and WKB at $V_b = 0.1$ V for $Nb/Nb_2O_5$-$Ta_2O_5/$ Nb MIIM diode. The inset shows the energy band diagram of the device.

## 2.3.2. Current voltage characteristics and comparison with experiments

In this subsection, we compare the results of the AFTMM and NEGF with three fabricated MIM diodes in order to validate our simulators.

First, we compare the I-V characteristics calculated using both the AF-TMM and NEGF with an I-V characteristics previously measured by K. Choi et. al. [77]. The measured I-V characteristics of a polysilicon/$SiO_2$/polysilicon is shown in Fig. 2.4 for a 60 $nm^2$ area, 2.9 eV $SiO_2$ barrier height and 1.38 nm $SiO_2$ thickness. The polysilicon used here was sufficiently doped such that it acts as a conductor. Our calculations using the AF-TMM and the NEGF methods show close matching to their measured ones. In the range of biasing $|V_b| > 0.25$ V our simulations show better matching to their experimental result than those calculated using Simmon's model [77]. This acceptable matching shows the acceptable accuracy of our models.

In addition, the results of the AFTMM and the NEGF is also compared with another Metal-Single Layer- Metal diode. The fabricated MIM is made up of a 2 nm $Nb_2O_5$ insulator layer sandwiched between two metal contacts made up of the Nb layer [66]. The results of comparing the AFTMM, NEGF and the experimental data is presented in Fig, 2.5. It is clear that the results of both the two simulators match with the points of the experimental data.

Moreover, another comparison is done with fabricated MIIM structure presented by P. Maraghechi et. al. [78]. The MIIM is made up of Cr/Al$_2$O$_3$-HfO$_2$/Cr. The insulator stack is made up of a 2 nm Al$_2$O$_3$ layer and 2 nm HfO$_2$ layer. The barrier heights of the two layer are given by 3.05 *eV* and 1.95 *eV*, respectively [78]. Fig. 2.6 shows the results of the fabricated MIIM diode, AFTMM model, and the NEGF.

As shown in Fig. 2.4, Fig. 2.5 and Fig. 2.6, both the two simulators present a complete matching with each other and a reasonable matching with the three presented experiments. However, ($t_{total}$, $t_{bias}$) for the single MIM depicted in Fig. 2.4 and Fig. 2.5 are (3 minutes, 1 second) using the AFTMM, while (52 minutes, 2 minutes) for the NEGF. Also, for the double MIIM depicted in Fig. 2.6, AFTMM takes (5.5 minutes, 2 seconds), while the NEGF takes (8 hours, 5 minutes).

These three examples, clearly illustrate that the AFTMM model can be used to simulate different MIM and MIIM structures with the advantage of its short simulation time.

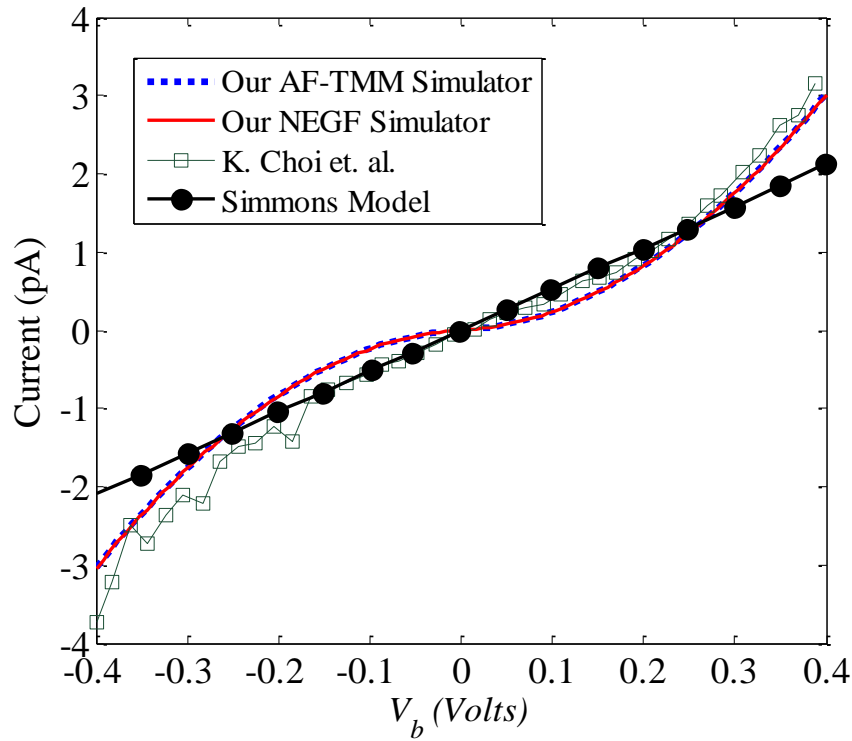

**Figure 2.4: I-V Characteristics of fabricated polysilicon/SiO2/polysilicon, Simmon's model simulation [79], AFTMM, and NEGF.**

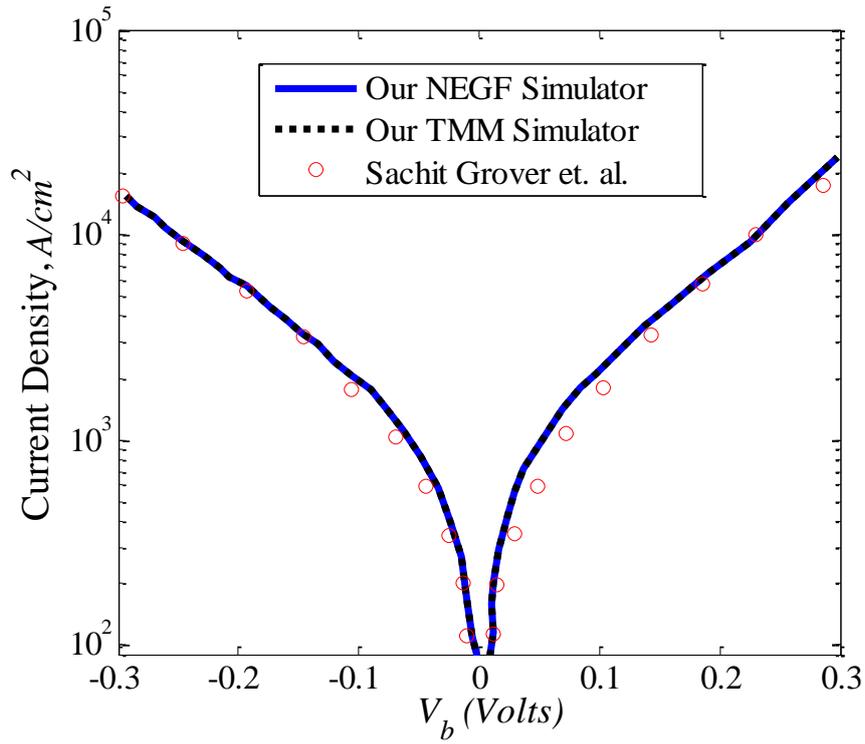

**Figure 2.5: J-V characteristics of fabricated Nb/Nb$_2$O$_5$/Nb [66], AFTMM and NEGF.**

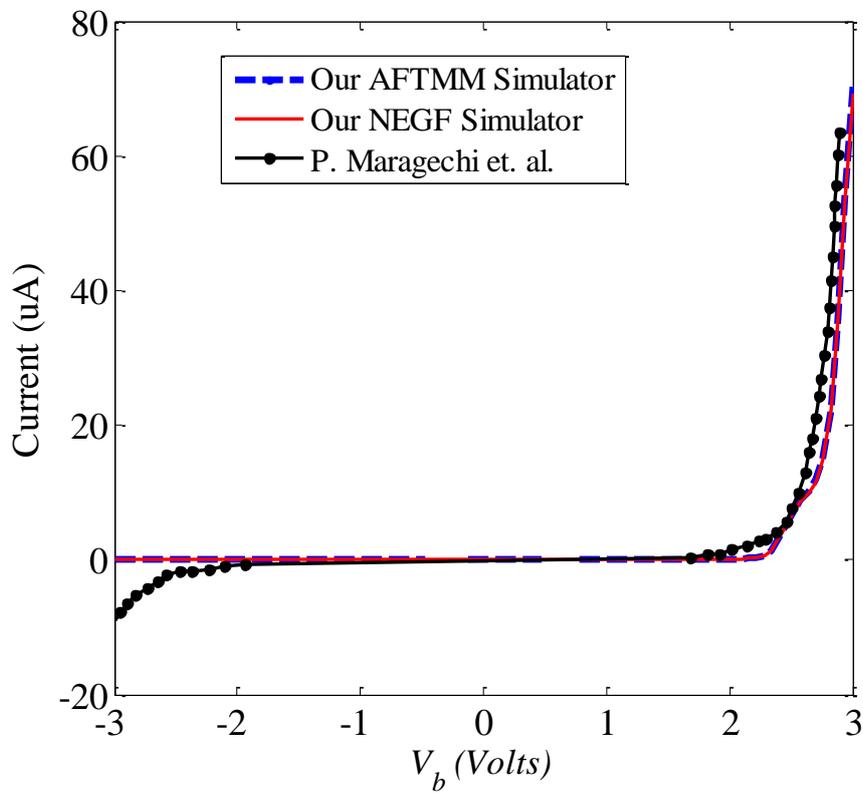

**Figure 2.6: I-V Characteristics of fabricated Cr/Al$_2$O$_3$-HfO$_2$/Cr [78], AFTMM, and NEGF.**

## 2.4. MIM performance

In this simulation study, the effect of the work function difference and insulator thickness on the MIM diode performance is presented. Nb, whose work function is 4.3 eV, is used as the left metal electrode and Nb$_2$O$_5$, whose electron affinity is 4 V, is used as the insulator [25]. The work function of the right metal is swept from 4.3 *eV* to 5.65 *eV* for different insulator thickness. Metals of work function scanning that range are: Nb (4.3 eV), Cr (4.5 eV), Cu (4.65 eV), NbN (4.7 eV), Au (5.1 eV), and Pt (5.65 eV) [73]. Fig. 2.7 shows the resistance, responsivity, nonlinearity, and asymmetry of Metal-Single insulator- Metal diode versus the barrier asymmetry $\phi_{R-L} = (\phi_R - \phi_L)$ for different insulator thicknesses at $V_b = 0.1$ V.

In general, these diode figures of merit show a higher dependence on $\phi_{R-L}$ for thick insulators than those of thinner ones. Fig. 2.7(a) shows that the resistance increases as each of the insulator thickness and barrier asymmetry $\phi_{R-L}$ increases. The rate of increase of the resistance with the increase of $\phi_{R-L}$ is very sensitive for thick insulators. Fig. 2.7(b) shows that the diode responsivity is almost constant and is independent of $\phi_{R-L}$ for thin insulator thicknesses; the responsivity is almost constant for insulator thicknesses 1 nm, and 1.5 nm. However, as the insulator thickness increases, the responsivity dependence on the barrier asymmetry increases. For a thicker insulator, the responsivity increases with the increase of $\phi_{R-L}$ till it reaches a peak value that depends on both the thickness and $\phi_{R-L}$, and then it decreases with the increase of $\phi_{R-L}$. Fig. 2.7(c) shows the diode nonlinearity, where its dependence on $\phi_{R-L}$ and the insulator thickness is similar to that of the responsivity. That is the dependence on $\phi_{R-L}$ is weak for insulator thickness less than 2 nm and is strong for thicker ones.

Also, the effect of the insulator thickness and $\phi_{R-L}$ on the asymmetry is shown in Fig. 2.7(d). It is clear that for small insulator thickness, the asymmetry is approximately 1 with low dependence on $\phi_{R-L}$. However, the increase of the thickness results in higher values of the asymmetry and higher dependence on $\phi_{R-L}$. It is worth noting here that it is recommended to use metals of work function difference $\phi_{R-L}$ less than 0.5 eV, as for metals of higher work function difference, a high resistance value will be reached. For the diode integration with the antenna, this might degrade their matching efficiency.

As mentioned before, the thickness of the insulator should be kept less than 4 nm in order for the tunneling current to dominate. Although the use of the 4 nm thickness results in better figures of merit values (except for the resistance), the DC current obtained using this device is less than those of thinner insulator thicknesses. Among the range of our simulation study, we choose an insulator thickness of 3 nm. That is a good tradeoff point between the diode figures of merit and the DC current. The maximum value of responsivity achieved at this thickness is 5.82 A/W at $\phi_{R-L} = 0.2$ eV. The feasible material of $\phi_R$ of that value is Cr, whose $\phi_R = 4.5$ eV. The figures of merit of this device at $V_b = 0.1$ V are as follows: resistance*area $\approx 3.6*10^{-6}$ $\Omega.m^2$, responsivity $\approx$ 5.82 A/W, nonlinearity $\approx 3.1$, and asymmetry $\approx 0.72$. In Section 2.5, this device will be compared to MIIM diode.

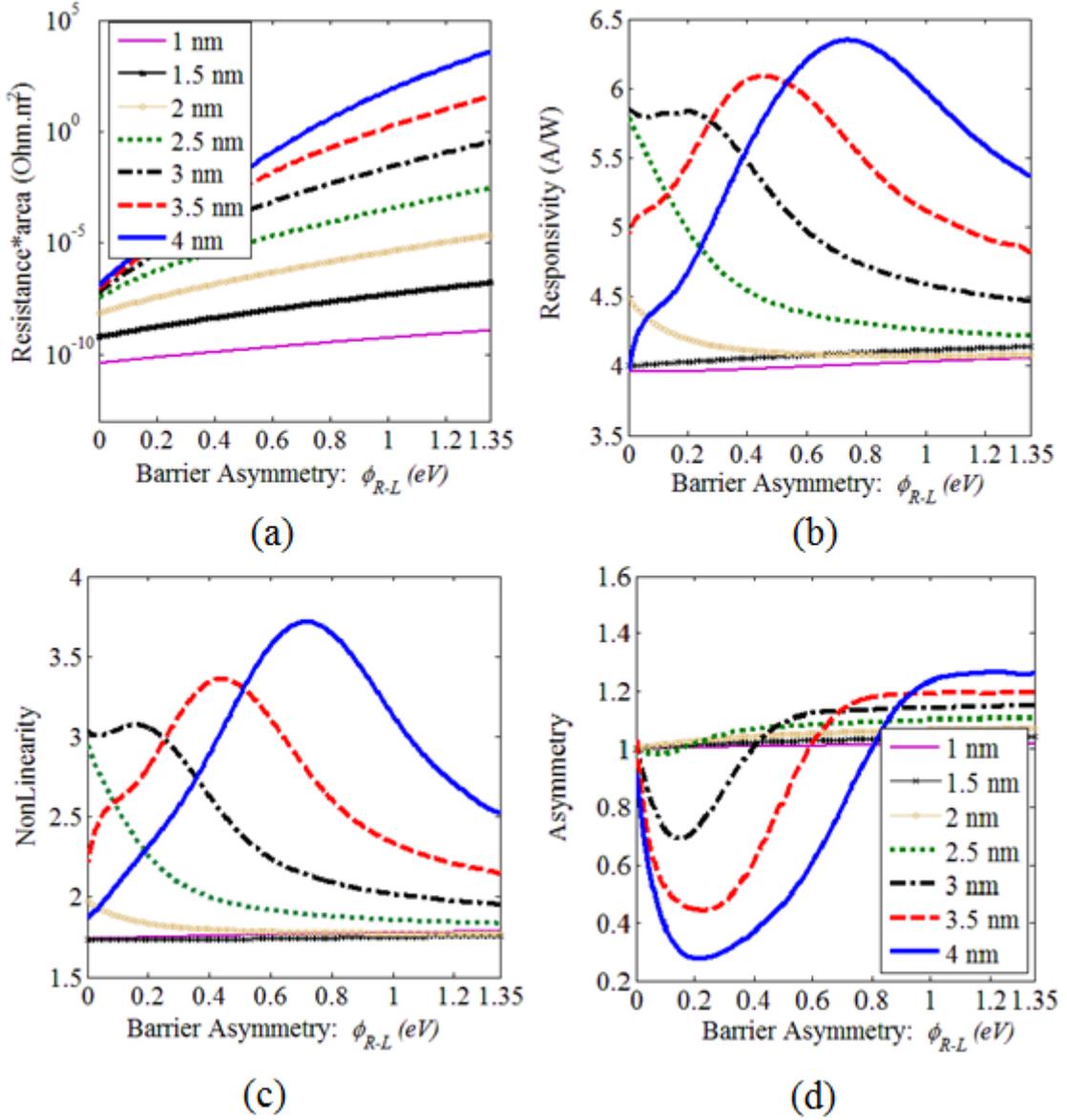

**Figure 2.7: Figures of merit of single insulator MIM diode for different insulator thickness at $V_b= 0.1$ V. (a) Resistance of the diode multiplied by area, (b) responsivity, (c) nonlinearity, and (d) diode asymmetry.**

The same calculations for the MIM diode figures of merit are done at different bias voltages. Fig. 2.8 shows such figures of merit at $V_b = 0.3$ V. It is clear that they follow the same behavior discussed at $V_b = 0.1$ V with much higher values specially for the diode nonlinearity.

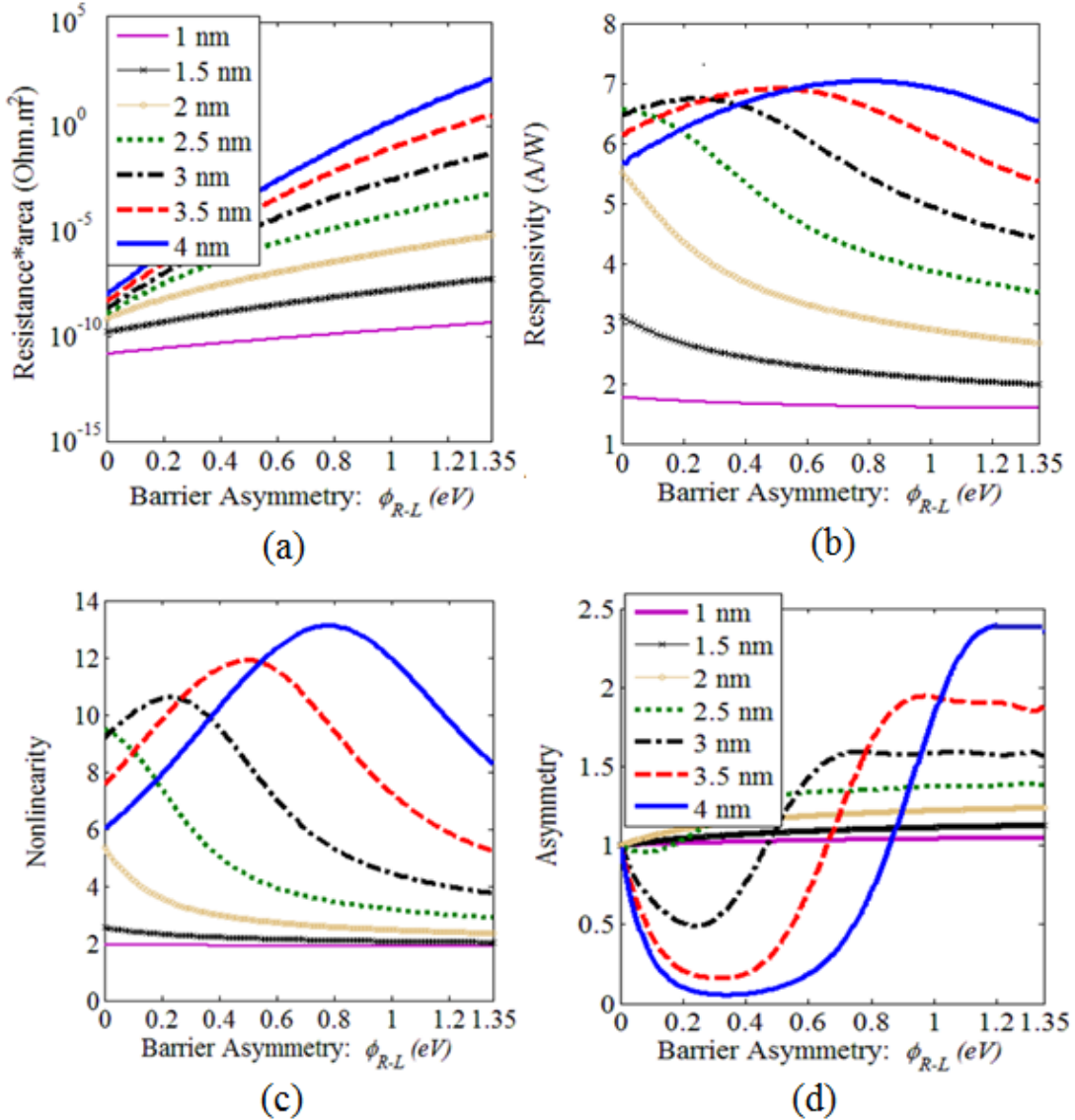

**Figure 2.8: Figures of merit of single insulator MIM diode for different insulator thickness at $V_b = 0.3$ V. (a) Resistance of the diode multiplied by area, (b) responsivity, (c) nonlinearity, and (d) diode asymmetry.**

The effect of the bias voltage on the maximum values of the responsivity and nonlinearity is shown in Fig. 2.9 and Fig. 2.10 for 3 nm, 4 nm, and 5 nm insulator thickness. It is clear from Fig. 2.9 that increasing the bias voltage results in a small improvement in the responsivity for different insulator thicknesses. However, the simulation shows that the nonlinearity, shown in Fig. 2.10, increases monotonically with the bias voltage for different insulator thicknesses. A conclusion can be drawn from here that the increase of $V_b$ might be cost ineffective in case of applications such as energy harvesting as the increase of $V_b$ will not be accompanied by a substantial increase in the MIM diode quantum efficiency.

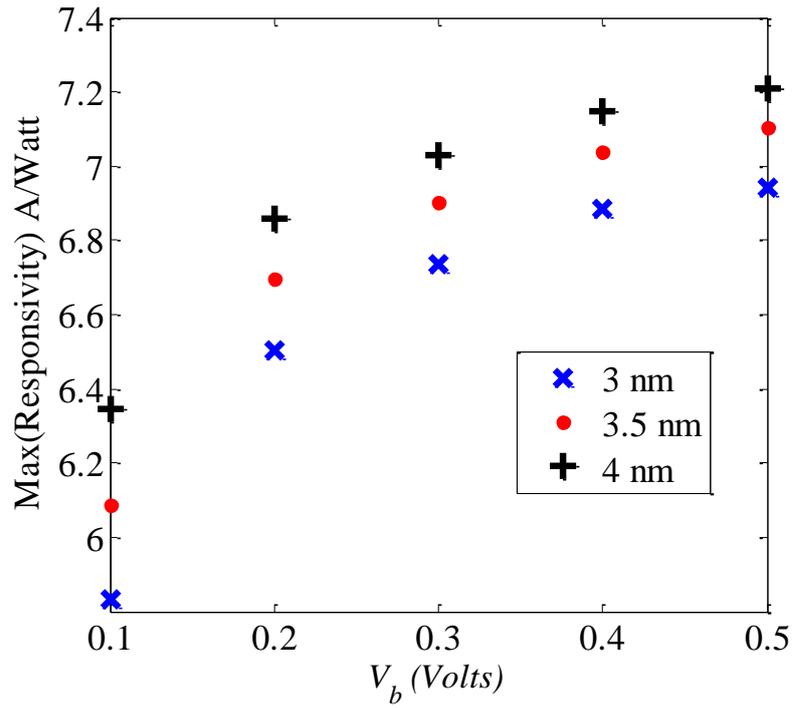

**Figure 2.9:** The maximum value of the diode responsivity versus the bias voltage, $V_b$ for different insulator thicknesses: 3 nm, 3.5 nm, 4 nm.

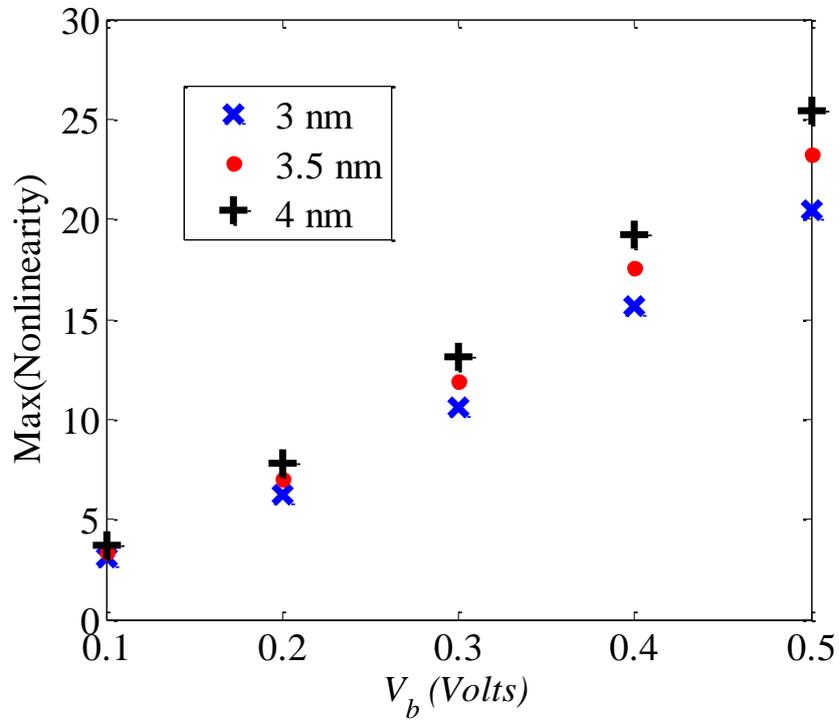

**Figure 2.10:** The maximum value of the diode Nonlinearity, versus the bias voltage, $V_b$ for different insulator thicknesses: 3 nm, 3.5 nm, 4 nm.

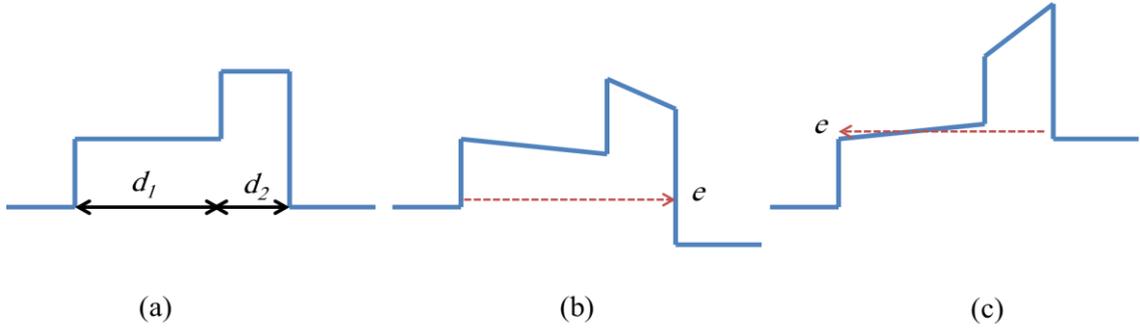

**Figure 2.11:** Energy band diagram of MIIM at (a) zero, (b) backward, (c) forward bias voltages.

## 2.5. Metal Insulator Insulator Metal (MIIM)

MIIM offers a high degree of flexibility on optimizing the parameters of the rectifiers as a high responsivity, high nonlinearity, high asymmetry, and low resistance might be achieved. At zero bias the potential barrier stack is asymmetric even using the same metal for both sides. Thus, at a certain polarity bias, the electron tunnels through a thickness less than the total thickness of the whole stack. While at the opposite polarity electron tunnels through the whole insulators thickness. This obviously corresponds to variation of the tunneling probability for the positive and negative bias voltage, as shown in the energy band diagram of Fig. 2.11. Hence, this alteration in the tunneling mechanism between different bias voltages, leads to the increase of the diode non-linearity.

On the contrary, this phenomena is not feasible in single MIM as the potential barrier is symmetric for positive and negative bias voltage in case of using the same metal at the left and the right contacts. Thus, to improve the MIM diode performance, different metal electrodes have to be used. Even though, this enhancement is less than that achieved by adding another insulator layer to the device.

Here, we propose Nb/Nb$_2$O$_5$-Ta$_2$O$_5$/Nb diode. The insulating potential barrier is made up of Nb$_2$O$_5$ layer of thickness $d_1$ and a Ta$_2$O$_5$ layer of thickness $d_2$. Fig. 2.12 shows the figures of merit of this MIIM versus $d_1$, for different values of $d_2$ simulated at $V_b = 0.1$ V. It is worth noting here that the higher asymmetry and the higher nonlinearity obtained here are due to the use of the second insulator layer as the two metal electrodes are symmetric. The responsivity, the nonlinearity, and the asymmetry can reach values greater than 7 A/W, 5 and 1.8 respectively by just adding 1 nm or thicker Ta$_2$O$_5$ layer. The best performance is obtained by optimizing the first and second insulator thickness, where the increase of $d_1$ and $d_2$ results in the increase of the figures of merit. As shown in Fig. 2.12, better values than that obtained by the MIM figures of merit could be easily achieved here by engineering MIIM through the appropriate selection of $d_1$ and $d_2$.

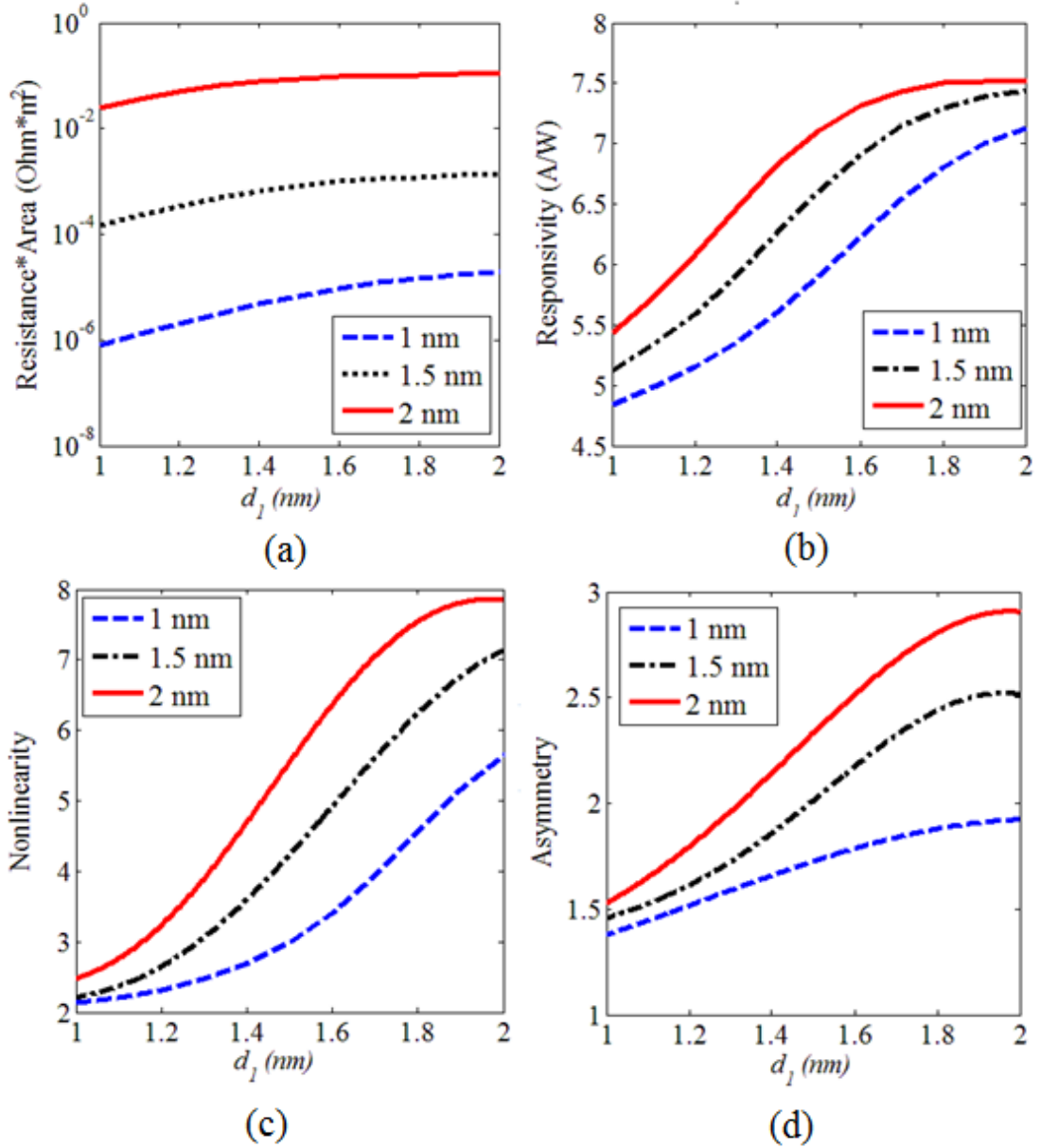

**Figure 2.12: Figures of merit of MIIM diode (Nb/Nb$_2$O$_5$-Ta$_2$O$_5$/Nb) versus the thickness of the first insulator, $d_1$, for different second insulator thickness, $d_2$, at $V_b$= 0.1 V. (a) Resistance of the diode multiplied by area, (b) responsivity, (c) nonlinearity, and (d) diode asymmetry.**

Even though the MIIM can offer better efficiency than the MIM diode, MIIM is not widely used as the single MIM. A reason for that is because, till now, most of the experimental work is based on a random selection of metals and insulators. That may result in unintended interfacial layer and degrade the device performance. So, more analysis is required to check the stability of the structures.

## 2.6. **Chapter summary**

In this chapter, a one dimensional simulator to solve Schrödinger equation for Metal-Insulator(s)-Metal diode is presented. The solution of the Schrödinger equation is based on the Airy functions due to gradient in potential along insulator layer/layers. The transfer matrix method (TMM) is used to couple all diode layers and a Matlab code is written to simulate the behavior of such diodes. The AF-TMM simulator results show complete matching with our code based on the well known Non Equilibrium Green Function method taking into consideration the great difference in calculation time between the AF-TMM analytical simulator and the NEGF numerical one. Also, the results of the simulator show reasonable matching with the fabricated MIM and MIIM diodes. This simulator is used to study the performance of the MIM diodes. The effect of work function differences and insulator thickness on the diode performance is studied. Nb/$Nb_2O_5$-$Ta_2O_5$/Nb diode shows better performance than Nb/$Nb_2O_5$/Metal diodes.

# Chapter 3 Plasmonic Transmission Lines

## 3.1. Introduction

When the size of the conventional optical devices decreases to the nanoscale, they do not allow light to propagate through them due to the diffraction limit [80]. However, the MIM plasmonic structures allow light to propagate through them, overcoming the diffraction limit, via surface plasmons. Surface plasmons are the cylindrical waves whose fields are guided by the dielectric/metal interfaces in the layer structure. Metals at such high operating frequencies behave like plasma, whose free electrons interact with the electromagnetic fields of the propagating waves [81].

In the previous chapter, it has been demonstrated that the product of the MIM junction resistance and overlapping area is in the order of $10^{-10}$-$10^{+2}$ $\Omega m^2$ depending on the selection of the metal and insulator types, and the thicknesses of the insulators. Since the nominal diode overlapping area is in the order of 100 $nm^2$, the diode junction resistance is in the order of $10^6$-$10^{18}$ $\Omega$. The nantenna impedance is in the order of hundreds $\Omega$. If the nantenna is connected to a *lumped* MIM diode, the significant impedance mismatch results in a strong reflection, which means great loss in the harvested energy. In order to overcome this problem, the *traveling wave* MIM structure, shown in Fig. 1.7, was proposed in 2006 [10].

Here, we propose three new traveling wave structures, as shown in Fig. 3.1, with the aim of integrating these lines to nantennas. The received AC voltage difference by the nantenna forces an electromagnetic wave to propagate through the MIM transmission line as shown in Fig. 3.1, independent of its topology. The voltage difference of the pure travelling wave along the line appears between the two metal sides of the distributed MIM rectifier. By making the thickness of the insulator between the two metals as small as few nano-meters, electrons can tunnel from one side to the other, creating tunneling current.

The traveling wave structure has two advantages that make it superior over the lumped element one. These advantages are listed below:

1- The impedance of the transmission line can be matched to the antenna impedance by the appropriate choice of the waveguide's cross-sectional transversal dimensions. By choosing the longitudinal length, *L* of the waveguide greater than the surface plasmon decay length ($\alpha^{-1}$) of the waveguide, no reflected wave is generated at the termination of the line, where $\alpha$ is the line's attenuation constant. Consequently, the nantenna sees the MIM waveguide as an impedance, whose value equals the characteristic impedance of the waveguide which is in the same order as the nantenna impedance, and matching can be easily achieved.

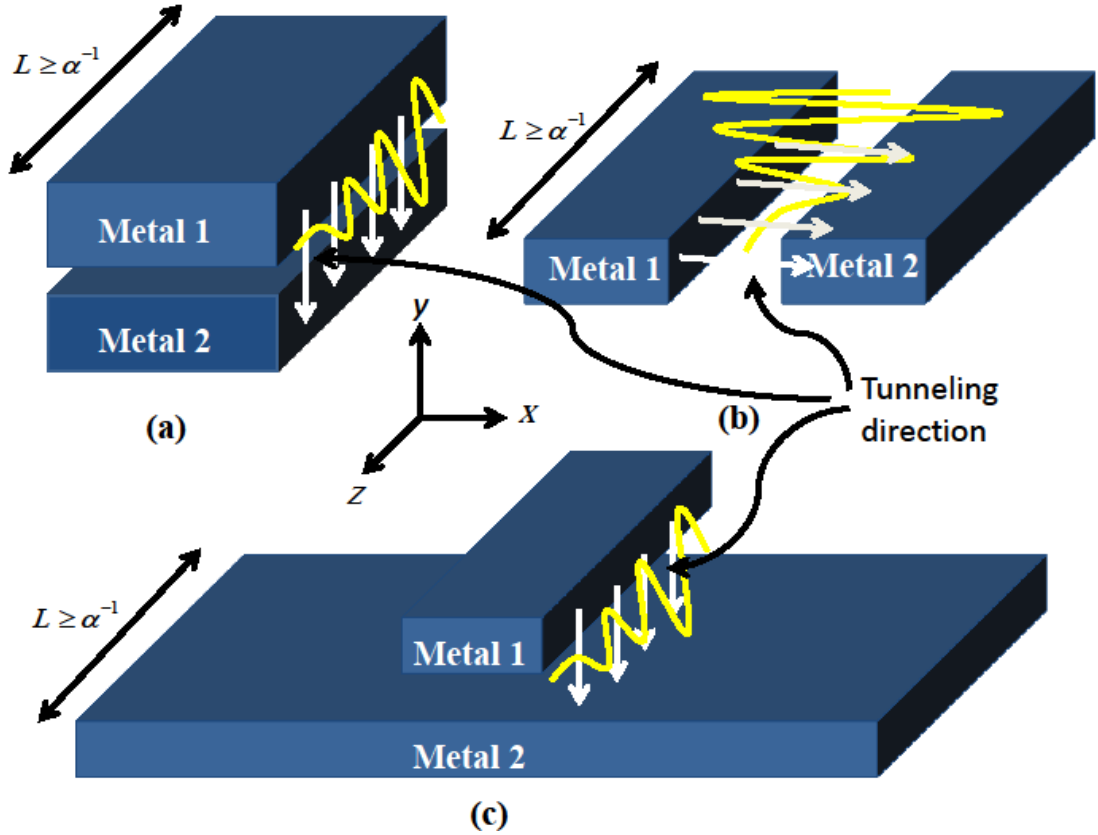

**Figure 3.1: The THz AC wave propagating through various transmission line topologies. (a) Vertical coupled strips, (b) lateral coupled strips, and (c) nanostrip waveguide.**

At distance equals to $\alpha^{-1}$ from the source, the fields should have been decayed by a factor of ($1/e$), which corresponds to a power decay by a factor of ($1/e^2$), which is equivalent to one-tenth of the incident source power.

2- Due to the absence of the equivalent capacitance, the traveling wave structure has no sharp lower cut-off frequency. However, the limitation on the operating frequency of the travelling wave MIM comes from the fact that the line should be electrically larger than plasmon decay length. This sets a mild higher cut-off frequency. Unlike the low-pass filter formed from the MIM junction in the lumped structure, the distributed high-pass filter formed from travelling wave transmission line possesses smooth transition from the pass-band to the stop-band. This results in wider operating bandwidth, which is interpreted as wider energy collection window from the infrared solar energy spectrum.

In this chapter, three plasmonic MIM waveguides shown in Fig. 3.1 are considered, namely: Vertical Coupled Strips (VCS), Lateral Coupled Strips (LCS), and the NanoStrip (NS) waveguide. For each transmission line type, the following characteristics are obtained: effective refractive index, attenuation, propagation length, and characteristic impedance. The length of these waveguides is assumed greater than the surface plasmon decay length, which results in total decay of the wave when it arrives at the open-circuit termination of the line. Hence, no reflection is expected at this termination, and a pure attenuated traveling wave is present along the line. Consequently, the input impedance of the transmission lines is equal to their characteristic impedances. Each of these three structures will be integrated to a suitable nantenna and used as solar rectenna. In order to successfully integrate the nantenna and the waveguide with maximum coupling efficiency, the nantenna impedance should be matched to the characteristic impedance of the associated waveguide.

## 3.2. Metals behavior at infrared frequencies

In order to enhance the efficiency of the MIM diode rectifier, two different metals with different work functions should be used as illustrated in Chapter 2. In this study, Gold (Au) and Silver (Ag) are selected as they lead to relatively high nantenna efficiency [82]. Metals at the optical and infrared frequency ranges behave like plasma [83]. They allow wave propagation within their volume, and they show high degree of frequency dependency, i.e. highly dispersive. The relative permittivity of Au and Ag can be expressed using the following Drude's model [84]:

$$\varepsilon_r = \varepsilon_{re} + i\varepsilon_{im} = \varepsilon_\infty - \frac{f_p^2}{f^2 + if\Gamma} \tag{3.1}$$

where $\varepsilon_\infty$ is the contribution of bound electrons to the dielectric constant, $f_p$ is the plasmon frequency and $\Gamma$ is the damping collision frequency. $\varepsilon_\infty$ for Au and Ag are 1, while $f_p$ for Au and Ag are 2.184 PHz and 2.181 PHz, respectively. The collision frequency $\Gamma$ for Au and Ag are 6.45 THz and 4.35 THz, respectively [84]. Fig. 3.2 shows the real and imaginary parts of the relative permittivity of both gold and silver versus frequency. The strong dependence on frequency can be easily noticed. The imaginary part, which is proportional to the conductivity, of gold is higher than that of silver. This indicates that gold is more lossy than silver within the selected range of frequency.

Fig. 3.3 shows the skin depth, $\delta$, for both the gold and silver, where $\delta$ is calculated as follows [85]:

$$\delta = k_0/Im(n) \tag{3.2}$$

where $k_0$ is the free space wave-number and $Im(n)$ is the imaginary part of the refractive index, n. where $n = \sqrt{\varepsilon_r}$.

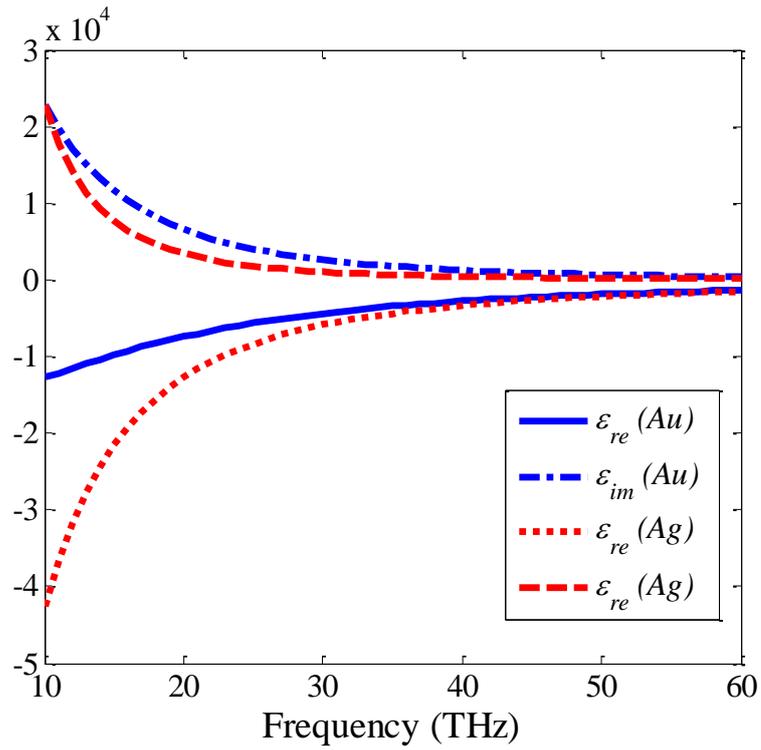

**Figure 3.2: Real and imaginary part of the permittivity of Gold and Silver.**

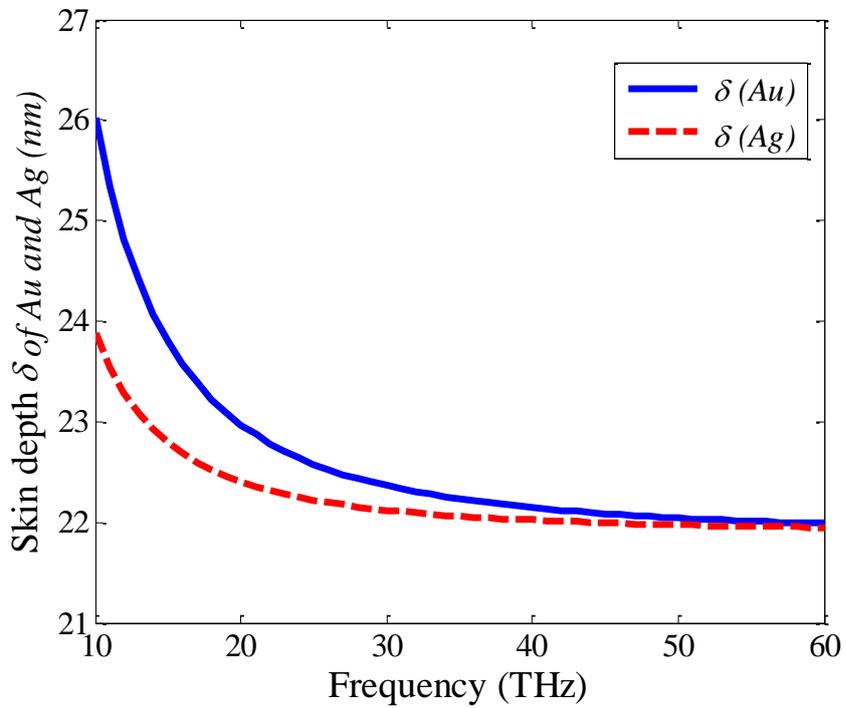

**Figure 3.3. Skin depth, $\delta$, of both the Gold (Au) and Silver (Ag).**

## 3.3. Transmission line structures

The three transmission line structures of interest in this research are shown in Fig. 3.1. These structures are known as: vertical coupled strips, lateral coupled strips, and the nanostrip line. The cross sections of these lines are presented in Fig. 3.4. Fig 3.4(a) shows the vertical coupled strips, where two metal strips are placed on the top of each other. Each of these two strips, is of width $w_1$ and height $h$. The insulating spacing between them is $d$. Fig. 3.4(b) shows the lateral coupled strips, where the same strips are located beside each other. Fig. 3.4(C) shows the nanostrip transmission line, where a metallic strip of width $w$ and height $h$ is located at distance $d$ above an unbounded metal plate.

The state-of-the-art fabrication technology allows deposition of layers with thickness down to the atomic scale, via Atomic Layer Deposition (ALD). This makes it possible to go with the insulator thickness down to 2 nm as long as insulator is stacked vertically with the two metals. This is the case for the vertical coupled strips and the nanostrip lines, while not feasible for the lateral coupled strips line as its insulator is stacked horizontally with the metals and hence the insulator is realized via electron beam lithography, whose resolution cannot go below 20 nm. Consequently, in the parametric study presented in Section 3.4, the dimensions of the three lines are swept as follows:

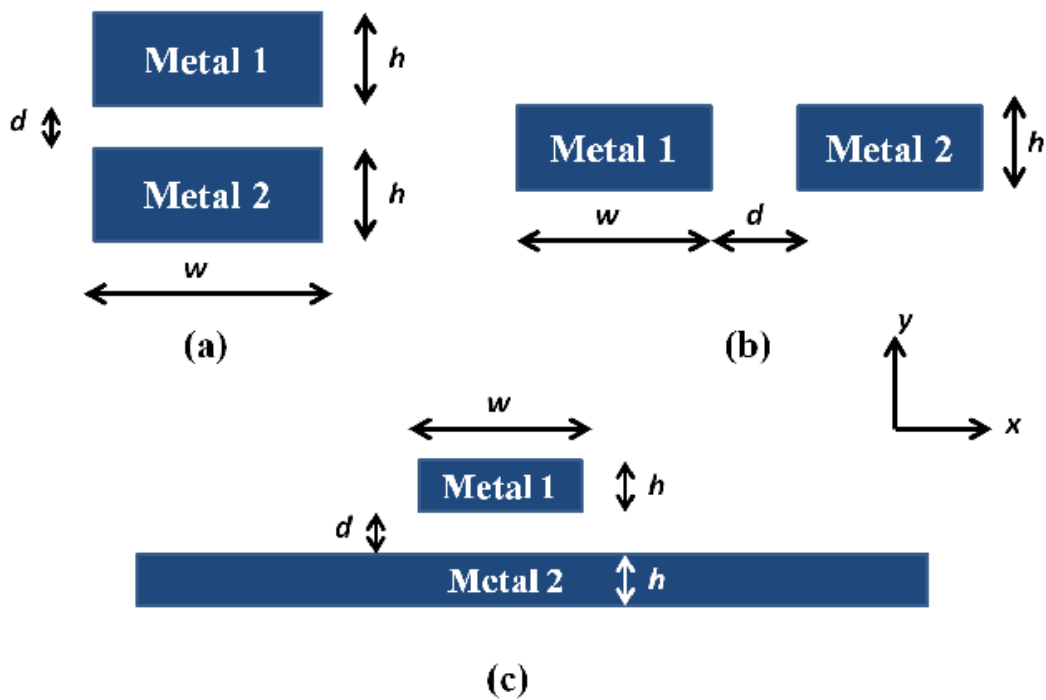

**Figure 3.4: Cross section of a) Vertical coupled strips, b) Lateral coupled strips, and c) Nanostrip transmission line.**

1. *Vertical Coupled Strips Line:* insulator thickness (*d*): 2 nm till 30 nm, strip width (*w*): 40 nm till 500 nm, and strip height (*h*): 40 nm till 200 nm.
2. *Lateral (Horizontal) Coupled Strips Line:* insulator thickness (*d*): 20 nm till 50 nm, strip width (*w*): 40 nm till 500 nm, and strip height (*h*): 40 nm till 200 nm.
3. *Nanostrip Line:* insulator thickness (*d*) 2 nm till 30 nm, strip width (*w*): 40 nm till 500 nm, and strip height (*h*): 40 nm till 200 nm.

The vertical coupled strips and lateral coupled strips surrounded by fee-space are basically the same with the strip's width and height are swapped. However, they are treated as two different transmission lines. This is due to the fabrication technology that allows for vertical spacing between the strips much smaller than the lateral spacing. This results in significant difference in the characteristics of the two lines, as illustrated in the following sections.

## 3.4. Modal field and current distributions

Each of the above three structures is supporting one fundamental TM mode, as the number of metal objects forming each transmission line is two. The distributions of the electric field magnitude calculated at 30 THz along the cross-sections of the three lines are shown in Fig. 3.5. The field distributions are calculated using COMSOL FEM Solver [86]. The electric field is concentrated mainly within the insulator between the two strips, as expected. The total current flowing through the two metallic objects is proportional to the longitudinal electric field component, according to the following equation:

$$\underline{J}_z = j\omega\varepsilon_o\varepsilon_r\underline{E}_z \tag{3.3}$$

Fig. 3.6 shows the distributions of the longitudinal electric field component along the lines' cross-sections. It is clear that currents are flowing in opposite directions in the two metallic sides of each transmission line. Unlike their behavior in the microwave range of frequencies, currents are no longer confined very close to the circumference of metals but penetrate to some extent through metal's volume. It can be noticed from Fig. 3.6 that longitudinal currents are maximum at the metal dielectric interface and decay along the perpendicular direction to that interface. The decay rate is not as fast as in the microwave transmission lines.

## 3.5. Transmission line characteristics

In this section, parametric studies performed using COMSOL are presented. They show the impact of varying the cross-sectional dimensions of each transmission line type on its characteristics. The important characteristics of each transmission line are the following:

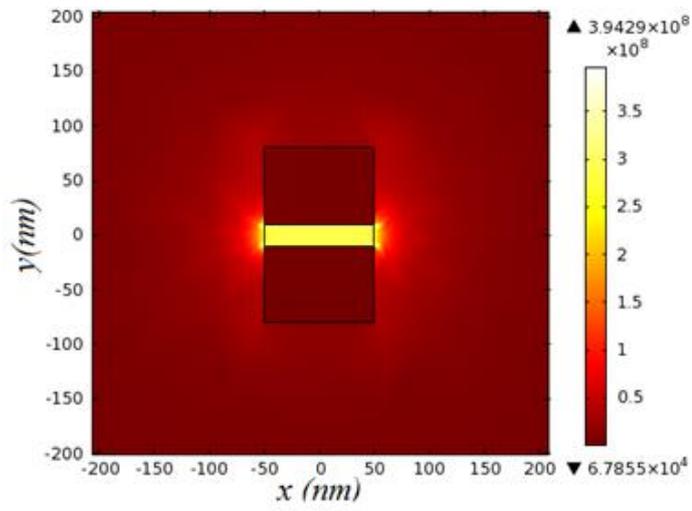

(a)

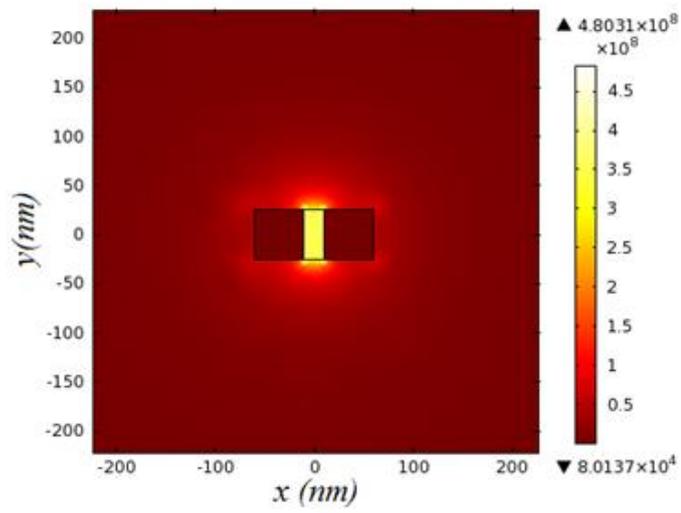

(b)

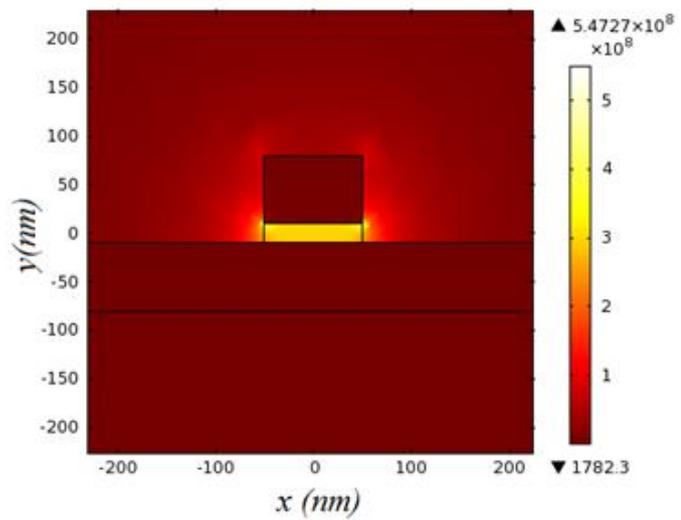

(c)

**Figure 3.5: Distribution of total electric field magnitude in: a) Vertical coupled strips, b) lateral coupled strips, c) nanostrip.**

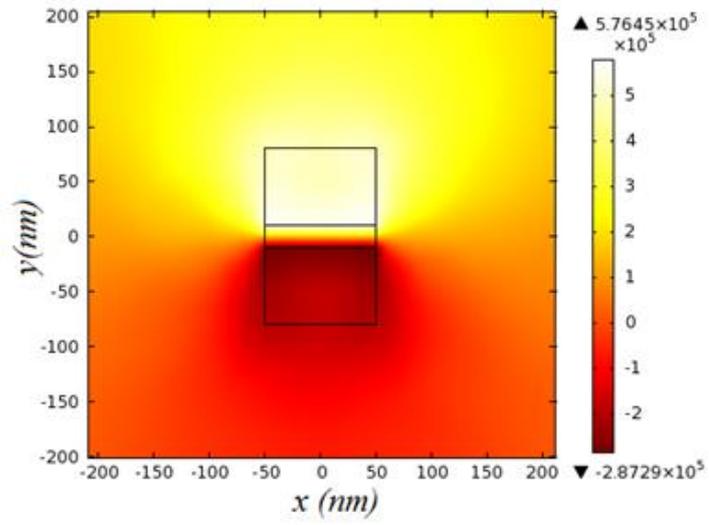

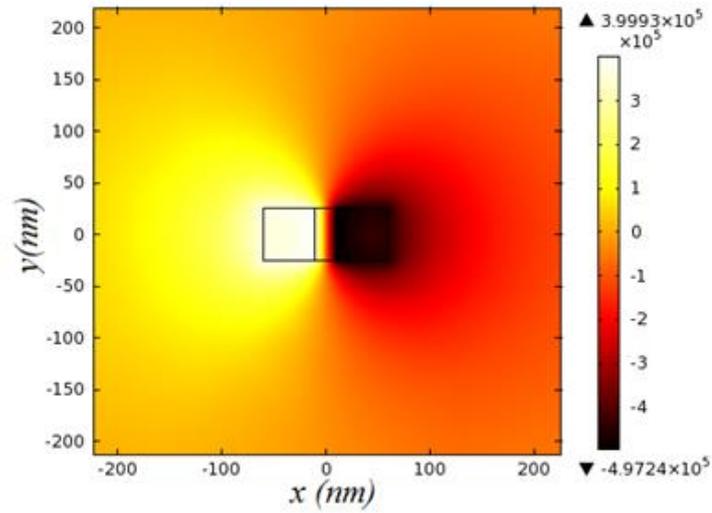

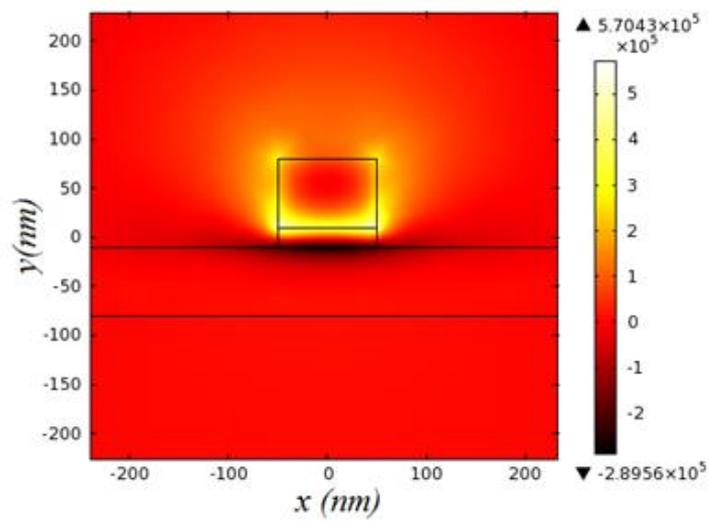

**Figure 3.6: Distribution of the electric field *z* component in: a) Vertical coupled strips, b) lateral coupled strips, c) nanostrip.**

1. Effective Refractive Index ($n_{\text{eff}}$): this quantity is defined as the refractive index of the equivalent homogenous medium that can replace the physical stack of layers within which the transmission line is immersed, without affecting the electric characteristics of the line. This can be considered as some sort of weighted average of the refractive indexes of the layers that contain the electromagnetic fields. The weights of this average are proportional to how much field is contained within the different materials. The velocity of propagation of the mode along the transmission line equals $c/n_{\text{eff}}$, where $c$ is speed of electromagnetic wave propagation in free-space, which equals $3\times10^8$ m/s.

2. Attenuation Constant ($\alpha$): this quantity specifies the rate of decay of electromagnetic fields as the guided wave is propagating along the line. All field are attenuated according to the factor ($e^{-\alpha z}$), where $e$ is Euler's constant and $z$ is the longitudinal direction. As $\alpha$ increases the wave decays more rapidly.

3. Propagation Length (plasmon decay length) ($L_{\text{prop}}$): defined as the length along which the fields decay by a factor ($1/e$). Since the attenuation factor of all fields along the line is $e^{-\alpha z}$, the propagation length equals ($1/\alpha$).

4. Characteristic Impedance ($Z_c$): it is defined as the ratio between the voltage and current associated with a pure traveling wave propagation along the line. The voltage can be calculated by integrating the component of the electric field along a path perpendicular to both metallic sides of the transmission line. The current can be calculated according to Ampere's law by integrating the magnetic field along a closed contour enclosing one of the two metallic objects forming the transmission line. The characteristic impedance of a transmission line is proportional to $\sqrt{L/C}$, where $L$ and $C$ are the equivalent inductance and capacitance of certain length from the line, respectively.

As the insulator thickness, $d$, increases the ratio of the dielectric to metal volumes carrying field increases. This results in reduction in the effective refractive index ($n_{\text{eff}}$), decrease in the attenuation constant ($\alpha$), and increase in the propagation length ($L_{\text{prop}}$), as confirmed by figures 3.(7)-3.(24). Here it should be noted that at relatively high frequency, the fields can penetrate the metal to some extent as the skin depth is not extremely small as in the microwave range of frequencies. The calculated skin depth of gold and silver at 30 THz are 22.3 and 22.2 nm, respectively. The variation of the dimensions of the strips, i.e. $w$ and $h$, has minor effect on the effective refractive index, as shown in Fig. 3.7, Fig. 3.8, Fig. 3.13, Fig. 3.14, Fig. 3.19 and Fig. 3.20, as long as the strip's height, $h$, is significantly higher than the skin depth of the metal.

As mentioned before, Fig. 3.6 reveals that the longitudinal currents are maximum at the metal/dielectric interface and decay along the perpendicular direction on that interface. The variation in the current value along the direction parallel to the interface is not as significant as along the perpendicular direction. As such, the strip's dimension along the perpendicular direction is expected to affect the attenuation constant more than the dimension parallel to the metal/dielectric interface. This is validated in Figs. 3.9 and 3.10, where the increase in strip width ($w$) has minor effect on attenuation constant ($\alpha$), while it can be seen that as the strip's height ($h$) increases, the attenuation constant decreases, as the strip gains thicker skin for the current flow. As the attenuation constant decreases with the increase in $h$, the propagation length $L_{prop}$ increases, as shown in Fig. 3.12, while it remains almost constant as $w$ varies, see Fig. 3.11.

Similar behavior of $\alpha$ and $L_{prop}$ can be observed for the lateral coupled strips and the nanostrip lines. The effective strip's dimension on $\alpha$ and $L_{prop}$ is $w$ and $h$, which is along the perpendicular to the dielectric/metal interface of the lateral coupled strips and the nanostrip line, respectively. As $w(h)$ increases, the attenuation constant decreases and the propagation length increases, while $h(w)$ has no significant effect on $\alpha$ and $L_{prop}$ for the lateral coupled strips (nanostrip) line. These expectations are validated by Figs. 3.(13)-3.(24)

As the insulator thickness ($d$) increases the equivalent capacitance ($C$) of the three transmission lines decreases as the spacing between the two metals forming the capacitance increases, which means more voltage difference between the two metals. This results in an increase in the characteristic impedance ($Z_c$), which is proportional to $\sqrt{L/C}$. This trend can be clearly seen in Figs 3.25-3.30. Similarly, the increase in the strip width ($w$), strip height ($h$), strip width ($w$) of the vertical coupled strips, lateral coupled strips, nanostrip line, respectively, results in an increase in the line's equivalent capacitance due to the increase in the total charge for the same charge density. This leads to a consequent decrease in the line's characteristic impedance, as shown in Figs. 3.25, 3.28 and 3.29.

On the other hand, the variation in the strip's dimension perpendicular to the metal/dielectric interface slightly affects the equivalent inductance ($L$) of each transmission line. As this dimension increases, the current will increase which results in a decrease in the inductance and a corresponding decrease in line's characteristic impedance. This behavior is confirmed by Figs. 3.26, 3.27 and 3.30.

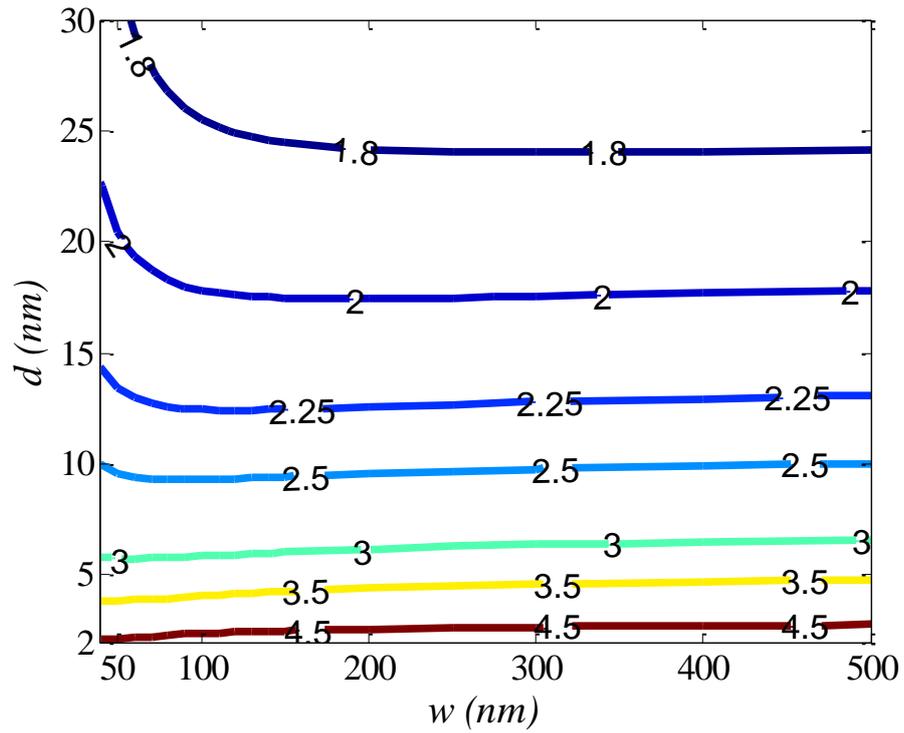

**Figure 3.7:** Contour plot showing the effect of the geometrical parameters $w$ and $d$ on the $n_{eff}$ in the vertical coupled strips at 30 THz. $h = 70$ nm.

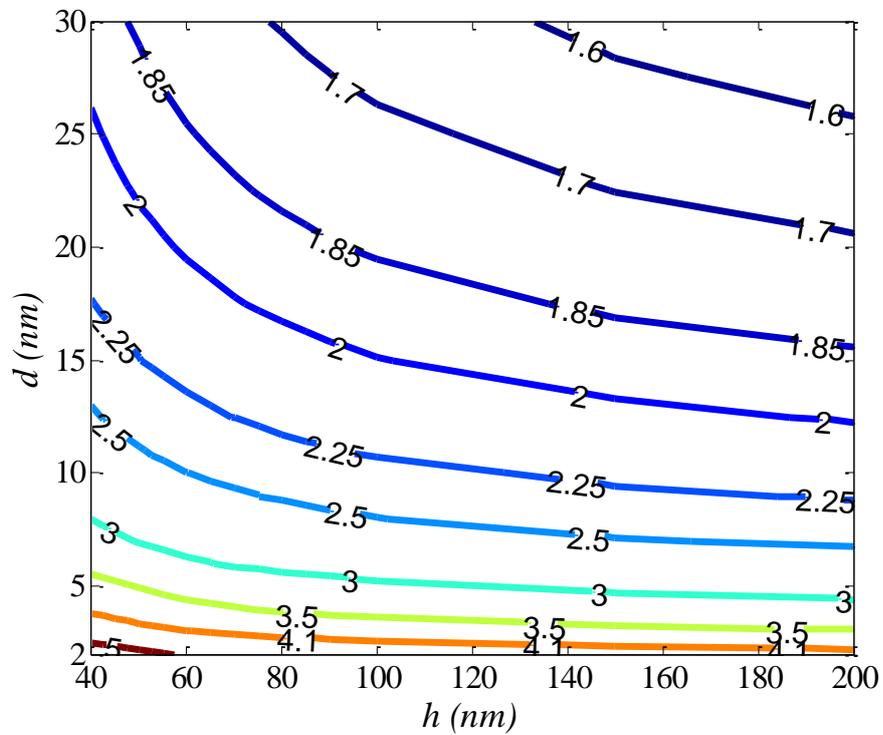

**Figure 3.8:** Contour plot showing the effect of the geometrical parameters $h$ and $d$ on the $n_{eff}$ in the vertical coupled strips at 30 THz. $w = 100$ nm.

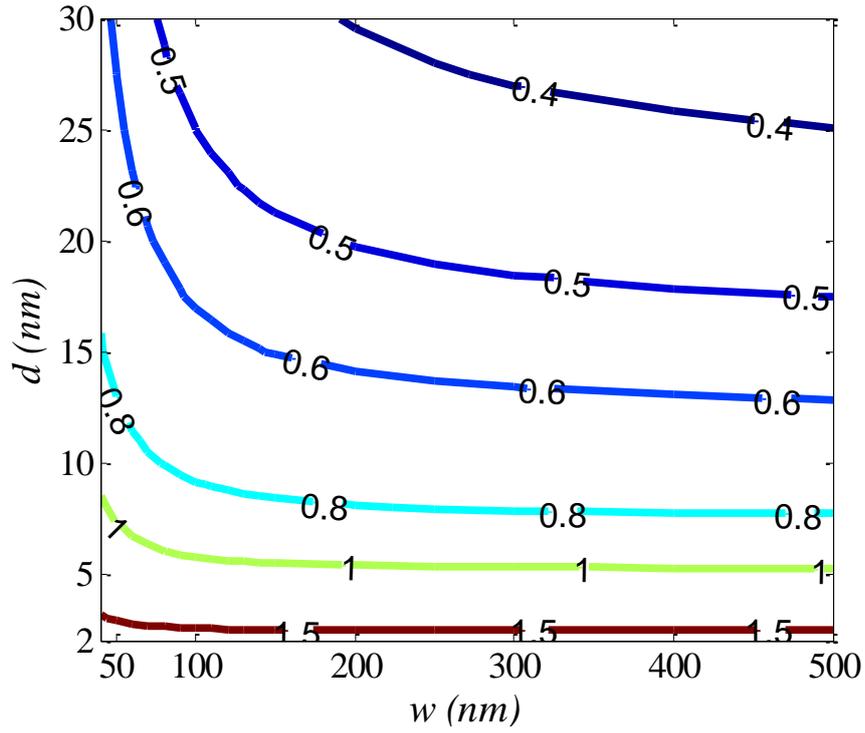

**Figure 3.9:** Contour plot showing the effect of the geometrical parameters *w* and *d* on the α *(dB/μm)* in the vertical coupled strips at 30 THz. *h = 70* nm.

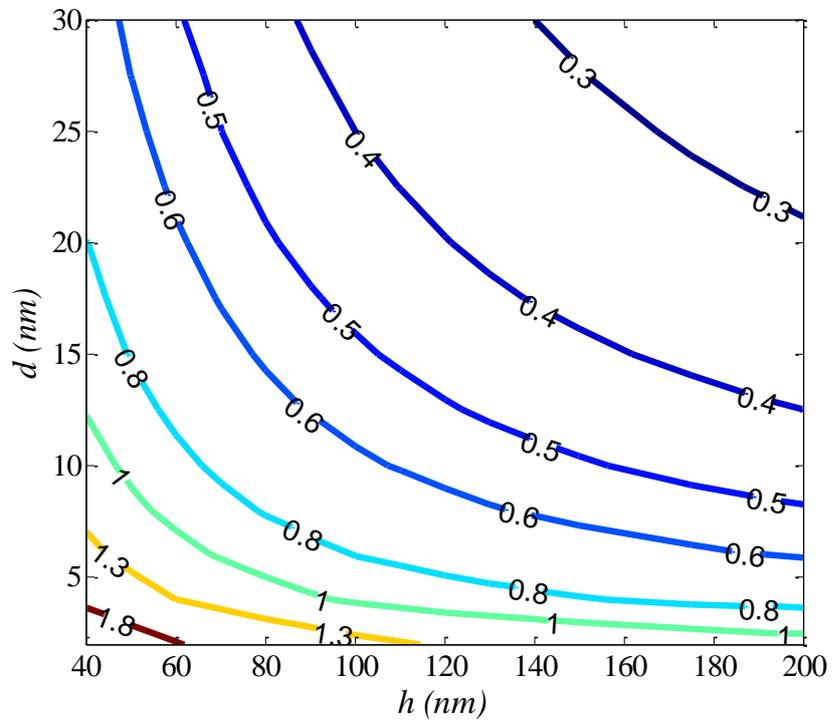

**Figure 3.10:** Contour plot showing the effect of the geometrical parameters *h* and *d* on the α *(dB/μm)* in the vertical coupled strips at 30 THz. *w =* 100 nm.

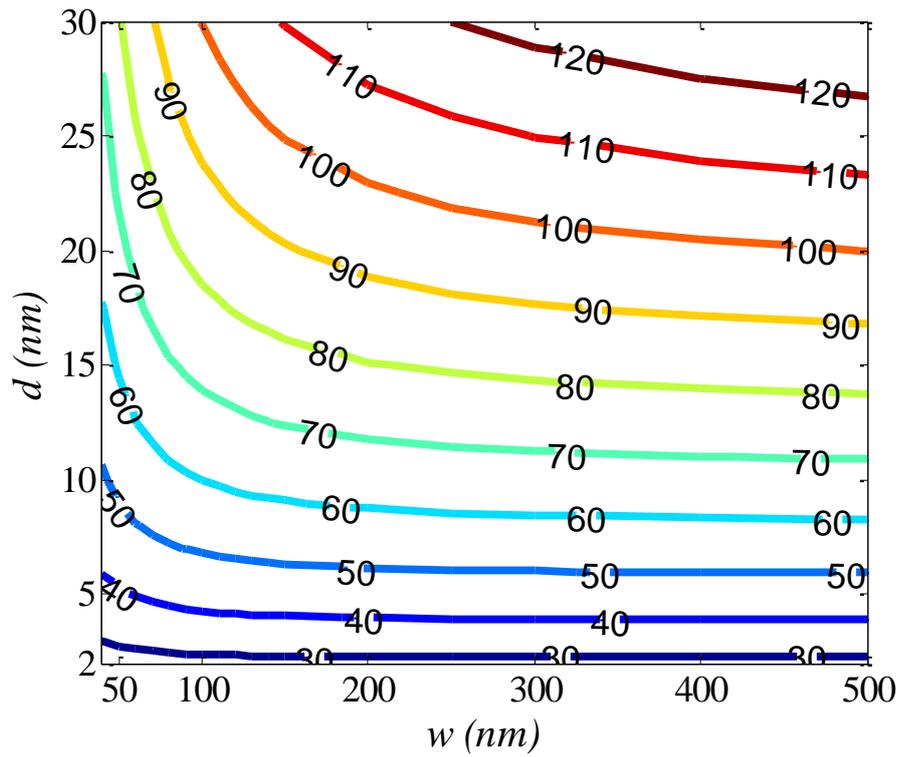

**Figure 3.11:** Contour plot showing the effect of the geometrical parameters *w* and *d* on the $L_{prop}$ (*μm*) in the vertical coupled strips at 30 THz. *h* = 70 nm.

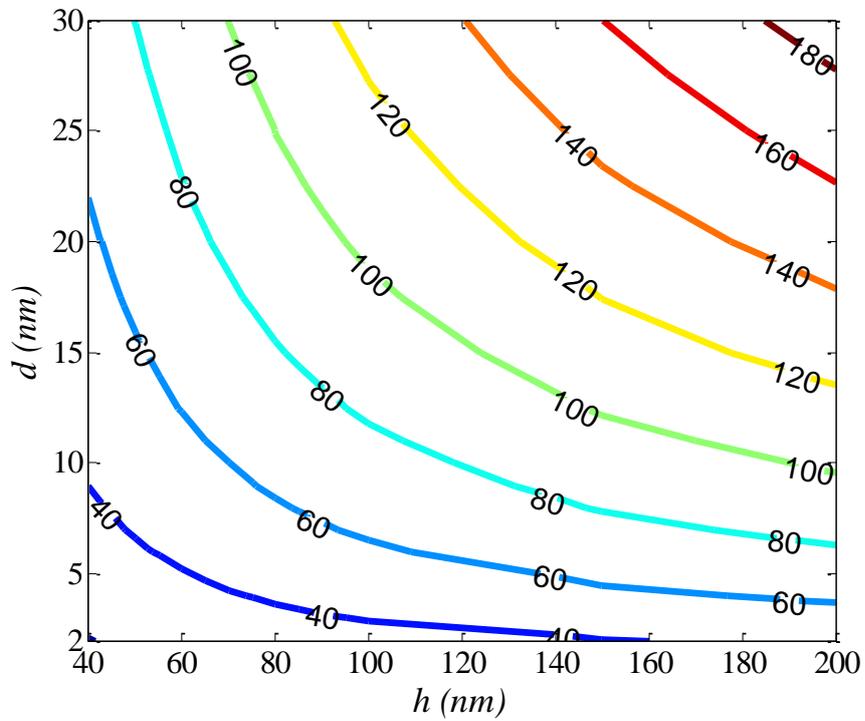

**Figure 3.12:** Contour plot showing the effect of the geometrical parameters *h* and *d* on the $L_{prop}$ (*μm*) in the vertical coupled strips at 30 THz. *w* = 100 nm.

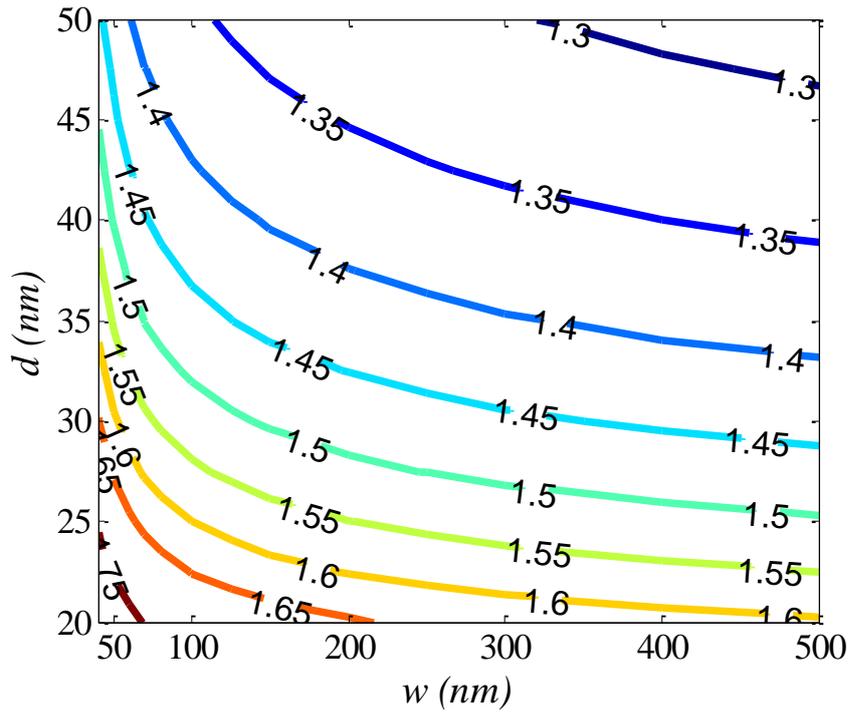

**Figure 3.13:** Contour plot showing the effect of the geometrical parameters *w* and *d* on the $n_{eff}$ in the lateral coupled strips at 30 THz. *h* = 180 nm.

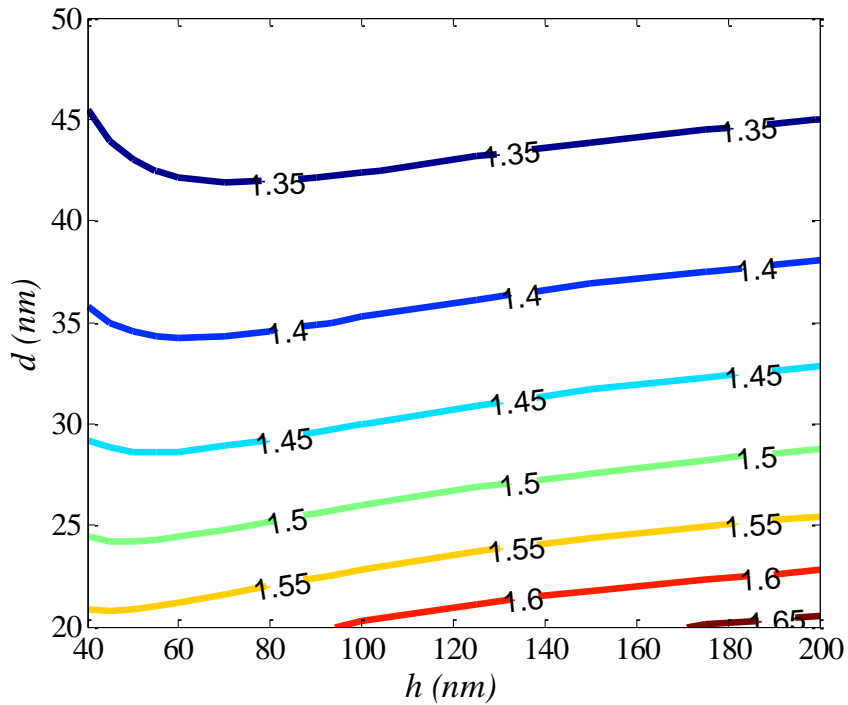

**Figure 3.14:** Contour plot showing the effect of the geometrical parameters *h* and *d* on the $n_{eff}$ in the lateral coupled strips at 30 THz. *w* = 200 nm.

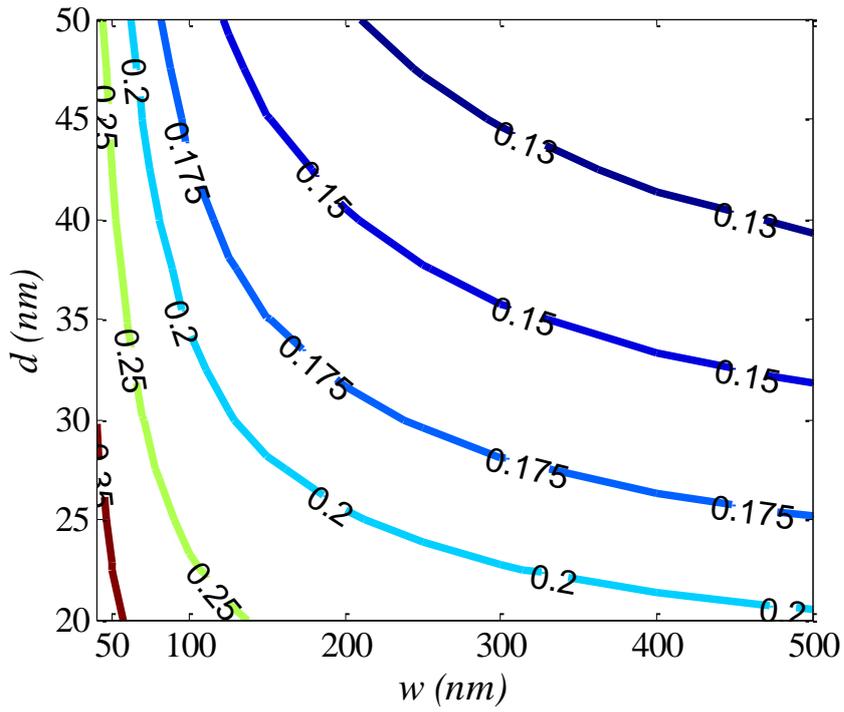

**Figure 3.15:** Contour plot showing the effect of the geometrical parameters *w* and *d* on the $\alpha$ (dB/µm) in the lateral coupled strips at 30 THz. *h* = 180 nm

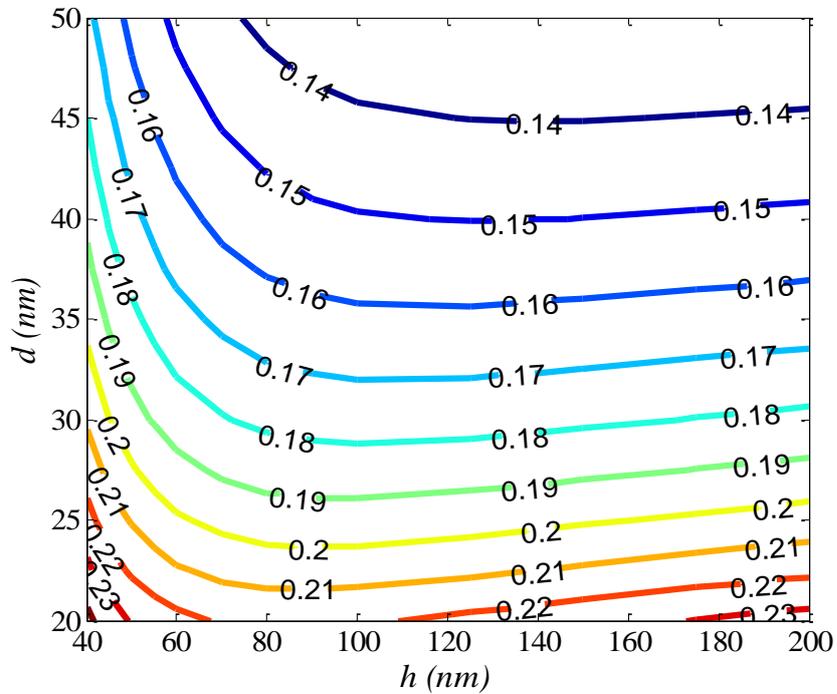

**Figure 3.16:** Contour plot showing the effect of the geometrical parameters *h* and *d* on the $\alpha$ (dB/µm) in the lateral coupled strips at 30 THz. *w = 200* nm.

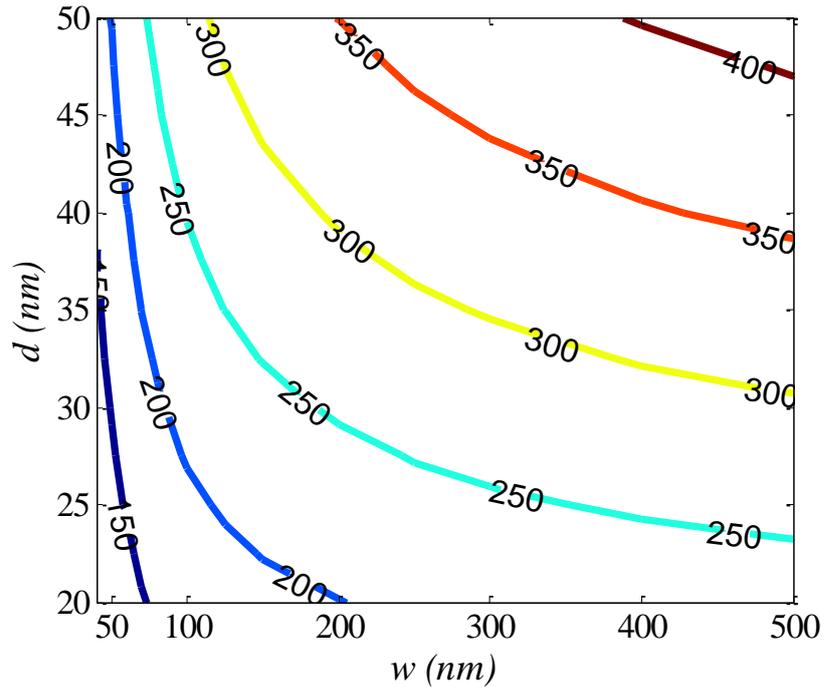

**Figure 3.17:** Contour plot showing the effect of the geometrical parameters *w* and *d* on the $L_{prop}$ (*μm*) in the lateral coupled strips at 30 THz. *h* = 180 nm

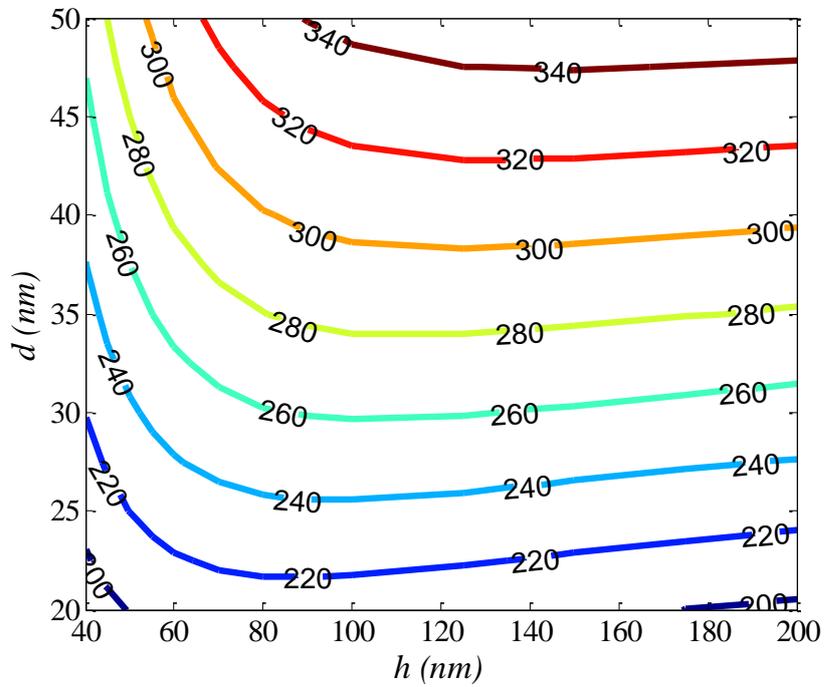

**Figure 3.18:** Contour plot showing the effect of the geometrical parameters *h* and *d* on the $L_{prop}$ (*μm*) in the lateral coupled strips at 30 THz. *w* = 200 nm.

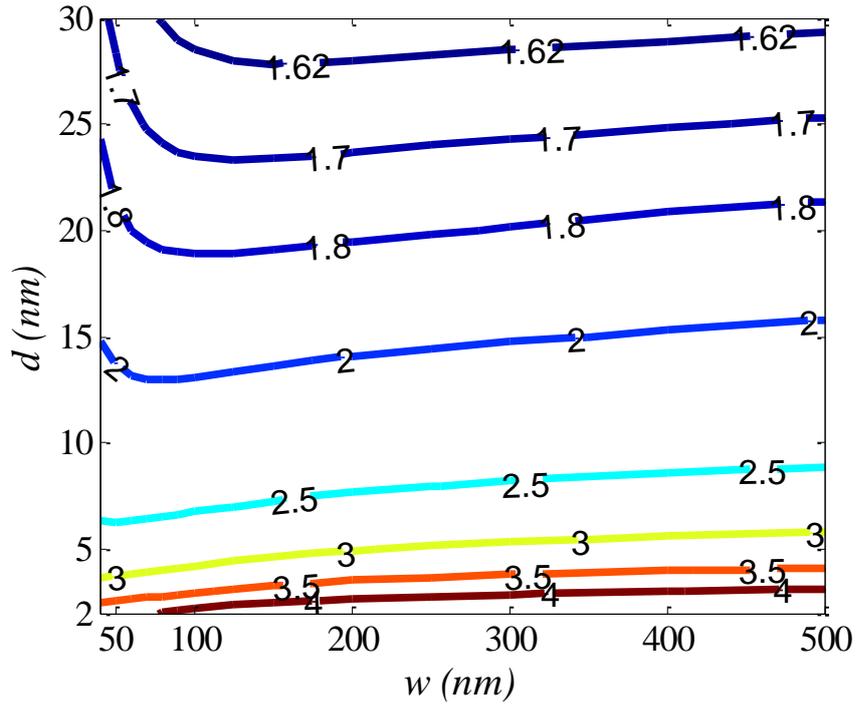

**Figure 3.19: Contour plot showing the effect of the geometrical parameters *w* and *d* on the $n_{eff}$ in the nanostrip waveguide at 30 THz. *h* = 100 nm.**

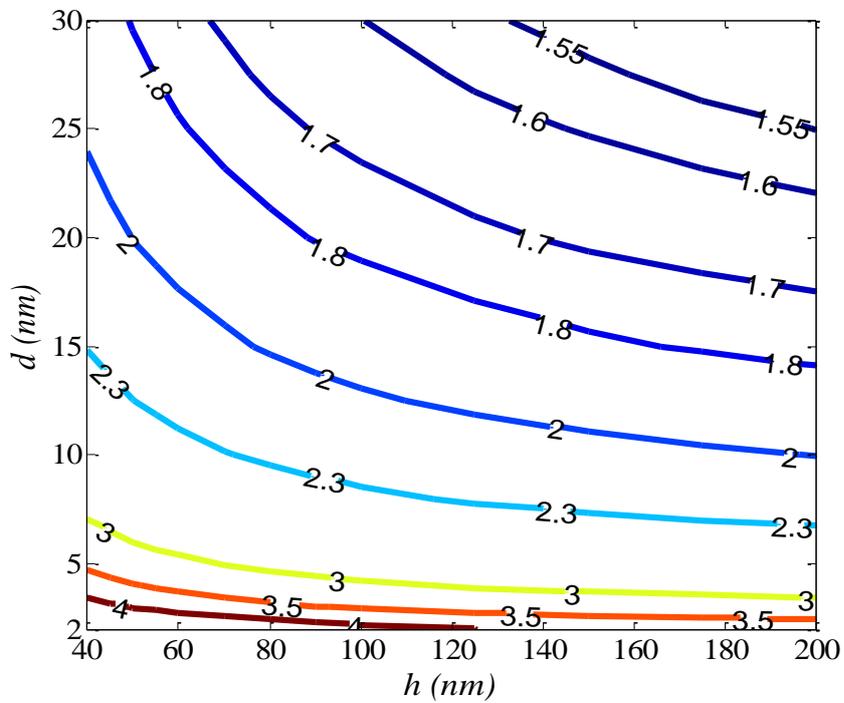

**Figure 3.20: Contour plot showing the effect of the geometrical parameters *h* and *d* on the $n_{eff}$ in the nanostrip waveguide at 30 THz. *w* = 100 nm.**

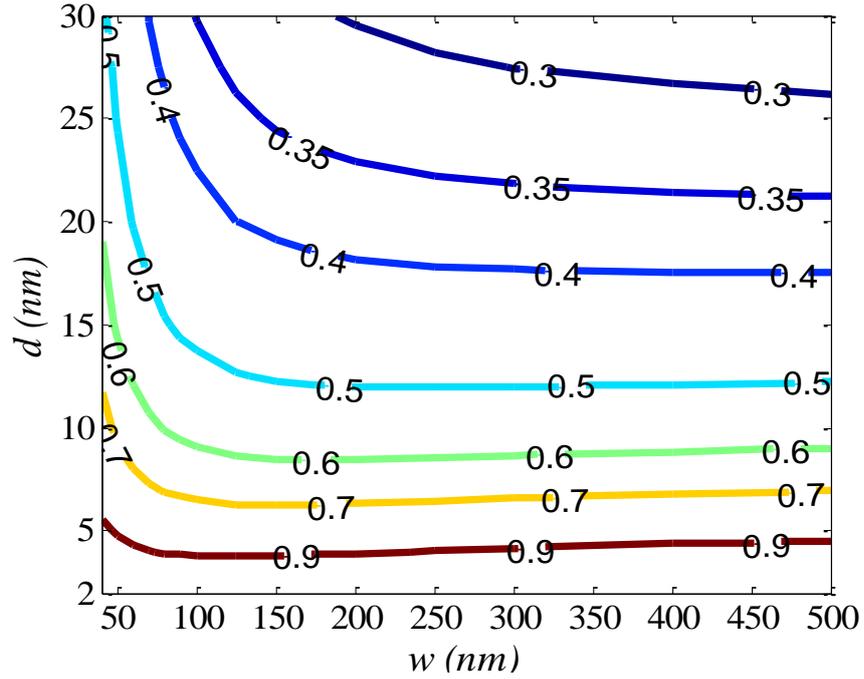

**Figure 3.21:** Contour plot showing the effect of the geometrical parameters *w* and *d* on the α *(dB/μm)* in the nanostrip waveguide at 30 THz. *h = 180 nm.*

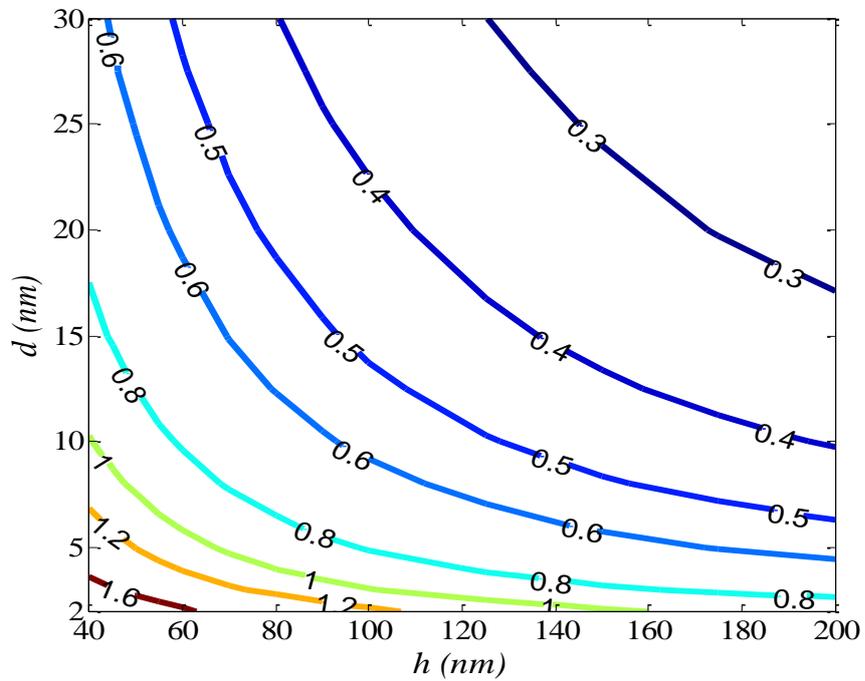

**Figure 3.22:** Contour plot showing the effect of the geometrical parameters *h* and *d* on the α *(dB/μm)* in the nanostrip waveguide at 30 THz. *w = 100 nm.*

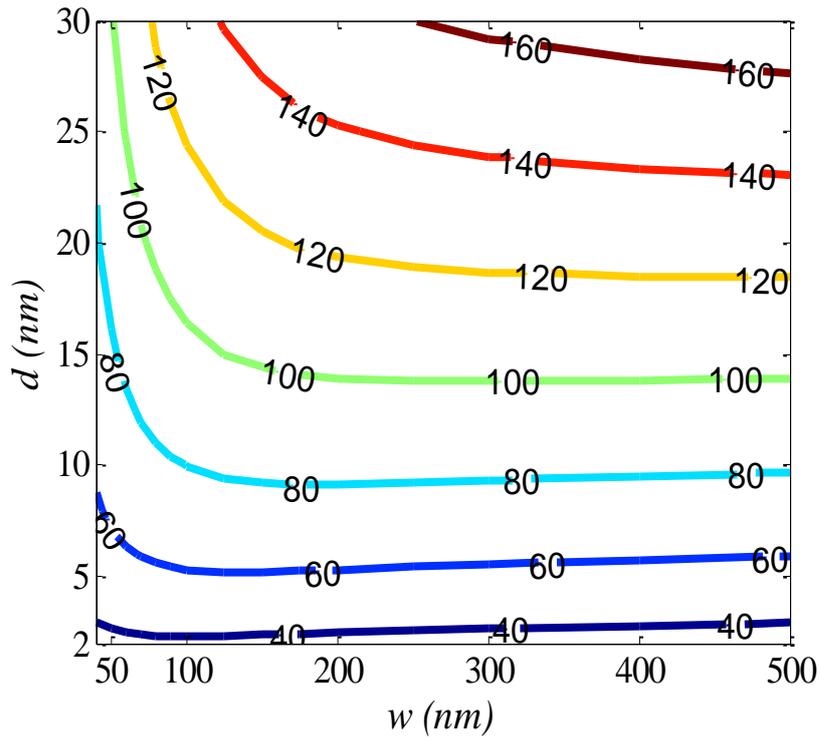

**Figure 3.23: Contour plot showing the effect of the geometrical parameters *w* and *d* on the $L_{prop}$ (µm) in the nanostrip waveguide at 30 THz. *h* = 100 nm.**

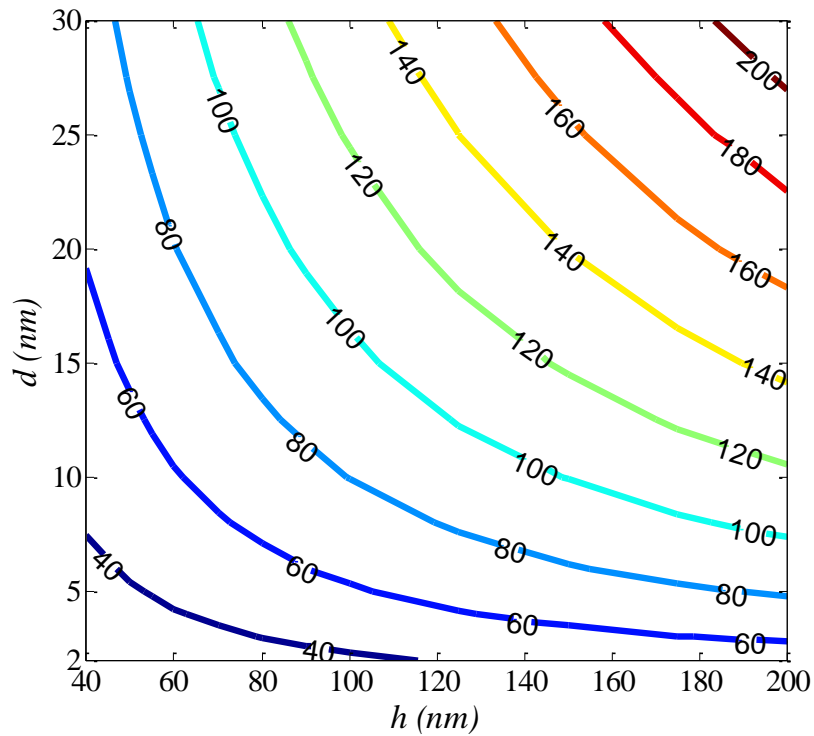

**Figure 3.24: Contour plot showing the effect of the geometrical parameters *h* and *d* on the $L_{prop}$(µm) in the nanostrip waveguide at 30 THz. *w* = 100 nm.**

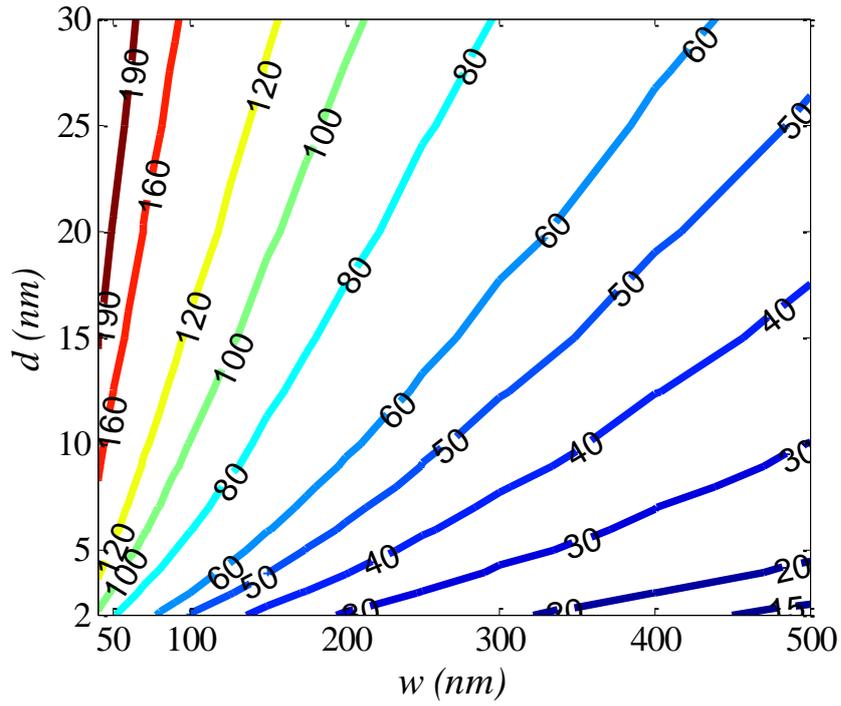

**Figure 3.25:** Contour plot showing the effect of the geometrical parameters *w* and *d* on the $Z_c$ (Ω) in the vertical coupled strips waveguide at 30 THz. *h* = 70 nm.

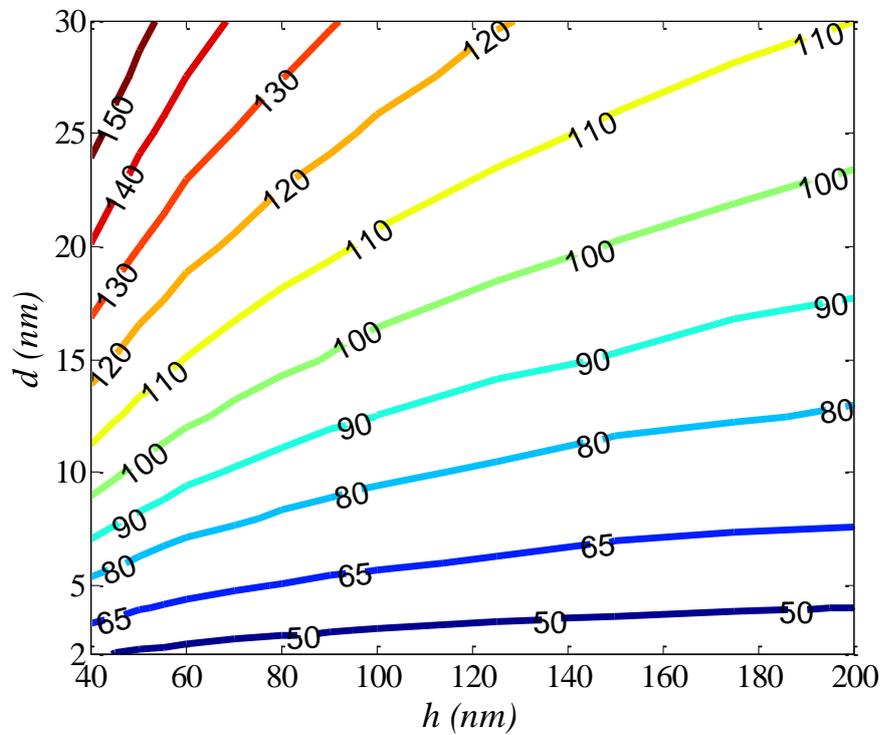

**Figure 3.26:** Contour plot showing the effect of the geometrical parameters *h* and *d* on the $Z_c$ (Ω) in the vertical coupled strips waveguide at 30 THz. *w* = 100 nm.

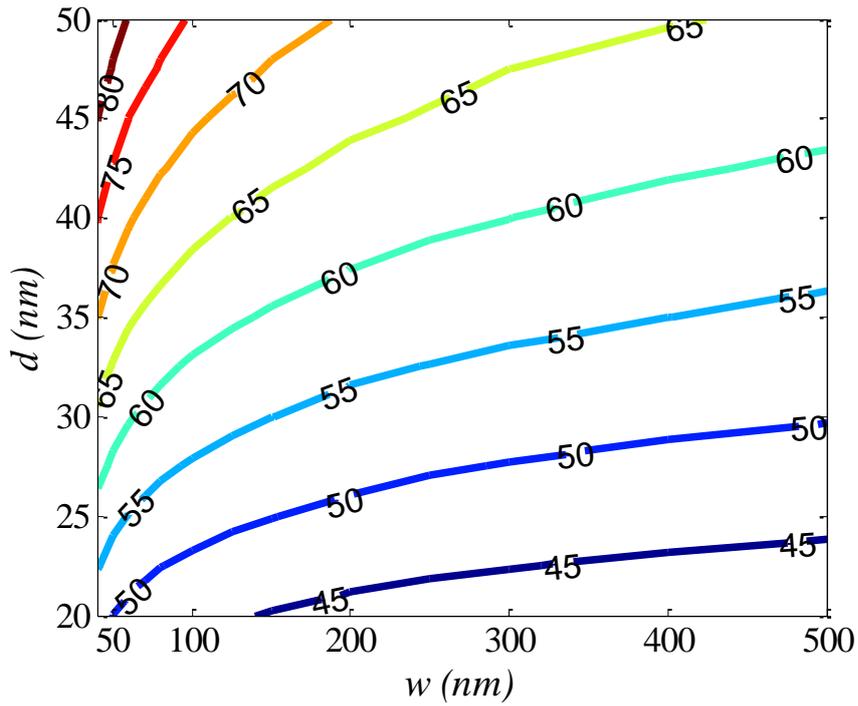

**Figure 3.27: Contour plot showing the effect of the geometrical parameters *w* and *d* on the $Z_c$ (Ω) in the lateral coupled strips waveguide at 30 THz. *h* = 180 nm.**

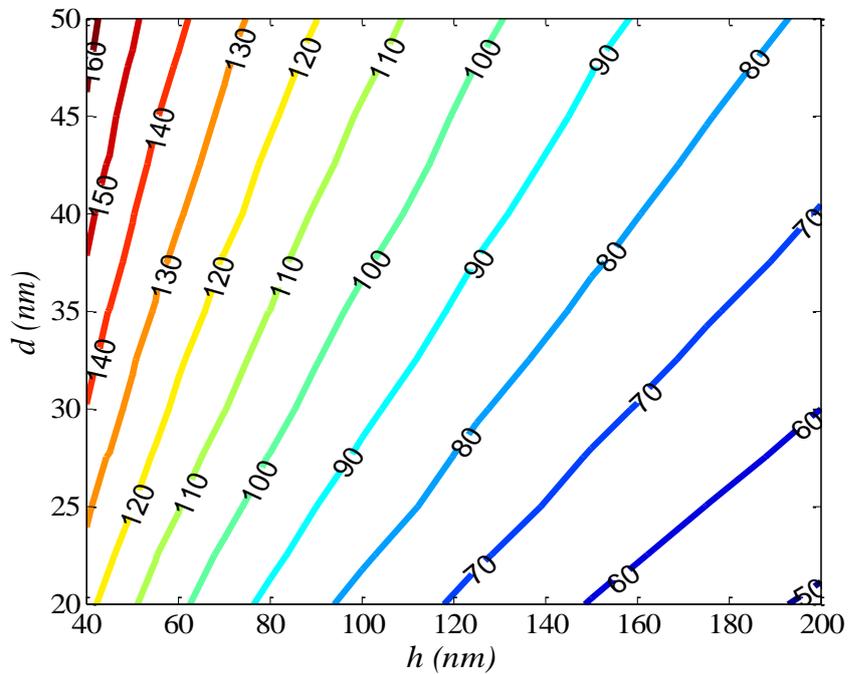

**Figure 3.28: Contour plot showing the effect of the geometrical parameters *h* and *d* on the $Z_c$ (Ω) in the lateral coupled strips waveguide at 30 THz. *w* = 200 nm.**

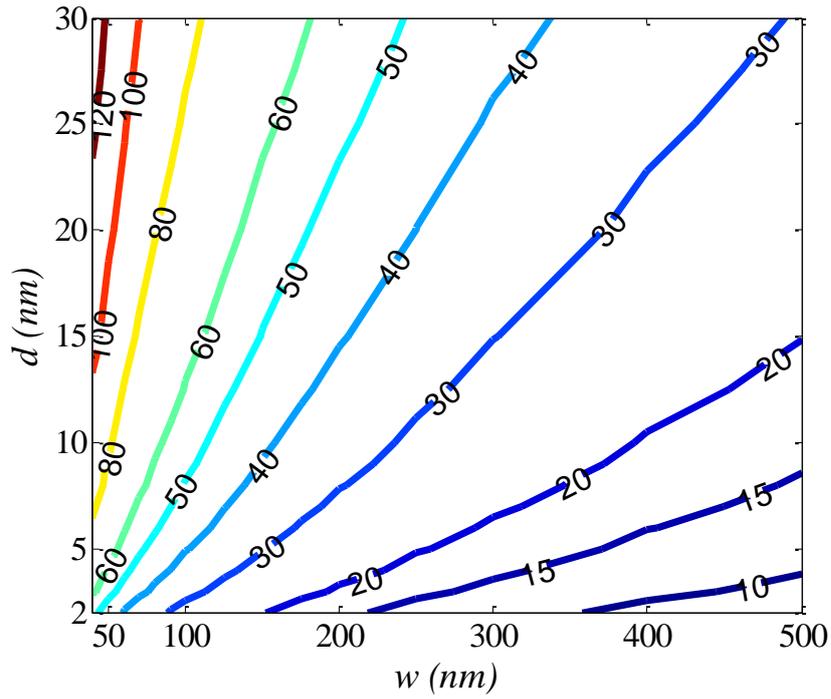

**Figure 3.29: Contour plot showing the effect of the geometrical parameters *w* and *d* on the $Z_c$ (Ω) in the nanostrip waveguide at 30 THz. *h = 100* nm.**

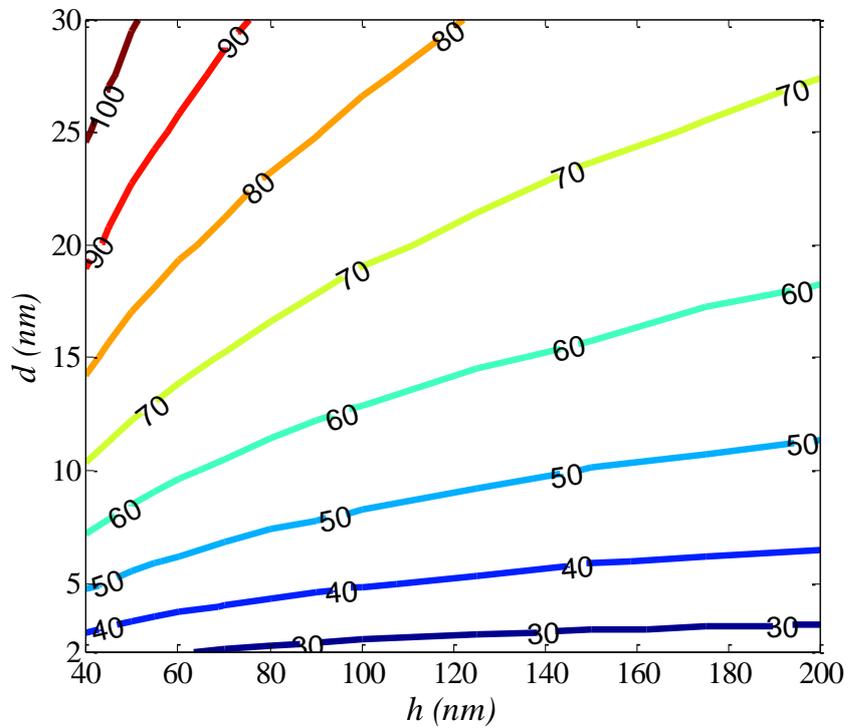

**Figure 3.30: Contour plot showing the effect of the geometrical parameters *w* and *d* on the $Z_c$ (Ω) in the nanostrip waveguide at 30 THz. *w = 100* nm.**

# Chapter 4 Dipole Nantennas Terminated by Traveling-Wave Rectifiers for Infrared Solar Energy Harvesting

## 4.1. Introduction

In this chapter, a complete rectenna operating along 10 THz − 60 THz range is presented. A conventional dipole nantenna is used for this rectenna system. The approach to maximize the coupling efficiency between the nantenna and rectifier is described. Since the dipole nantenna is a balanced structure with its two sides are identical to each other, it should be directly fed by a balanced transmission line, such as the coupled strips. Unbalanced transmission line, such as nanostrip cannot be directly connected to the dipole. A Balanced-to-Unbalanced, Balun, circuit should be inserted in between [87]. The two versions of the coupled strips, namely the vertical and lateral versions, are used to terminate the dipole nantenna which is operating in the receiving mode. A detailed comparison between the two types of termination is performed and presented in this chapter. The traveling wave rectifier is adopted, rather than the lumped rectifier, owing to its superior coupling efficiency with the nantenna.

A parametric study showing the impact of different geometrical parameters on the impedance of the dipole nantenna and rectifier is presented. The nantenna input impedance and the transmission line characteristic impedance are adjusted to match each others. Due to fabrication technology limitations, major difference is expected between the vertical and lateral dielectric spacing that isolates the two metallic sides of both the nantenna and transmission line. This results in huge difference between the overall efficiency of the two rectennas.

Different types of efficiencies that are characterizing the rectenna system are identified and briefly discussed in Section 4.2. Then, the limitations of the lumped (localized) rectennas are presented and analyzed in Section 4.3. The proposed traveling wave rectennas are presented in Section 4.4. The dimensions of the vertical and lateral coupled strips transmission lines operating as travelling wave rectifiers are adjusted to selected reference characteristic impedances, takes place also in this section. In addition, the dependence of the dipole nantenna input impedance on its geometrical dimensions is investigated also in this section. In Section 4.5, the coupling efficiency between the nantennas and travelling wave rectifiers is calculated, and the rectifiers responsivities together with the whole system efficiency are obtained for the proposed types of rectennas. The chapter is concluded in Section 4.6.

## 4.2. Efficiency of the rectenna system

The nano-rectenna system consists of a nantenna and a rectifier. The overall efficiency of the system, $\eta$, is determined by a combination of several factors [88]:

$$\eta = \eta_{ant}.\eta_c\eta_D \qquad (4.1)$$

where $\eta_{ant}$ is the efficiency of the nantenna is converting the incident infrared solar radiation energy into an AC voltage difference across its gap. This nantenna efficiency depends on both the infrared incident spectrum and the behavior of the nantenna radiation efficiency with wavelength, according to the following formula [11]:

$$\eta_{ant.} = \frac{\int P(\lambda,T)\,\eta_{rad.}(\lambda)d\lambda}{\int P(\lambda,T)d\lambda} \qquad (4.2)$$

where $P(\lambda,T)$ is the Earth infrared radiation spectrum, which can be expressed according Planck's irradiance for Blackbody radiation and $\lambda$ is the wavelength. $\eta_{rad.}(\lambda)$ is the radiation efficiency of the nantenna, which is by reciprocity, the same as the efficiency of converting the Earth's radiation into THz voltage difference.

$\eta_c$ is the coupling, or matching, efficiency between the nanoantenna and the rectifier. This efficiency is maximum when the nantenna input impedance equals the complex conjugate of the rectifier impedance. The rectifier's impedance depends strongly on whether it is of lumped or traveling wave type.

$\eta_D$ is the responsivity of the diode, which represents the efficiency of converting that THz AC waveform into DC current.

Throughout the rest of this chapter, these efficiencies are obtained and the parameters affecting them are discussed.

## 4.3. Limitations of lumped element rectennas

The dipole nantenna terminated by a *lumped (localized)* rectifier, shown in Fig. 4.1(a), can be modeled by the equivalent circuit in Fig. 4.1(b). The nantenna is represented by its Thevenin's equivalent that consists of a voltage source in series with an antenna impedance, $Z_A$. This impedance represents the input impedance of the antenna which depends strongly on frequency. At resonance, $Z_A$ has zero reactance, which results in zero reactive power and maximum possible active power that matches the overall apparent received power. The overlapping area between the arms the nanodipiole represented the MIM diode, which is modeled as parallel combination of a resistance, $R_D$, and a capacitor, $C_D$. The rectenna of Fig. 4.1 has two main problems that degrade the overall system efficiency.

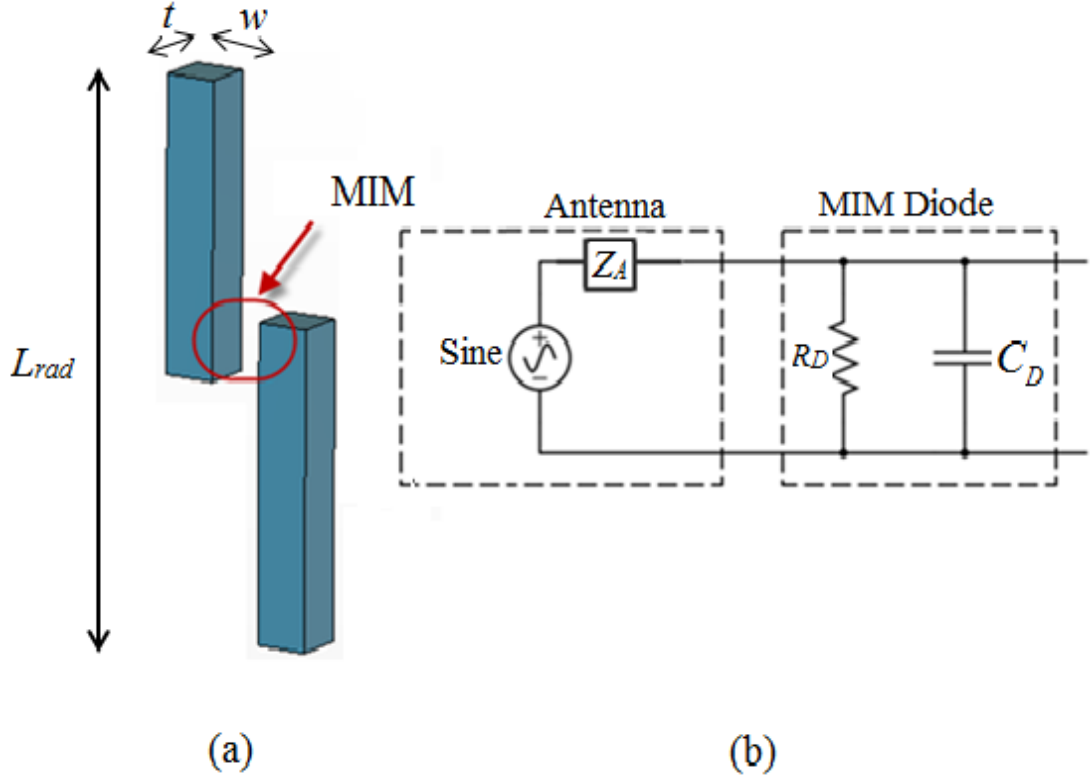

**Figure 4.1: Lumped Element Rectenna. a) Nano-dipole rectenna where the overlapped region represents the MIM diode. b) Small signal AC circuit of the lumped rectenna.**

First, the MIM diode has a sharp cut-off frequency which is related to the inverse of the $R_D C_D$ time constant. This cut-off frequency can be expressed as follows:

$$f_c = \frac{1}{2\pi R_D C_D}, \qquad (4.3)$$

where the capacitance, $C_D$, can be considered as a parallel plate capacitor, and is given by:

$$C_D = \frac{\varepsilon_o \varepsilon_r A}{d}, \qquad (4.4)$$

where $\varepsilon_r$ is the relative permittivity of the used insulator, $\varepsilon_o$ is the free space permittivity, $A$ is the diode overlapping area, and $d$ is the thickness of the insulator layer. To harvest the solar infrared energy, whose spectrum is centered at 30 THz, the cut-off frequency should be adjusted to be higher than 60 THz. In order to increase that cut off frequency, the insulator thickness, $d$, should be increased or the overlapping area, $A$, should be decreased. Increasing the insulator thickness will decrease the tunneling probability and hence, will decrease the output current. While, decreasing the overlapping area that much is impossible from the fabrication point of view.

Second, the antenna and the diode cannot be conjugately matched **[89]**. Assuming that the antenna's impedance has no reactive part, the coupling efficiency between the nantenna and *localized* rectifier, $\eta_c$, is calculated as follows:

$$\eta_c = 1 - \left|\frac{Z_A - Z_D^*}{Z_A + Z_D}\right|^2, \tag{4.5}$$

where $Z_D$ is the equivalent diode's impedance,

$$Z_D = \frac{R_D}{1 + j\omega C_D R_D} \tag{4.6}$$

where $R_D$ is the diode resistance. At such THz frequencies, semiclassical analysis should be applied as the MIM tunneling diode cannot be considered as a classical rectifier. Hence, the semiclassical resistance, $R_D$ is given by [56]:

$$R_D = \frac{1}{I'} = \frac{(2\hbar\omega/q)}{I(V_b + \hbar\omega/q) - I(V_b - \hbar\omega/q)} \tag{4.7}$$

where the current, *I*, is tunneling current calculated using the AFTMM and NEGF methods presented in Chapter 2. The dependency of $R_D$ on the frequency is presented in Fig. 4.2 and it is shown that the increase of the frequency of operation results in the decrease of the diode resistance as it is clear from the figure.

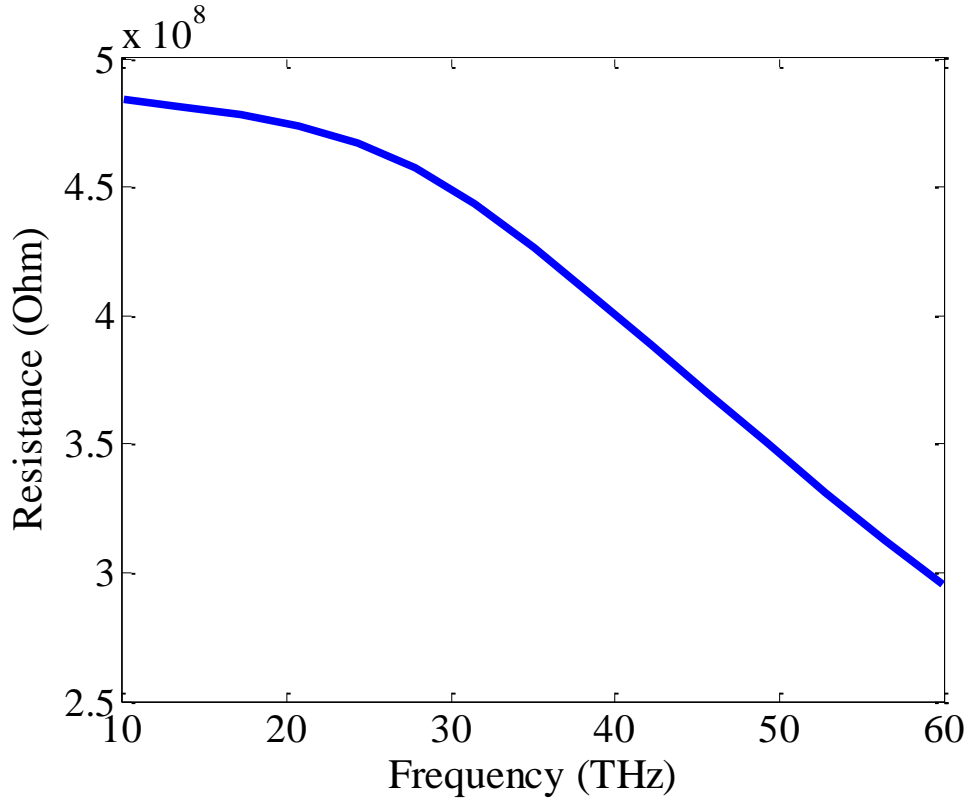

**Figure 4.2: Diode resistance, $R_D$, versus frequency calculated semi-classically using eqn. (4.7), with $V_b = 0.1$ Volts.**

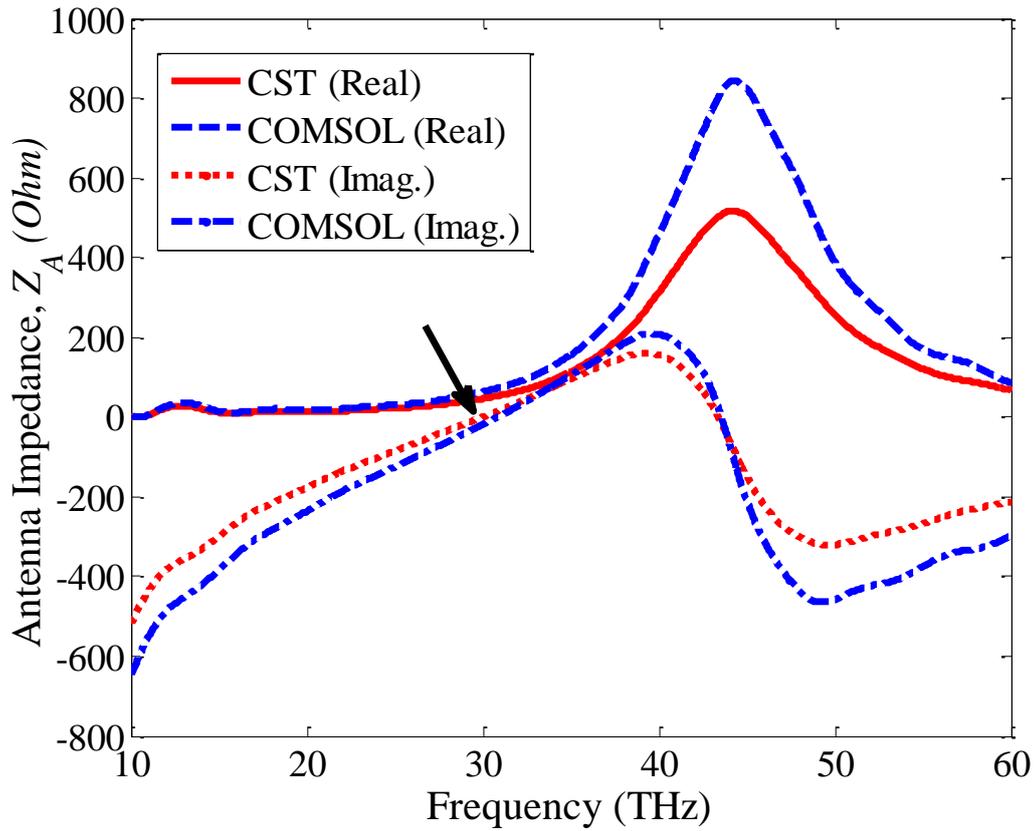

**Figure 4.3: Antenna impedance, $Z_A$, for dipole made of Gold and Silver, and of length $L_{rad}$ = 3700 nm, $w$ = 50 nm, and $t$ = 50 nm. The results are calculated using COMSOL FEM Solver and CST Microwave Studio.**

The antenna's impedance, $Z_A$ is calculated by feeding the nantenna at its gap. The $Z_A$ is the ratio of the driving voltage difference across the gap to the displacement current flowing from the source to the nantenna. This is not the same as the classical RF dipole antenna, at which the current is flowing at the antenna's surface. Using the aforementioned definition, the antenna impedance is calculated using CST Microwave Studio and the results are validated with COMSOL FEM Solver. Fig. 4.3 shows the $Z_A$ calculated for dipole of $L_{rad}$ = 3700 nm, $w$ = 50 nm, and $t$ = 50 nm. The presented results using the commercial packages are close enough such that they can be used to analyze the IR nanodipoles presented in this Chapter. For low frequencies, the impedance is strongly capacitive, which is analogous to the short RF dipoles. By increasing the frequency, the total flux of the induced displacement current increase and thus, the reactive part starts to increase. The antenna hits its first resonance, close to $\lambda_g/2$, as indicated by the arrow in the figure, where $\lambda_g$ is the guided wavelength experienced by the nanodipole. The resonance of the antenna is defined when the antenna's reactive part crosses zero from negative to positive values. The value of the input resistance at this frequency is very low. So, it is commonly used to match the antenna to any feeding device at this frequency. As the frequency increases, the reactance crosses the zero again from positive to negative values, but the resistance at this frequency is greater than that the first resonance. As the frequency increases, the impedance becomes capacitive again and the antenna's reaches its maximum value, which corresponds to the maximum power radiated from the antenna.

The coupling efficiency, $\eta_c$, for the lumped element rectenna is calculated using eqn. (4.5) and presented in Fig. 4.4. It is clear that there is a *big mismatch* between the nantenna and MIM diode here. The reason behind that as the diode resistance is in the order of $10^8$ $\Omega$, while the antenna's resistance is in hundreds $\Omega$. Also it is clear that the maximum coupling efficiency is achieved at the frequency at which the antenna's resistance is maximum as it is the closer resistance's value to the diode's semiclassical resistance, $R_D$. In order to bring the antenna's maximum coupling efficiency at 30 THz, the antenna's length should be scaled up.

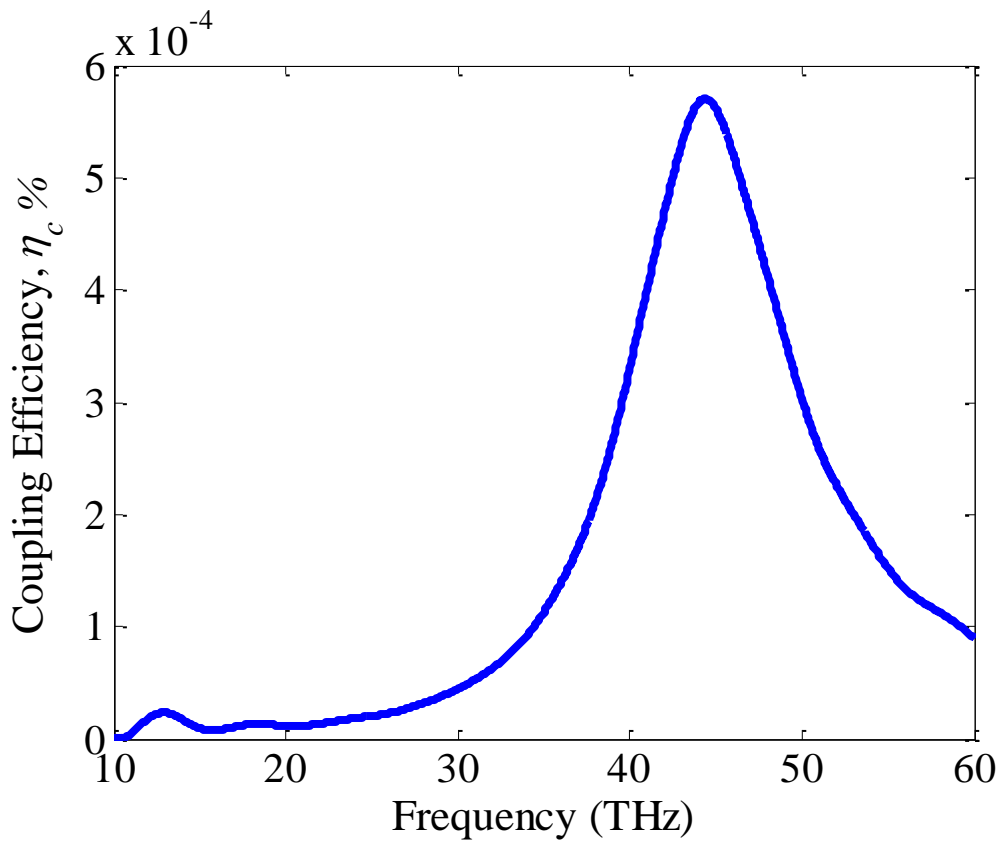

**Figure 4.4: Coupling efficiency, $\eta_c$, versus frequency for Gold/ Al$_2$O$_3$/ Silver rectenna, where the overlapping area, $A = 50 \times 50$ *nm$^2$*, and *d = 2 nm***

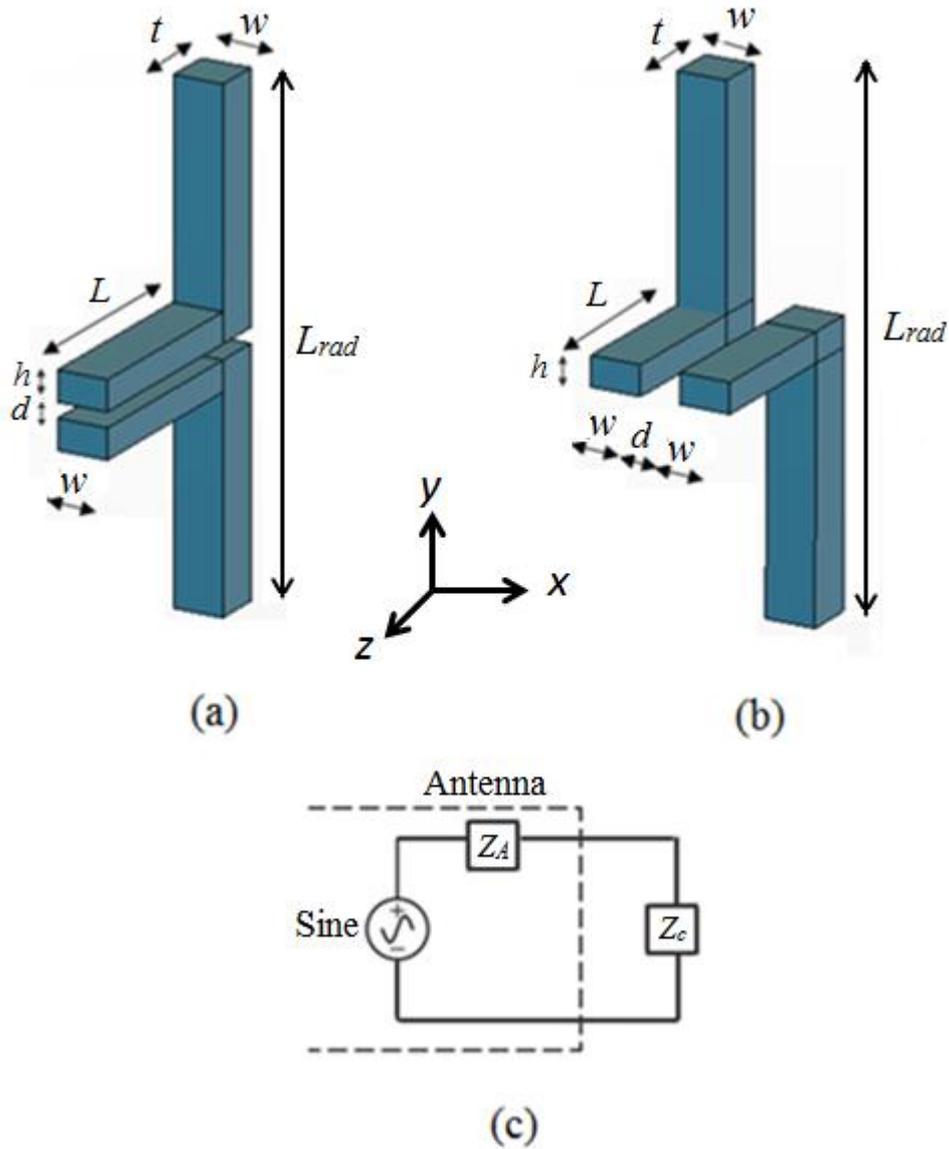

**Figure 4.5: Traveling Wave (TW) rectenna. a) Nanodipole antenna fed by vertical coupled strips. b) Nano-dipole fed by lateral coupled strips. c) Circuit model of the TW rectenna.**

## 4.4. Traveling wave rectennas

To overcome the MIM cut-off frequency limitation and the mismatching between the nantenna and the MIM dilemma, the Traveling Wave (TW) structure was proposed in 2006 [10]. As illustrated in Chapter 3, the traveling wave rectenna consists of a plasmonic MIM waveguide integrated to an antenna. Here, we propose two new TW rectennas. The nantenna used in both the two structures is the conventional nanodipoles. In the first structure, the nanodipole is terminated by Vertical Coupled Strips (VCS), as shown in Fig. 4.5(a). In second rectenna, the nano-dipole is terminated by Lateral Coupled Strips (LCS), as shown in Fig. 4.5(b). The characteristics of these two lines were extensively discussed in Chapter 3.

As mentioned in Chapter 3, the length of the transmission line is longer than the plasmon decay length, such that the nantenna sees the MIM line as an impedance whose value is exactly the same as the characteristic impedance of the plasmonic line. The circuit model of the TW rectenna is shown in Fig. 4.5(c), where the nantenna is modeled by its Thevenin's equivalent and the traveling wave rectifier is modeled with the corresponding transmission line characteristic impedance $Z_c$.

The two difficulties associated with the localized rectifier are avoided if the traveling wave version is used. First, the detection bandwidth is limited primarily by the plasmon decay length ($\alpha^{-1}$) of the MIM waveguide, where $\alpha$ is the attenuation constant of the fundamental mode of the waveguide. Any frequency component with plasmon decay length smaller than TW rectifier length, can be coupled from the nantenna to the rectifier. Second, the nantenna can be easily matched to the rectifier as its characteristic impedance is in the same order as the nantenna input impedance. The coupling efficiency between the nantenna and the plasmonic waveguide is as follows:

$$\eta_c = 1 - \left| \frac{Z_A - Z_c}{Z_A + Z_c} \right|^2 \quad (4.8)$$

where $Z_c$ is the coupled strips line characteristic impedance.

In order to enhance the efficiency of the MIM diode rectifier whose operation is based on tunneling of electrons in unified direction from one metal to the other, two different metals with different work functions should be used [90]. In this study, Gold (Au) and Silver (Ag) are selected as they lead to relatively high nantenna radiation efficiency with reasonable difference in their work functions [82].

The characteristics of each MIM line and its associated nantenna are presented and discussed in the following sub-sections.

### 4.4.1. Coupled strips plasmonic waveguides

In this sub-section, the vertical and lateral coupled strips associated with the nanodipoles in Fig. 4.5 are considered. In order to maximize the tunneling probability of the electrons from one metallic strip to the other, these strips should be brought as close as possible to each other. For the VCS rectifier, the spacing between the strips is vertical, which can be made as small as 2 nm using Atomic Layer Deposition (ALD) technology. As for the LCS, the minimum spacing that can be made with good accuracy using e-Beam Lithography (EBL) is 20 nm. For this reason, the spacing between the two strips is kept constant at 2 nm and 20 nm for the VCS and LCS, respectively, as shown in Fig. 4.6.

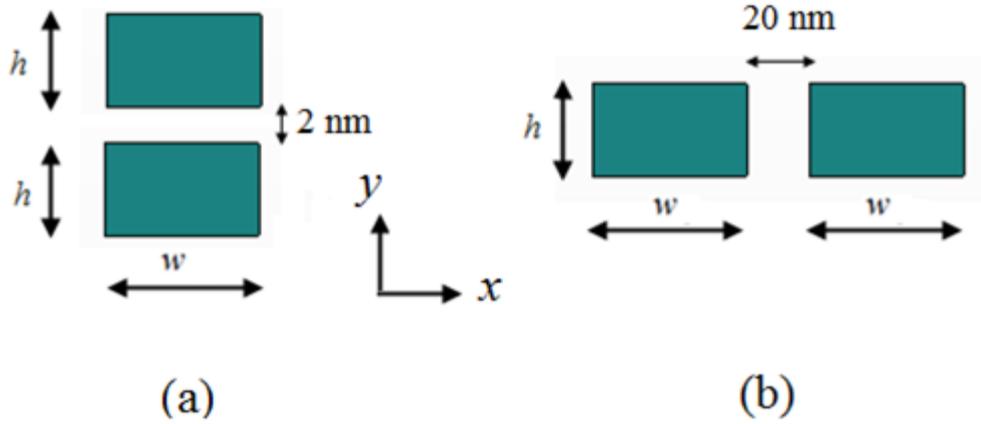

**Figure 4.6:Cross sections of the 2D travelling wave plasmonic transmission lines. a) Vertical coupled strips, b) Lateral coupled strips.**

Successful traveling wave rectenna system should possesses transmission line with length longer than the plasmon decay, and having characteristic impedance matching the input impedance of the antenna. The behavior of the characteristic impedances of the VCS and LCS versus strip's width, $w$, and height, $h$, are shown in Fig. 4.7. For the LCS, the 20 nm spacing between the two strips requires the height of the strips to be less than 80 nm. This keeps the etching aspect ratio to be less than 4, which is reasonable for the e-beam patterning. As for the VCS, using strip's width and height of 100 nm and 75 nm, respectively, leads to a characteristic impedance of 50 Ω. According to Fig. 4.7(b), if the strip's height is 78 nm, a characteristic impedance of 85 Ω can be obtained using a reasonable strip width of 100 nm in the LCS rectifier. Normally, the line width should not be relatively large in order to avoid parasitic radiation at discontinuities.

It is very important at this point to study the variation of the coupled strips plasmonic lines' characteristics with frequency, i.e. the dispersion curves. The strip's width for both lines is 100 nm. The strip's height for the VCS and LCS are 75 nm and 78 nm, respectively. The spacing between the two strips are 2 nm and 20 nm for the VCS and LCS, respectively, and the results are presented in Figs. 4.8-4.11. The results are calculated using COMSOL FEM Solver and validated with LUMERICAL Mode Solver [91]. In general, the variation of the $n_{eff}$, $α$, $L_{prop}$, and $Z_c$ with the frequency is very low, as the electric field does not penetrate the metal volumes to high extent (see Fig. 3.5). So, the line's characteristics are less sensitive to the metal's properties. As it is clear from Fig. 4.8, the $n_{eff}$ is almost constant for both the VCS and LCS along the studied frequency band. As expected, the $α$ ($L_{prop}$) increases (decreases) with the frequency for both the VCS and LCS, as depicted in Fig. 4.9 and 4.10, respectively. The lengths of the transmission lines are adjusted to allow frequencies down to 10 THz to couple efficiently to the rectifier. These lengths are adjusted to be equal to the plasmon decay length of the fundamental mode of associated transmission line at 10 THz. The higher the frequency, the longer the line will be and the better the coupling -

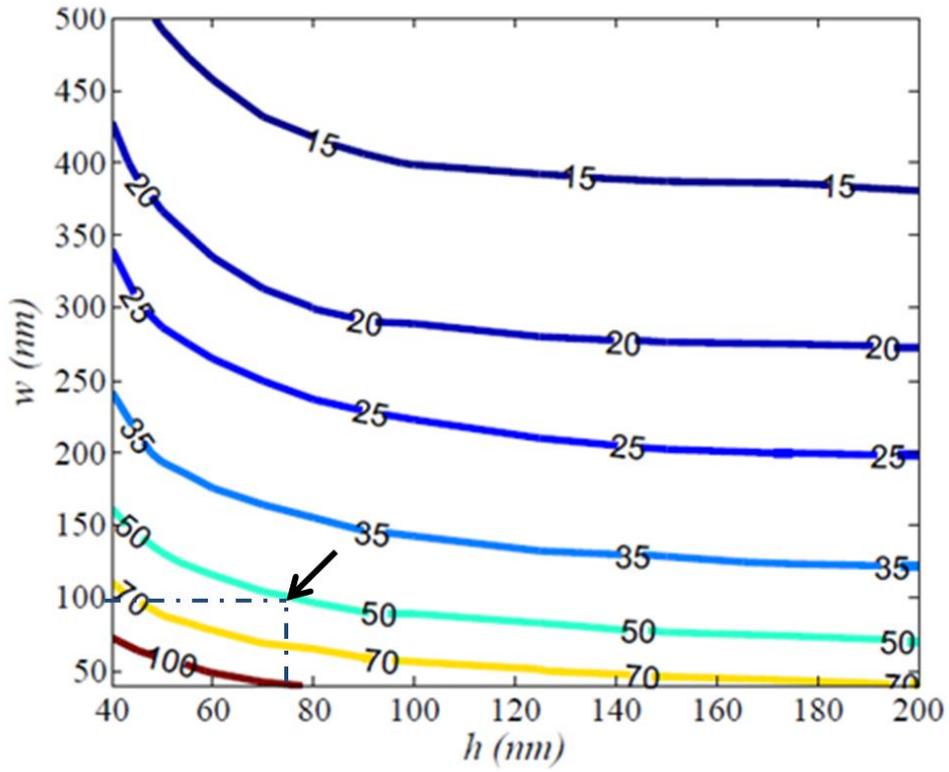

(a)

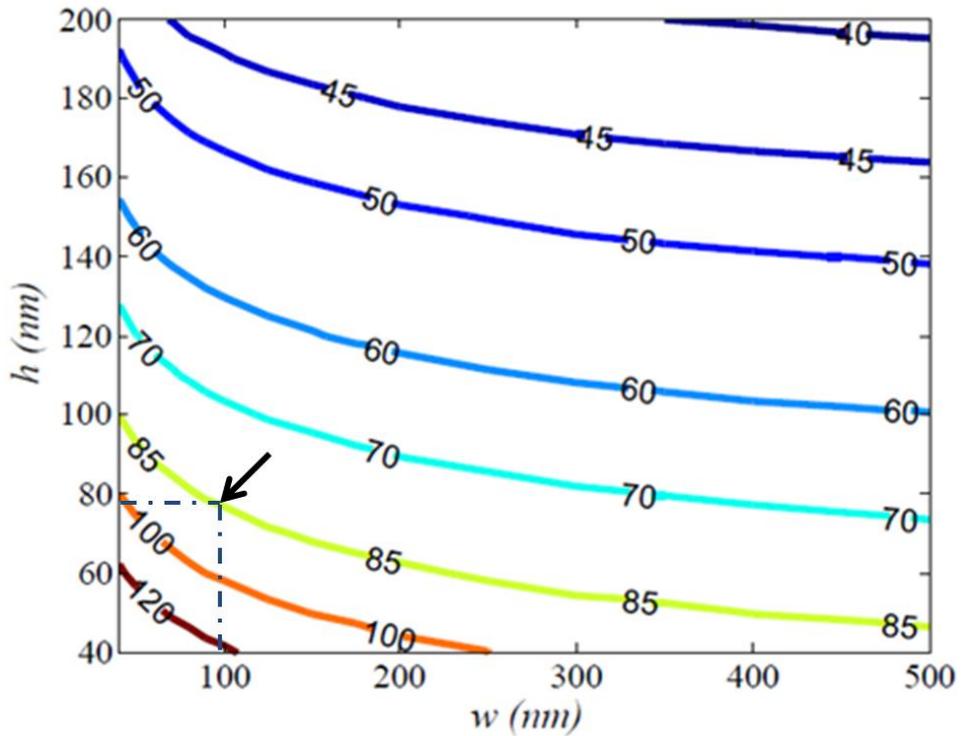

(b)

**Figure 4.7: Contour plots showing the effect of *h* and *w* on: a) characteristic impedance, $Z_c$, (Ω) of vertical coupled strips ($d = 2$ nm), b) characteristic impedance, $Z_c$, (Ω) of lateral coupled strips ($d = 20$ nm).**

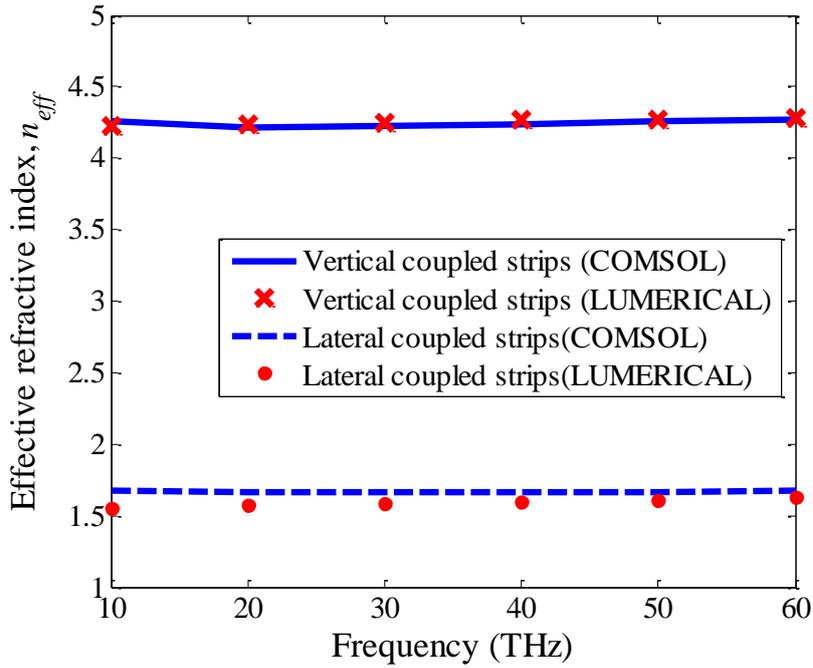

**Figure 4.8:** Effective refractive index, $n_{eff}$, versus frequency for vertical coupled strips ($w$ = 100 nm, $h$ = 75 nm, and $d$ = 2 nm) and lateral coupled strips ($w$ = 100 nm, $h$ = 78 nm, and $d$ = 20 nm), as calculated using COMSOL FEM Solver and LUMERICAL Mode Solver.

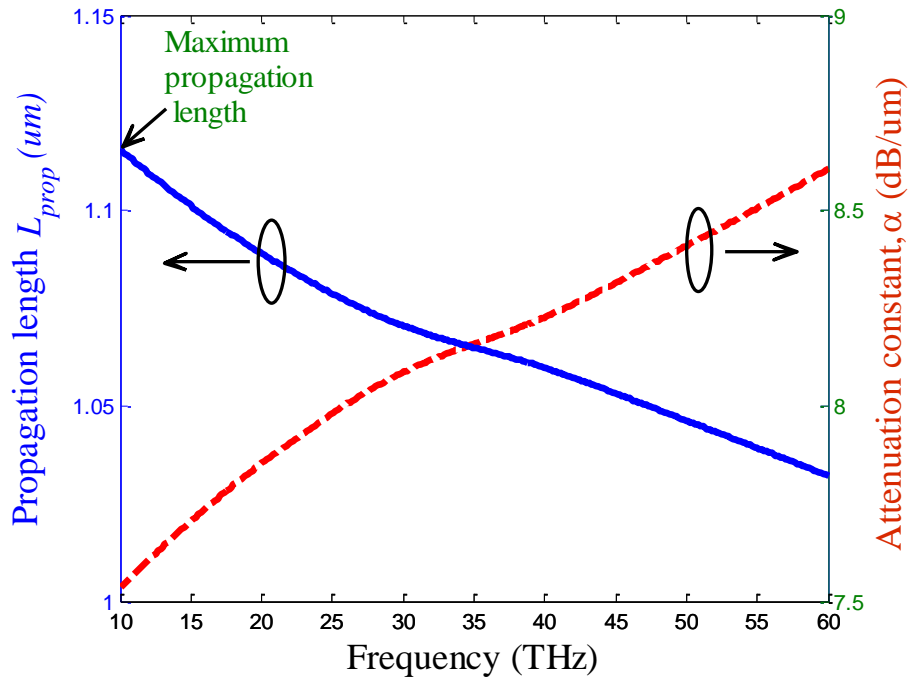

**Figure 4.9:** Propagation length, $L_{prop}$, and attenuation constant, $\alpha$, versus frequency for the vertical coupled strips ($w$ = 100 nm, $h$ = 75 nm, and $d$ = 2 nm).

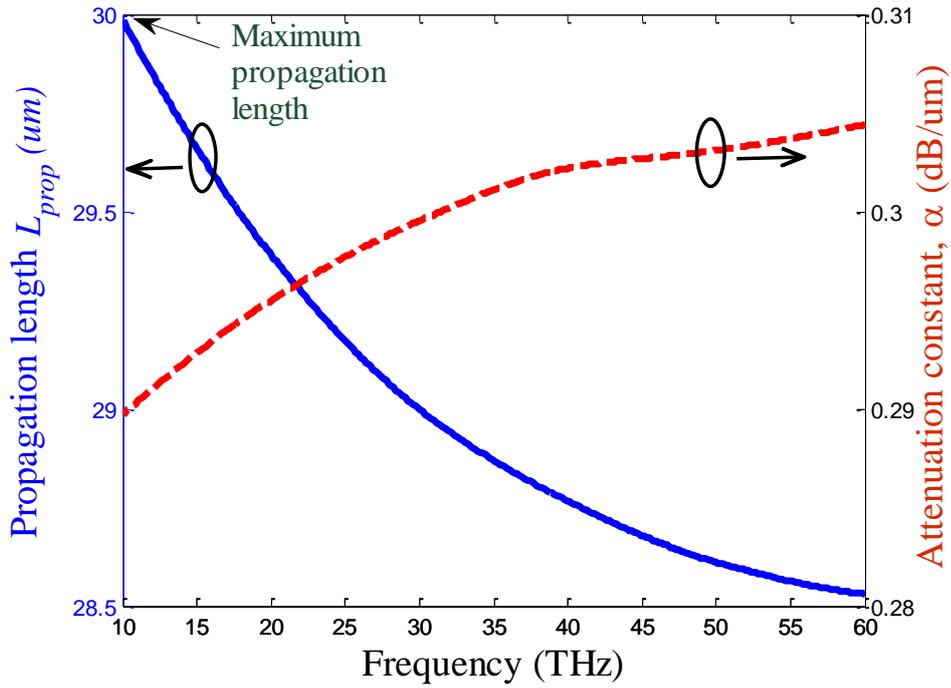

**Figure 4.10:** Propagation length, $L_{prop}$, and attenuation constant, $\alpha$, versus frequency for the lateral coupled strips ($w = 100$ nm, $h = 78$ nm, and $d = 20$ nm).

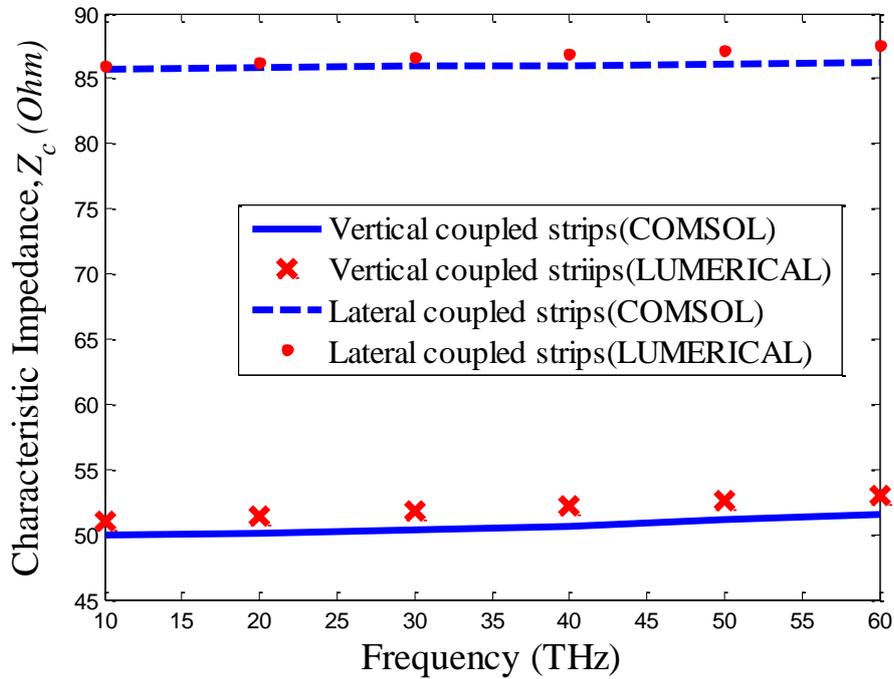

**Figure 4.11:** Characteristic impedance, $Z_c$, versus frequency for vertical coupled strips ($w = 100$ nm, $h = 75$ nm, and $d = 2$ nm) and lateral coupled strips ($w = 100$ nm, $h = 78$ nm, and $d = 20$ nm), as calculated using COMSOL FEM Solver and LUMERICAL Mode Solver.

from the nantenna to the transmission line. The only disadvantage of the travelling wave rectenna is the extra added transmission line. On the other hand, this rectenna is superior over the lumped rectenna as it offers much higher coupling efficiency.

### 4.4.2. Nano-dipoles in each proposed structure

In this subsection, the two nanodipole antennas associated with each transmission line are analyzed in terms of the antenna impedance. The two dipoles are very similar to each other, except for the little misalignment of the dipole's arms connected to the LCS line, as shown in Fig. 4.12. The polarization of the received electric field at the dipole gap is vertical for the VCS case, while horizontal for the dipole terminated by LCS, as shown in Fig. 4.12. The nantenna input impedance is the most important characteristic of the nantenna in such system, as it should match the characteristic impedance of the plasmonic line connected to the nantenna.

The main parameter of the nanodipole is its length, $L_{rad}$, which determines the antenna's resonance frequency. The resonance frequency of the antenna is the frequency at which the reactive part of the input impedance vanishes. This occurs at certain electrical length of the dipole, where the current and field distributions results in a balance of the inductive and capacitive effects. The cross-section of the each dipole arm has dimensions *w* and *t*, as shown in Fig. 4.12. As discussed in the previous section, the value of *w* and *h* are fixed at 100 nm and 80 nm, for both the VCS and LCS terminated dipoles. The other dimension, *t*, should be adjusted to achieve matching between the nantennas' input impedances at their first resonance and the characteristic impedances of their associated transmission lines. The length of each dipole, $L_{rad}$, is adjusted such that the nantenna first resonance occurs at 30 THz, which is the frequency that corresponds to the peak intensity of the incident solar infrared radiation.

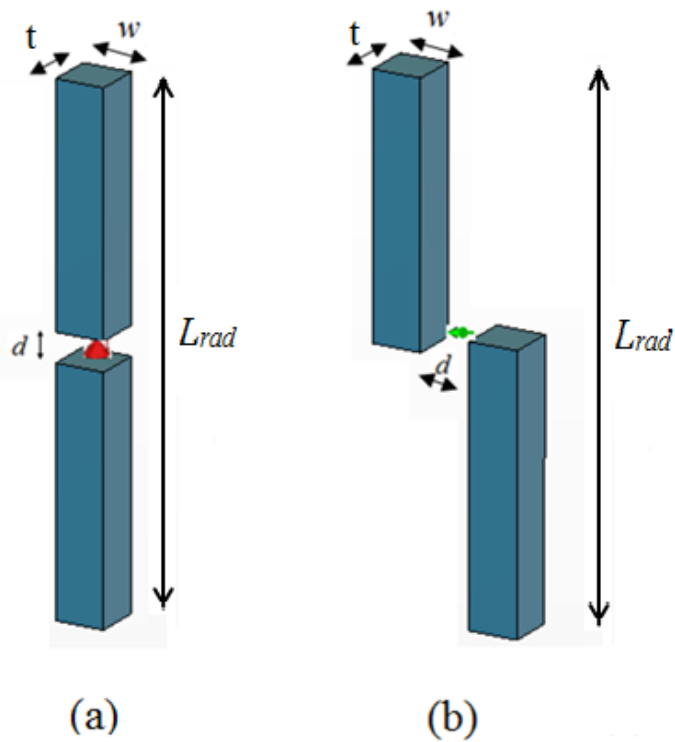

**Figure 4.12: a) The nanodipole associated with the vertical coupled strips. b) The nanodipole associated with the lateral coupled strips.**

Fig. 4.13 shows the input impedances of the nanodipole associated with the vertical coupled strips versus frequency for different nantenna lengths. As expected, the increase of the nantenna length results in decreasing the value of its resonance frequency. According to Fig. 4.13(b), the first zero crossing of the input reactance's is 30 THz when $L_{rad}$ of 3700 nm. This value corresponds to 0.37 of the free-space wavelength of 10 μm. If perfect electric conductor (PEC) is used to realize the dipoles, their electric length at the first resonance would be 0.5 of the free-space value. Since at infrared frequencies, gold and silver allow field to be present within their volume, the guided wavelength is smaller than the free-space one. It can be concluded that the value of the guided wavelength is double $L_{rad}$ of the first resonance, i.e. 7400 nm, which is less than the free-space value of 10,000 nm. Fig. 4.13(a) reveal that the input resistances at the first resonance are in the order of 100 Ω, while they go to their peak values of about 500 Ω. Consequently, it is much easier to match the nantenna at its first resonance frequency to the transmission characteristic impedance than the nantenna at the second resonance.

The impact of the free cross-sectional dimension, *t*, on the nantennas' input impedances is shown in Fig. 4.14 for the nano-dipole associated with the VCS line. The other dimension of the cross-section, *w*, is fixed as it is common with the transmission line dimensions. It can be easily seen that the parameter *t* has minor effect on the location of the resonance frequency. Its main effect is on the level of the resistive and reactive parts of the input impedance of the dipoles. Consequently, this parameter is used to match the nantenna to the feeding line without affecting the value of resonance frequency too much. The optimum value of the *t* that leads to 50 Ω input impedance matching of the VCS line is 40 nm. The same analysis is repeated for the LCS transmission line and a value of 35 nm should be assigned for *t* to achieve the required 85Ω impedance matching.

At this point, the two rectennas are complete. Table 4.1 shows the optimum dimensions together with the materials used in both systems. These optimum values will be used in order to calculate the antenna harvesting efficiency, coupling efficiency of the rectenna, diode responsivity, and the total system's efficiency.

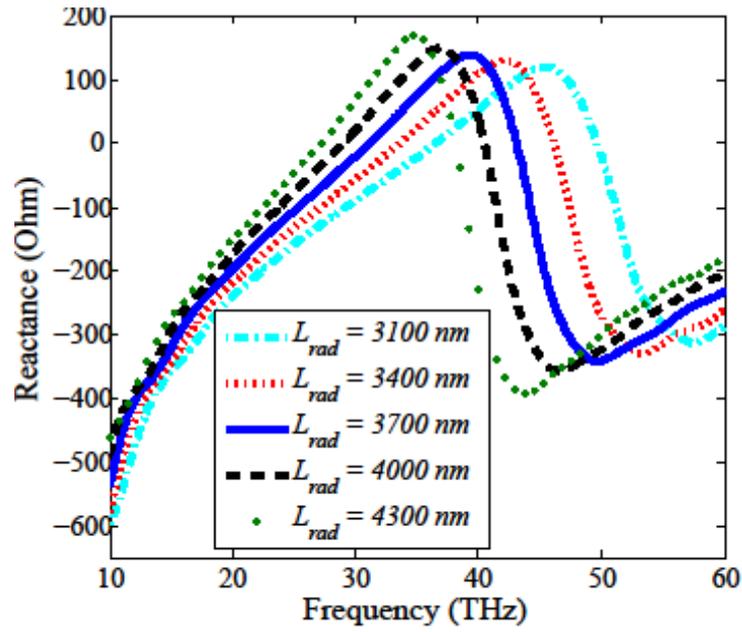

(a)

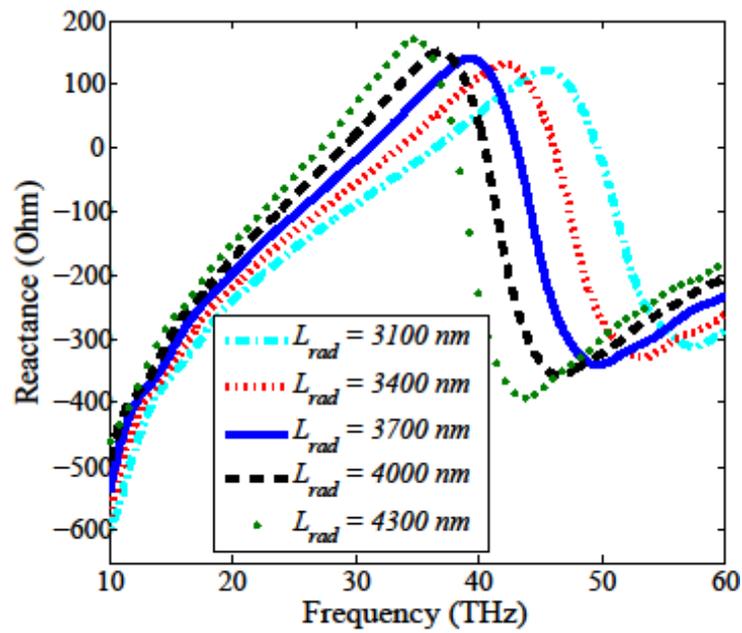

(b)

**Figure 4.13: Resistance and reactance of the nanodipole associated with the vertical coupled strips (*w* = 100 *nm*, *t* = 40 *nm* and *d* = 2 nm), versus frequency for various dipole lengths.**

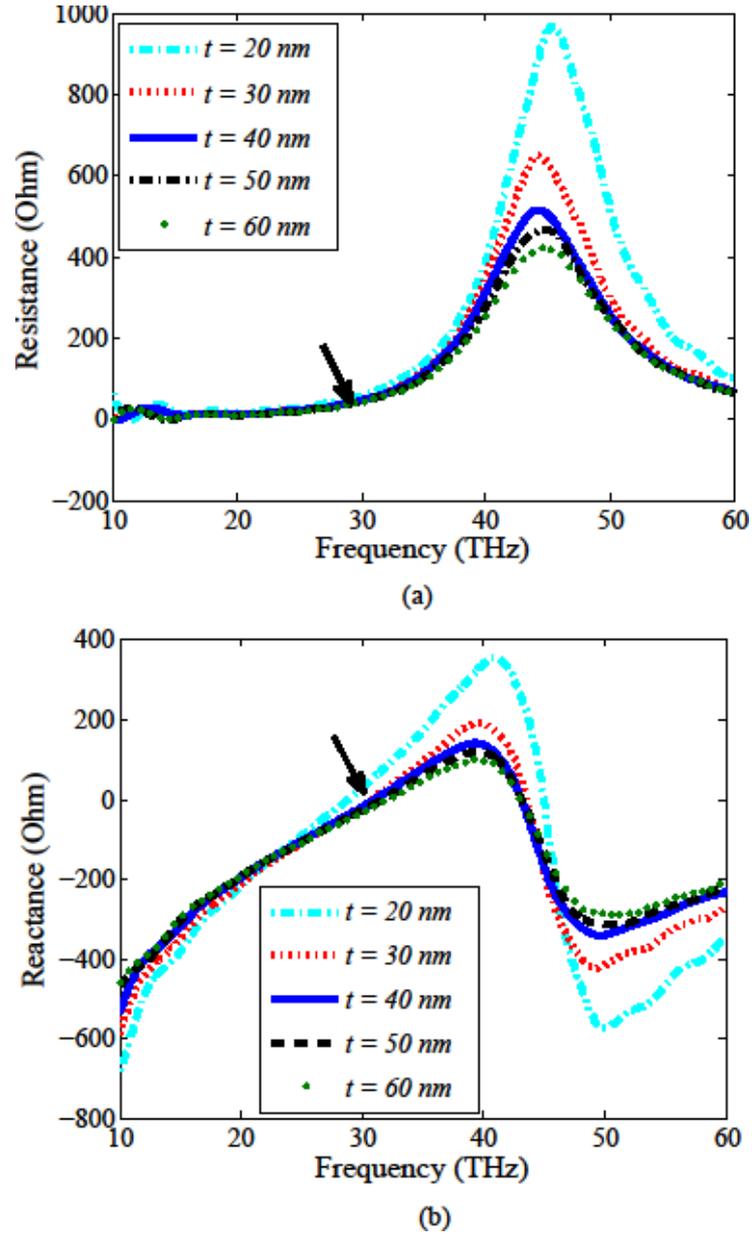

Figure 4.14: Resistance and reactance of the nanodipole associated with the vertical coupled strips ($w = 100$ $nm$, $L_{rad} = 3700$ $nm$, and $d = 2$ $nm$), versus frequency for various $t$ values.

| Rectenna design parameters | Nanodipole fed by vertical coupled strips | Nanodipole fed by lateral coupled strips |
|---|---|---|
| $L_{rad}$ (nm) | 3700 | 3700 |
| w (nm) | 100 | 100 |
| h (nm) | 75 | 78 |
| d (nm) | 2 | 20 |
| t (nm) | 40 | 35 |
| L (µm) | 1.12 | 30 |

Table 4.1: Rectenna design parameters values for both the two proposed structures.

## 4.5. Overall efficiency of the proposed travelling wave rectennas

In this section, the overall efficiencies of the optimized rectennas are obtained. The radiation efficiencies, $\eta_{rad}$, of the two nanodipoles terminated with VCS and LCS transmission lines are presented in Fig. 4.15. These efficiencies represent the ratio between the overall radiated power to the input power. According to reciprocity theory, the radiation efficiencies of an antenna in transmitting and receiving modes are the same. Taking the spectrum of the incident infrared solar energy into account, a more comprehensive efficiency in defined according to eqn. (4.2). This comprehensive efficiency is referred to as the antenna efficiency, $\eta_{ant}$, whose value is 83.61 % and 80.15% for the nano dipole terminated by VCS and LCS respectively. The coupling efficiencies between the nantennas and their associated transmission lines are calculated using eqn. (4.8), and plotted versus frequency in Fig. 4.16. It can be seen that perfect matching at 30 THz is achieved, which is characterized by a 100% coupling efficiencies.

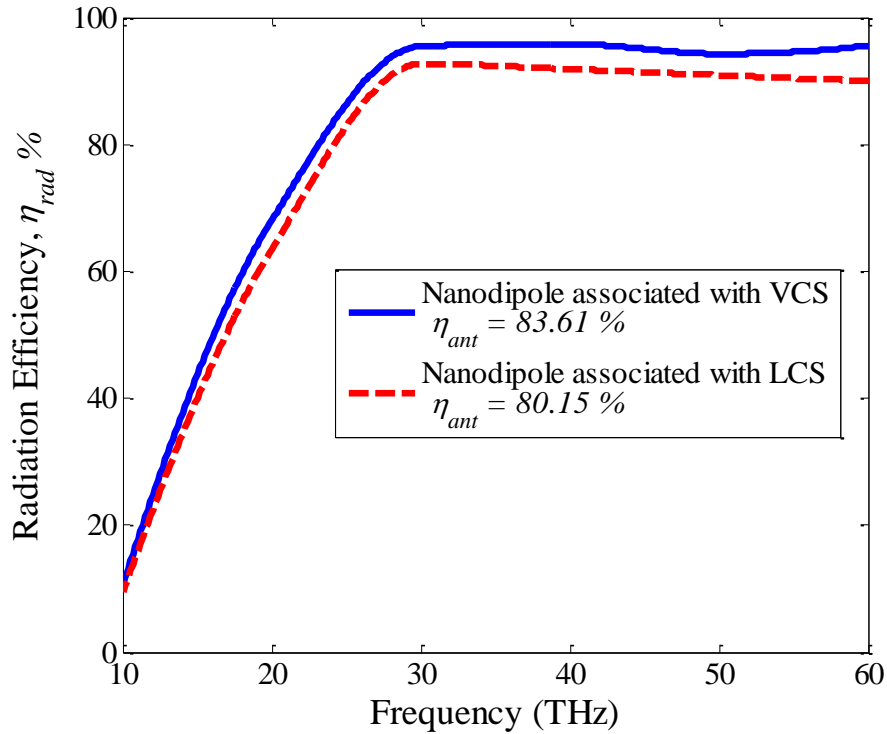

**Figure 4.15: Radiation efficiency of the two nanodipoles associated with each of the vertical coupled strips and the lateral coupled strips.**

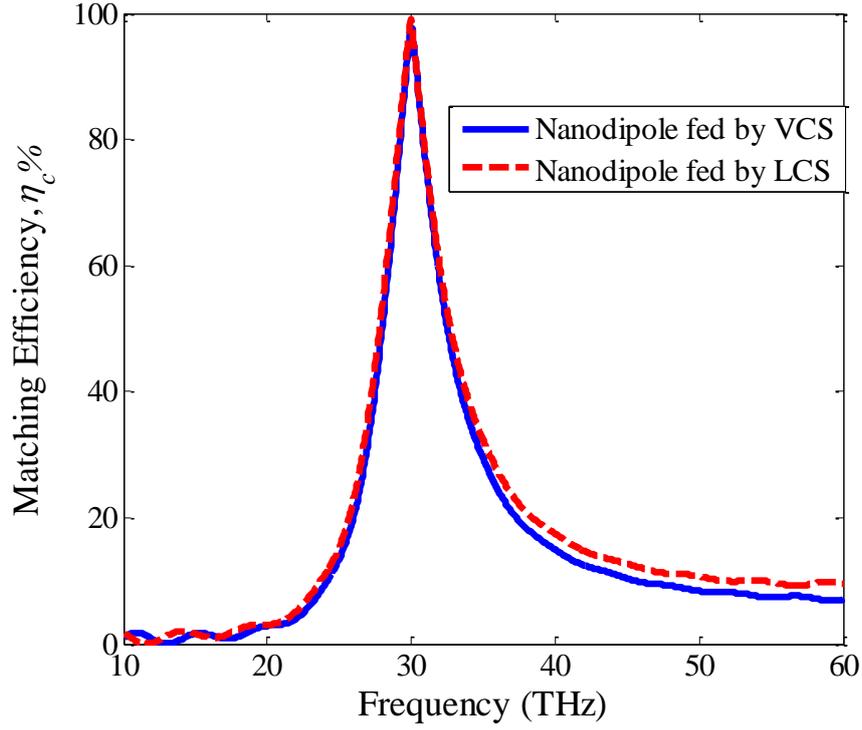

**Figure 4.16: Matching efficiency of the two proposed rectenna structures.**

The lengths of the VCS and LCS plasmonic transmission lines are equal to the propagation length at the minimum frequency within the interested band. Consequently, the voltage difference between the two metallic strips along the direction of propagation, $z$, takes the form: $V(z) = V_o\, e^{-\alpha z}$, where $V_o$ is the voltage at $z = 0$ (see Fig. 3.1a, 3.1b). Assuming a constant voltage distribution along the transversal direction, the responsivity of the vertical coupled strips diode rectifier, can be calculated as follows [10]:

$$\eta_D = \frac{\beta V_o^2}{4 R_{D/A} \alpha} w \quad (4.9)$$

where $V_o = \int E_y dy / \int P_z dz$. $P_z$ is the $z$ component of the power density across the cross section of the transmission lines, $R_{D/A}$ is the diode semiclassical resistance per unit area, and $\beta$ is the semi-classical current responsivity and is given by [56]:

$$\beta = \frac{1}{2}\frac{I''}{I'} = \frac{1}{2}\frac{q}{\hbar\omega}\left[\frac{J(V_b+\hbar\omega/q)-2J(V_b)+J(V_b-\hbar\omega/q)}{J(V_b+\hbar\omega/q)-J(V_b-\hbar\omega/q)}\right] \quad (4.10)$$

where $J$ is the tunnel current density. Similarly, the responsivity of the lateral coupled strips is calculated as follows:

$$\eta_D = \frac{\beta V_o^2}{4 R_{D/A} \alpha} h \quad (4.11)$$

where $V_o = \int E_x dx / \int P_z\, dz$, while $R_{D/A}$ and $\beta$ can be calculated using eqn. (4.7) and eqn. (4.10), respectively.

The calculated responsivities of the travelling wave rectifiers, calculated using eqn. (4.9) and (4.11), are depicted in Fig. 4.17. It is clear that the LCS rectifier possesees a very low responsivity (in order of $10^{-13}$ A/Watt). This due to the relatively large spacing between the two metals constituting the diode rectifier. On the other hand, the VCS rectifier with strips' spacing down to 2 nm, shows significantly higher responsivity in order of $10^{-4}$ A/Watt. Finally, the overall system efficiency - the product of radiation efficiency, matching efficiency and rectifier quantum efficiency - is presented in Fig. 4.18. It is clear that the nanodipole fed by VCS shows a superior efficiency than that fed by LCS. Though each of the two nanodipoles were successfully matched to its associated transmission line. the large insulator thickness of the LCS decrease its overall system efficiency a lot.

It can be noticed that the diode rectifier efficiency, $\eta_D$, is way below the antenna efficiency, $\eta_{ant}$, and the coupling efficiency, $\eta_C$. The extremely small value of $\eta_D$ is the main obstacle towards using rectennas as an efficient replacement for solar cells.

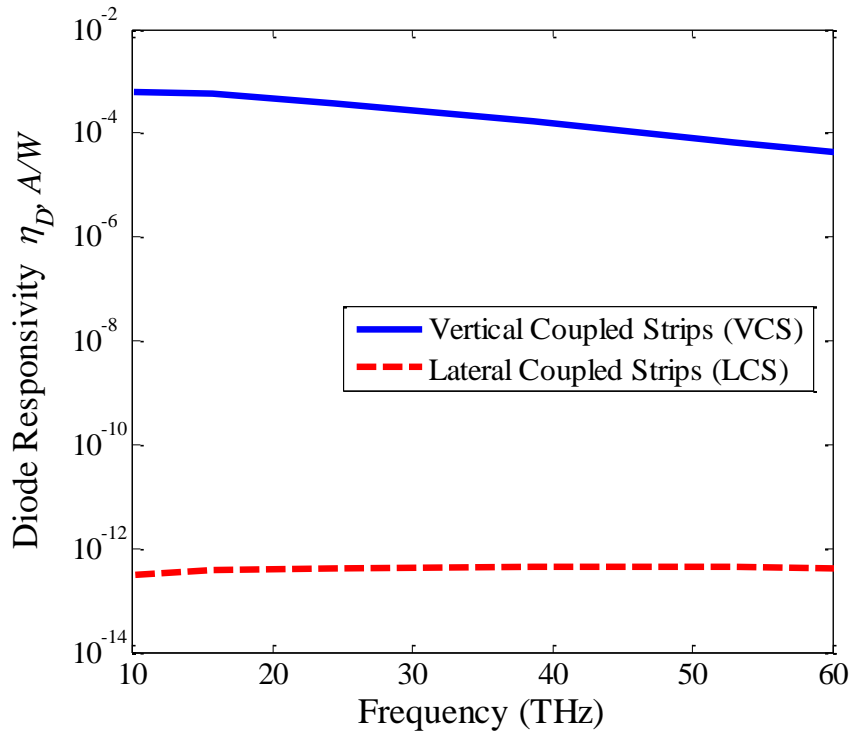

**Figure 4.17: Quantum efficiency, $\eta_D$, of the vertical coupled strips and lateral coupled strips rectifiers.**

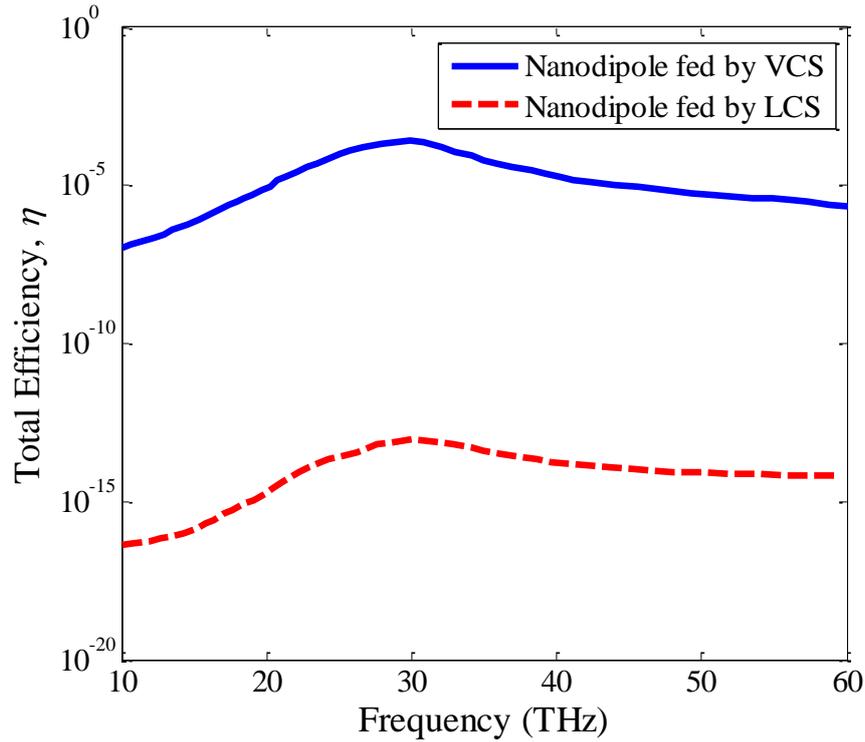

**Figure 4.18: Overall system efficiency of the two proposed rectennas.**

## 4.6. Chapter summary

Two complete travelling wave rectennas are presented in this chapter. A systematic approach to optimize the dimensions of these rectennas, is presented. To maximize the efficiency of the diode rectifier, the spacing between the two metallic strips should be as small as the fabrication technology allows. Specifically, the ALD used to realize the vertical spacing allows for 2 nm spacing. On the other hand, the e-beam lithography provides, with good accuracy, a lateral spacing of 20 nm. The strips' width and height are specified such that width does not exceed 100 nm in order to avoid parasitic radiation from transmission line discontinuities. The heights of the strips are specified such that to maintain reasonable strip's aspect ratio. Another factor that is kept in mind when specifying the dimensions of the strips is to have characteristic impedances in the order of tens of ohms. The transmission lines are kept longer than the plasmon decay length, such that no reflection at the open-circuit termination takes place, which allows modeling the transmission lines with their characteristic impedances. The length of the dipole nantenna is optimized characterizing the proposed rectennas are expressed and calculated. The one order of magnitude difference between the strips' spacing of the VCS and LCS travelling wave rectifiers, results in VCS diode rectifier efficiency nine order of magnitude higher than that of the LCS rectifier. However, the overall VCS travelling rectenna is as small as $10^{-4}$A/watt. This leads to the fact that the only available MIM diode rectifier at such high frequency of operation, is not a successful solution.

# Chapter 5 Nanocrescent Antenna as Transciever for Optical Communication Systems

Nano-antennas, or nantennas, can be used as transmitters or receivers in modern optical communication systems. Nantennas as optical transceivers have a number of inherent advantages over the conventional laser diode and photo detectors, such as: small size, fast response, high directivity, doubling the information capacity via sensitivity to polarization, and broadband behavior. The recent advances in fabrication technology permit the realization of these nantennas whose features are extremely small. In this section a nanocrescent antenna is proposed. It can be considered as a developed version of the nanocrescent presented in [5]. In that work, the patch and aperture forming the nanocrescent were located in the same plane, which made that nantenna suitable for light localization and overheating application. In this Chapter, the patch is slightly elevated above the aperture, where an insulating layer is inserted in between. This makes it possible to feed the nantenna with a nanostrip line, similar to that presented in Chapter 3. Consequently, the proposed nanocrescent is more suitable for optical communication systems. The design and optimization of the proposed nantenna in this Chapter are performed using Lumerical [92]. The results are validated using CST Microwave Studio.

## 5.1. Nantenna structure and optimum Design

The proposed nanocrescent antenna is shown in Fig. 5.1. It consists of a circular gold patch located on top of a circular aperture etched from a gold plate. The centers of both circles are not on top of each other, such that lateral spacing between them is 20 *nm*. A nanostrip line of 20 nm length and 20 nm width is connected to the patch. The length of the nanostrip is selected to be extremely small, such that it looks like an open circuit termination. Based on the transmitting or receiving circuit connected to the nantenna, the width and length of the nanostrip line should be specified in order to achieve impedance matching. The volume inside the aperture and in between the patch and aperture is filled with $Al_2O_3$. The thickness of both gold layers is 30 nm, and the spacing between them, $d$ is fixed at 20 nm. The footprint of the proposed nantenna is 420 nm × 420 nm. The nanocrescent antenna has two geometrical design parameters: diameter of the circular patch ($D_{out}$), and the diameter of the circular patch ($D_{in}$), as shown in Fig. 5.1.

The antenna is surrounded by free-space everywhere, with unity refractive index. The size of the computational domain is 4000 nm × 4000 nm × 6000 nm. This implies that more than quarter of a free-space wavelength at the maximum simulation wavelength of 3000 nm, is left in all directions between the nantenna and the Perfect Matched Layer (PML) boundary box used in the simulation.

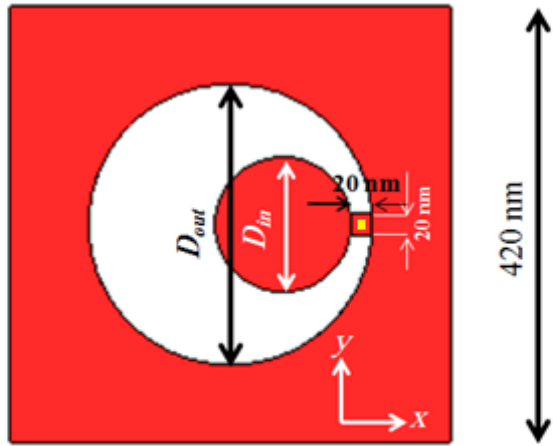

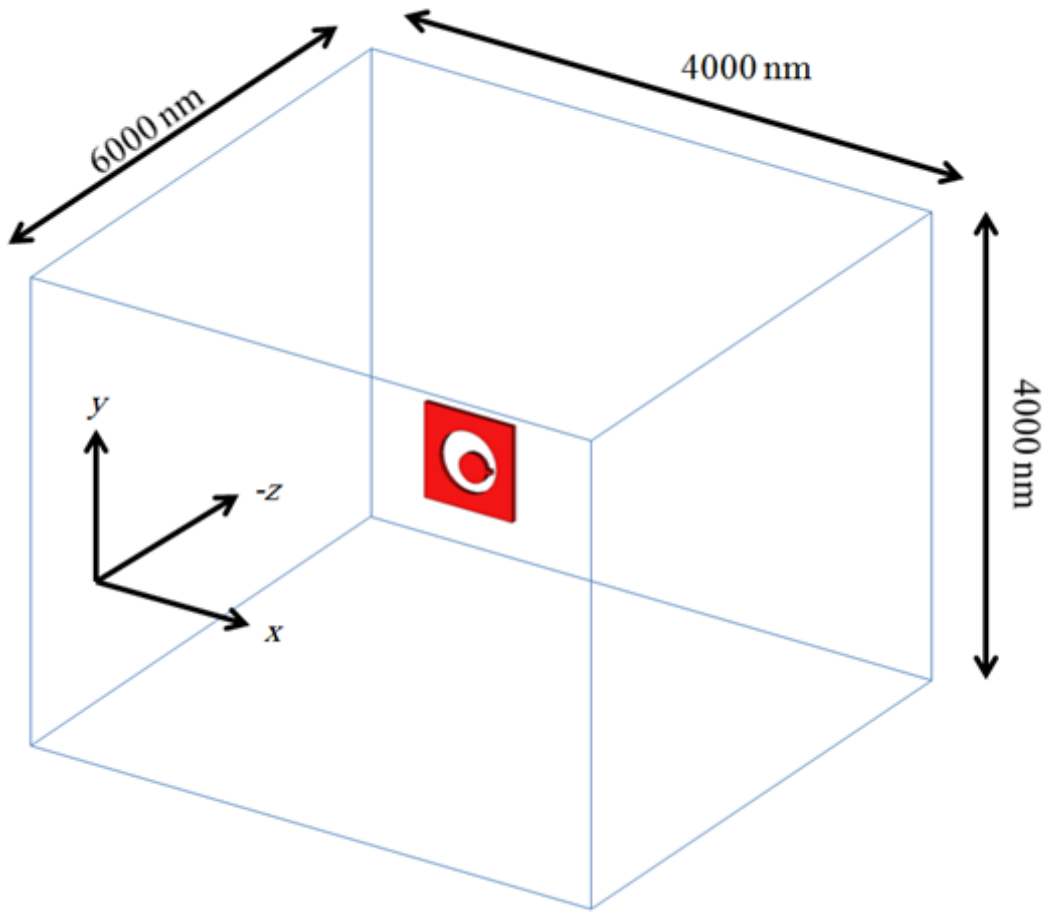

**Figure 5.1: The proposed nanocrescent antenna structure.**

Being the commonly used material for realizing nantennas, gold is adopted in the design of the proposed nanosrescent antenna. However, a comparison between the nantenna's response using the four metals (Gold, Silver, Aluminum, and Copper) is performed in Section 5.3. The geometrical design parameters of the nanocrescent antenna are optimized using Lumerical. This software is based on Finite-Difference Time-Domain (FD-TD) and designed specifically for simulating nano-photonic devices, such as nantennas. Convergence test for the used mesh is performed to arrive at the mesh size that leads to accurate results with reasonable memory and computation time requirements. The geometrical design parameters of the proposed nantenna are: diameter of the patch ($D_{in}$) and diameter of the aperture ($D_{out}$). The field intensity enhancement of the proposed nantenna is observed at the middle of the 20 nm gap between the patch and aperture, as shown in Fig. 5.1. The optimization goal is to maximize the full width half maximum (FWHM), and to center it at 1.55 μm wavelength, which is the commonly used wavelength in optical communications. The optimization process results in the following optimum values: $D_{in}$ = 130 nm and $D_{out}$ = 270 nm.

The field intensity enhancement at the gap of the optimum nantenna is plotted versus wavelength in Fig. 5.2. For the sake of validating the presented results, CST is used to simulate the optimum design. The gold permittivity function extracted from Palik and is supplied to CST [43]. The same mesh used by Lumerical is reused by CST, for the sake of having fair comparison. Reasonable agreement between the two software can be seen in Fig. 5.2. The two packages predict the two resonance wavelengths at almost the same locations. The proposed nanocrescent has two resonance wavelengths at which the intensity enhancement is maximum. The location of the first resonance is at 1120 nm and the normalized intensity enhancement is 766. On the other hand, the second resonance location is at 1920 nm with a normalized intensity of 788. These two resonances are linked to the diameters of the circular patch and aperture of the nantenna structure. Details of engineering the spectral response of the nanocrescent using $D_{in}$ and $D_{out}$ are provided in Section 5.2. Fig. 5.2 indicates that the optimization goal is met, as the FWHM is centered around 1.55 μm as required. The value of FWHM is about 61%, which significantly wider than that of conventional dipoles and bowties (35%). This significant enhancement in the FWHM increases the information capacity of the optical communication system that uses the nanocrescent as transceiver. The 3D radiation pattern at the two resonance frequencies is shown in Fig. 5.3. The directivity of this crescent nantenna at the two resonance frequencies is 2.95 and 2.97 dBi, respectively.

The distribution of the *x*-component of the electric field, at the middle of the dielectric layer, along the *xy* plane is shown in Fig. 5.4. The figure shows the field distribution at the two resonance wavelengths of the proposed nantenna. It can be seen in Fig. 5.4(a), that $E_x$ at the lower resonance frequency is confined mainly near the edge of the edge of the patch. On the other hand, the $E_x$ component extends to fill the whole space between the patch and the aperture at the upper resonance, as shown in Fig.

5.4(b). Fig. 5.5 shows the distribution of the $E_z$ component along the $yz$ plane that contains the nantenna gap, at the two resonance wavelengths. The electric field is concentrated mainly in the insulating layer below the line between the two gold plates, as expected.

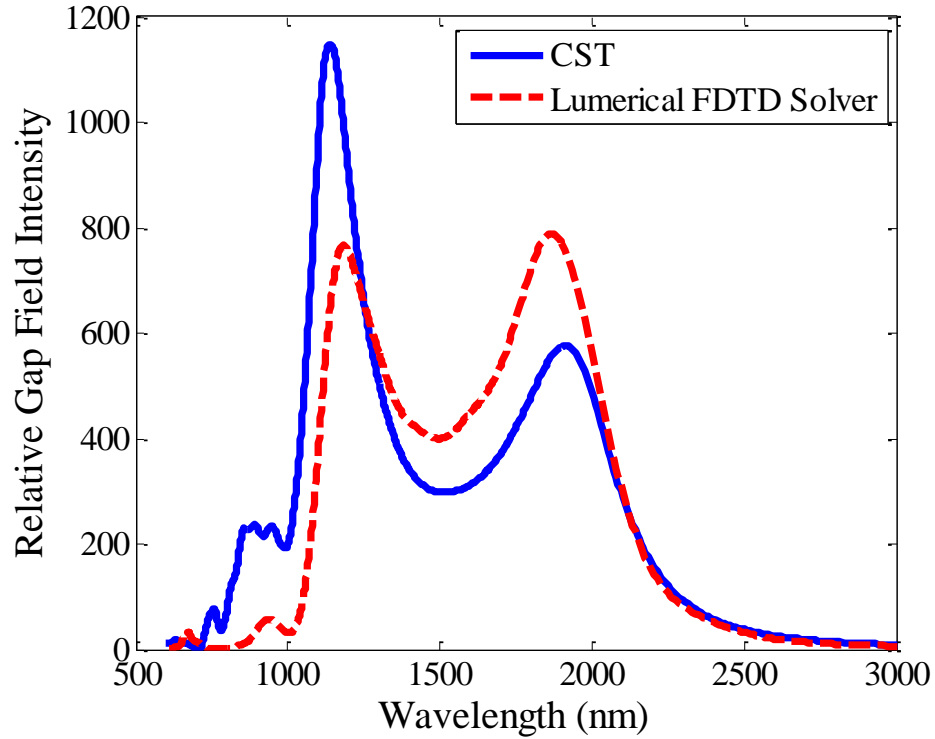

**Figure 5.2: Electric field intensity enhancement at the middle of the gap between the patch and the aperture versus the wavelength, as calculated using both Lumerical and CST.**

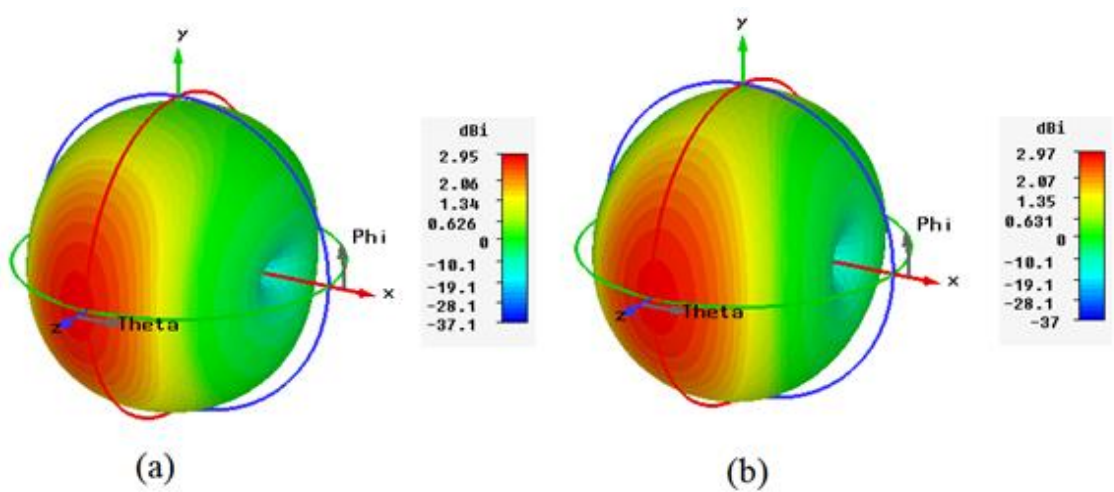

**Figure 5.3: 3D radiation pattern of the nanocrescent antenna, at its two resonance wavelengths, (a) $\lambda_1 = 1102\ nm,$ and (b)$\lambda_2 = 1920\ nm.$**

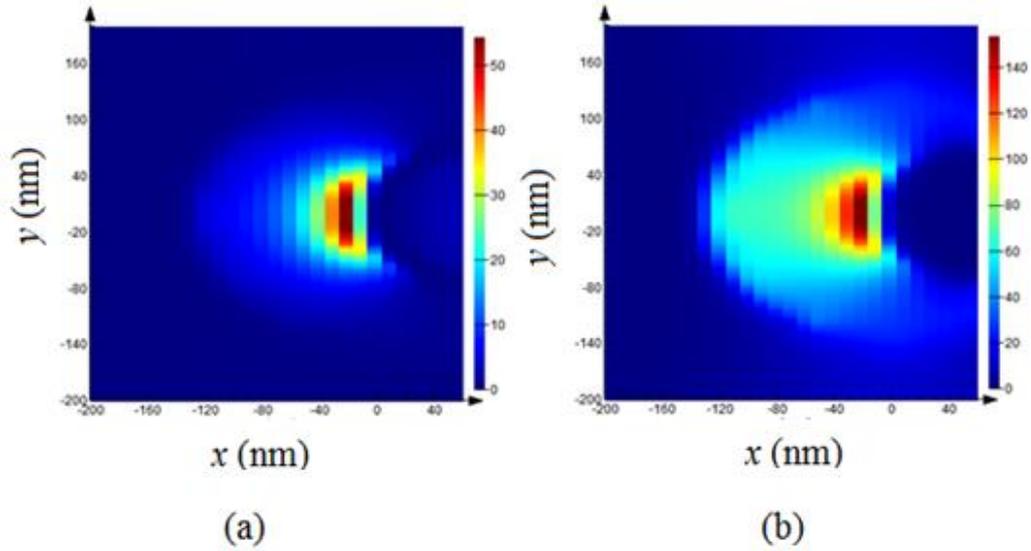

**Figure 5.4:** Intensity distribution along the central *xy* plane at the middle of the insulator layer of the optimum nanocrescent at its two resonance wavelengths, $\lambda_1 = 1102\ nm$, and $\lambda_2 = 1920\ nm$

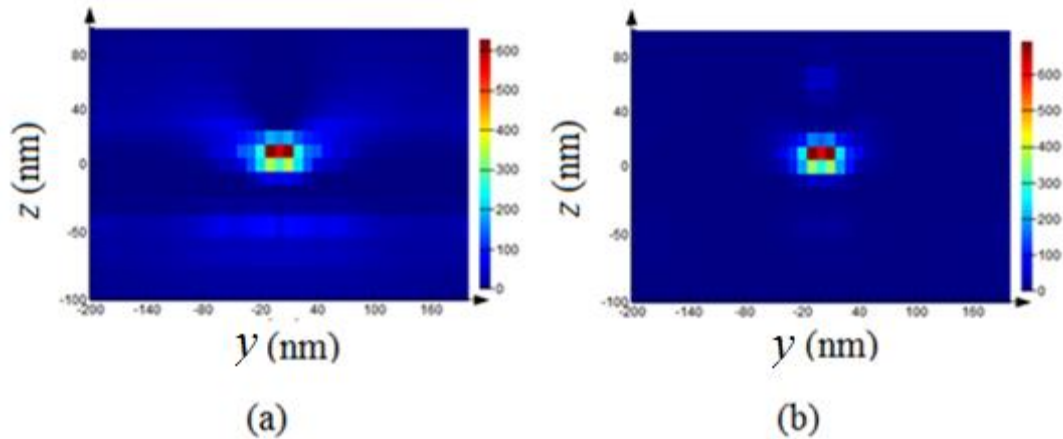

**Figure 5.5:** Intensity distribution along the *yz* plane at the middle of the gap of the optimum nanocrescent at its two resonance wavelengths, $\lambda_1 = 1102\ nm$, and $\lambda_2 = 1920\ nm$.

## 5.2. Geometrical parametric study

As mentioned in the previous section, the nanocrescent antenna has two resonance wavelengths. The first resonance is attributed to the small circular patch. Hence, the parameters $D_{in}$ can be used to control its location. Fig. 5.6(a) shows the intensity enhancement versus wavelength for different values of $D_{in}$: 90, 110, 130, 150, and 170 nm. For this figure, the parameter $D_{out}$ is kept constant at its optimum value of 270 nm. The location of the first resonance shifts up as $D_{in}$ increases, while the location of the second resonance is almost unaffected. It is also clear that the increase of $D_{in}$ results in an increase in the gap intensity enhancement at both resonances, which is more pronounced at the second resonance. This can be attributed to the fact that as $D_{in}$ increases, the spacing between the patch and aperture edges decreases, which results in an increase in the electric field intensity when the conducting edges get significantly closer to each other.

On the other hand, the large circular aperture whose diameter is $D_{out}$, gives rise to the second resonance wavelength. In Fig. 5.6(b), the impact of varying $D_{out}$ on the spectral response of the nanocrescent is investigated. Five different values of $D_{out}$ are considered: 230, 250, 270, 290, and 310 nm, while $D_{in}$ is held constant at its optimum value of 130 nm. It is clear that the location of the first resonance remains almost the same as $D_{out}$ varies. On the other hand, the increase in $D_{out}$ causes shift for the second resonance, as shown in Fig. 5.6(b). In addition, the increase of the $D_{out}$ is accompanied by a little increase in the first resonance gap intensity, while the intensity of the second resonance is almost constant.

As expected, the change of the thickness of the insulator, $d$, between the two metal plates affects the intensity enhancement level. The impact of this thickness on the spectral response of the nanocrescent is presented in Fig. 5.6(c) for various $d$ values: 20, 30, and 40 nm. The corresponding intensity enhancement at the nantenna gap is 800, 450, and 250, respectively, with almost no change in the locations of the resonance wavelengths.

## 5.3. **Materials Parametric Study**

The effect of using various materials on the nanocrescent antenna spectral response is investigated in this section. Silver, aluminum, and copper in addition to gold are the well-known materials used to realize nantennas. The comparison is made using the same $D_{in}$ and $D_{out}$, for the optimum design obtained using gold. Fig. 5.7 shows the intensity enhancement at the nantenna gap versus wavelength these four metals are used to realize the nanocrescent. As it is clear from the figure, silver shows the maximum electric field intensity enhancement. This can be attributed to the fact that silver has the smallest imaginary part of permittivity, according to Fig. 1.3(b), which means the least losses and the most intensity enhancement. This can be attributed to the fact that silver has the smallest imaginary part of permittivity, according to Fig. 1.3(b), which means the least losses and the most intensity enhancement. On the other hand, aluminum has the largest losses and the smallest intensity enhancement. Although silver provides the maximum electric field intensity enhancement, it is not recommended in fabricating nantennas. The reason behind this is that it is more likely to be affected by oxidation [11]. Consequently, gold is adopted in realizing the proposed nanocrescent antenna.

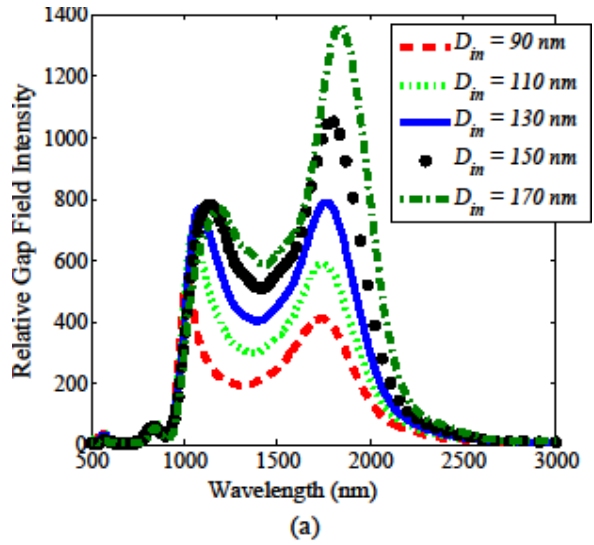

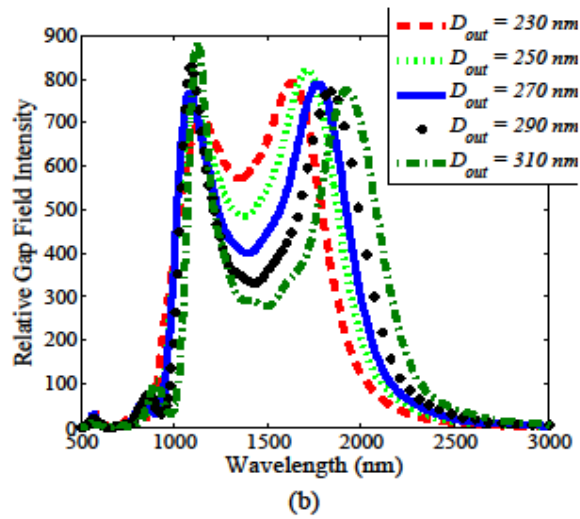

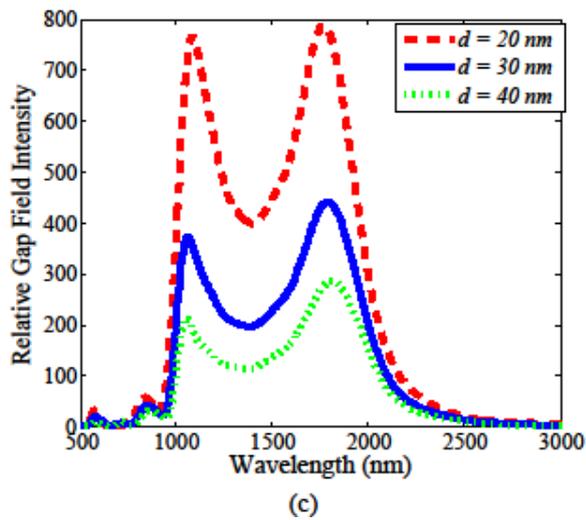

**Figure 5.6:** a) Relative gap field intensity enhancement versus wavelength for different values of $D_{in}$. $D_{out}$ = 270 nm and $d$ = 20 nm. b) Relative gap field intensity enhancement versus wavelength for different values of $D_{out}$. $D_{in}$ = 130 nm and $d$ = 20 nm. c) Relative gap field intensity enhancement versus wavelength for various insulator thickness, $d$. $D_{out}$ = 270 nm and $D_{in}$ = 130 nm.

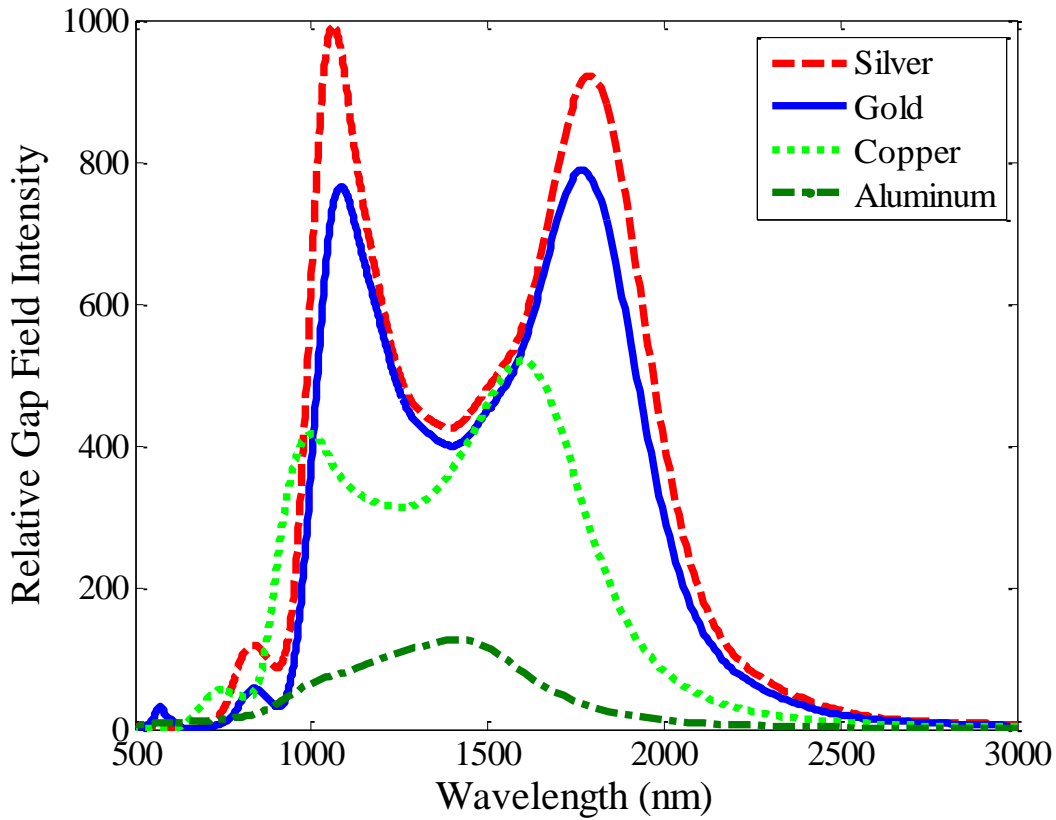

**Figure 5.7: Relative gap field intensity enhancement, versus the wavelength in case of using silver (Ag), copper (Cu), gold (Au), and aluminum (Al).**

## 5.4. Summary

A new nantenna design is presented in this Chapter, namely the nanocrescent. This design consists of metal dielectric metal forming a crescent shape. A circular patch is elevated on top of a non-concentric circular aperture. A parametric study is carried out showing the effect of the geometrical and material parameters on the spectral response of the nantenna. This study can be repeated to tune the nanocerescent antenna to any other wavelength of interest. As the diameters of the patch and aperture get close to each other, the two resonances approach each other which results in narrow FWHM. On the other hand, the two resonance wavelengths can be separated more from each other, as the difference between the two diameters increases. The widest FWHM centered at 1.55 µm can be obtained when the minimum intensity enhancement between the two maxima is located at 1.55 µm, with a value slightly higher than half of the maximum.

# Chapter 6 **Conclusion and Future Work**

## 6.1. **Summary of research**

In this research, traveling wave solar rectennas (rectifying nantennas) are investigated theoretically. The nantennas used are the conventional dipoles. The rectifier used is metal insulator metal (MIM) diode. Two different travelling wave rectifiers are used, namely the vertical coupled strips and the lateral (horizontal) coupled strips rectifiers. The proposed travelling wave rectifiers are compared with the frequently used lumped version. It has been demonstrated that the travelling wave rectennas are superior over the lumped rectennas due to the great enhancement in the coupling efficiency between the nantenna and rectifier.

The used MIM diode structure is modeled using three methods.
First, the Transfer Matrix Method based on the Airy Function (AF-TMM), where an analytical expression for the tunneling probability is derived. Second, the Non Equilibrium Green's Function (NEGF), where the tunneling probability is calculated as function of the Green's function and the escape rate of the electrons to the left and the right contacts. Finally, the WKB approximation to solve Schrödinger equation, where the wave function is expressed in exponential form.

The three aforementioned methods are compared to each other. The TMM results agree well with the NEGF results, while the WKB method over-estimate the tunneling current. By comparing the simulation time of both the AFTMM and the NEGF, the AFTMM is superior over the NEGF. The modeling of the MIM using the AFTMM and NEGF provide results that agree well with the previously published experimental data.

After successfully modeling the current in of the MIM rectifier, the figures of merit of this device are investigated: diode resistance, responsivity, nonlinearity, and asymmetry. The left metal and insulator used here was $Nb/Nb_2O_5$, while the right metal function was scanned from $4.3$ $eV$ to $5.65$ $eV$. An extensive discussion is presented showing the effect of the work function difference and the insulator layer thickness on the figures of merit. Also, the possibility of using two insulating layers to form what is called Metal Insulator Insulator Metal (MIIM) rectifier is discussed with highlighting the achieved enhancement of the diode performance.

Three plasmonic transmission lines that can be used as travelling wave rectifiers are thoroughly investigated
First, the vertical coupled strips transmission line, where the two strips are placed on top of each other and the spacing between them is filled with insulator. Second, the lateral coupled strips line, where the two strips are located besides each other and the lateral spacing between them is also filled with an insulator. Third, the nanostrip plasmonic line, where a strip is placed above an infinite metal layer, with an insulator in between.

For the three lines, design curves in the form of contour plots are presented, that show the effect of various geometrical parameters on the effective refractive index, attenuation constant, propagation length, and the characteristic impedance. This design curves are essential tools that facilitate the efficient coupling from the nantenna to these transmission lines.

Two new traveling wave rectenna systems are presented and simulated. In the first rectenna, a nanodipole is integrated with a vertical coupled strips rectifier, while a lateral coupled strips line is used as termination in the second rectenna. For both rectennas, a systematic approach for integration is presented, where the transmission lines are perfectly matched to their nantennas For the vertical coupled strips termination, the insulator layer is fixed at 2 nm, while the spacing between the two strips increases to be 20 nm. This significant difference in the spacing between the strips in the two transmission lines is due to limitations of the fabrication technologies. It results in huge difference in the rectifier efficiency. With the strips' spacing (insulator thickness) is kept constant at the minimum possible value, the width and height of the strips are swept, and design contour plots for the characteristic impedance are prepared. In addition, the tuning of the dipole nantennas to the desired resonance frequency and their matching to the plasmonic lines are performed. The length of the nantenna determines mainly the location of the resonance frequency, as expected, while the width of the dipole arms affects the value of the input impedance at resonance, and consequently plays a major in maximizing the coupling efficiency between the nantenna and rectifier.

The different sub-efficiencies of the proposed travelling wave rectennas are formulated and obtained. These efficiencies includes, the antenna efficiency, coupling efficiency, and the rectifier efficiency. The achieved values of the first two efficiencies are 83.6% and 100%, respectively. On the other hand, the MIM rectifier efficiency is found to be as small as $10^{-4}$ A/Watt and $10^{-13}$ A/Watt for the vertical and lateral coupled strips rectifier, respectively. Such extremely small rectifier efficiency reveals that the main obstacle behind commercializing the nano rectennas as infrared solar energy harvesters is the MIM rectifier. Modern transistor with channel length down to 10 nm, are expected to perform the rectification with dramatically higher efficiency, which overcomes the current obstacle and open the door for full utilization of the nano-rectennas [93, 94].

As an efficient replacement for the laser diode and the photo detector in optical communication systems, a novel nanocrescent antenna is introduced in this research. The commonly used 1.55 µm wavelength is adopted. The proposed nantenna enjoys wide band behavior with a bandwidth double the average value of the nantennas presented in literature. Such great enhancement in the bandwidth results in doubling the capacity of the optical communication systems. A comprehensive parametric study for the new antenna is performed. This study include both geometrical dimensions and material properties. The design methodology that can be used to engineer the spectral response of the new nantenna is outlined. Following this methodology, the nanocrescent can be easily to any other frequency band of interest.

## 6.2. Thesis contributions

The contributions of this thesis are as follows:

(i) Two quantum mechanical reliable models are presented and successfully compared to each other. In addition, the comparison of these models with the experimental work shows reasonable agreement.

(ii) The figures of merit of the MIM rectifier are extensively discussed. All the previous theoretical work on this device did not provide such deep discussion.

(iii) Design curves are presented for a variety of plasmonic transmission lines around 30 THz. These curves are essential aids to the realization of infrared plasmonic transmission lines.

(iv) Two new traveling wave complete rectenna systems are presented. This can be considered the first integration of the various components of the travelling wave rectenna system.

(v) Novel nantenna with crescent shape is introduced. The new nantenna offers a FWHM of 61% compared to 35% for other nantennas presented in literature. The proposed nantenna will be found attractive for the majority of optical communication systems.

## 6.3. Future work

The further works in the rectenna can be divided into five main directions: developing various nantennas with broader bandwidth, proposing nantennas with two orthogonal perpendicular polarizations that can receive the incident solar radiation whose polarization is random, covering both the visible and infrared range of solar energy, studying the effect of the substrate on the rectenna's performance, establishing a rectenna array for visible and infrared solar energy harvesting, and proposing a new structure for the rectifier that can rectify that fast THz signal with high quantum efficiency.

Most research works in the rectenna focus on developing rectennas using the conventional nantenna structures, such as dipoles, bowties, and spirals. However, other new types of nantennas, which might be superior over the conventional ones from the fractional bandwidth aspect, have not been utilized yet. A step toward this is performed and presented in Chapter 5 of this thesis, where a new nantenna design is introduced, namely a nanocrescent.. This nantenna has a remarkable wideband behavior, which make it very promising to be used not only for optical communication systems, but also at the infrared frequency range. Integrating this nantenna, tuned to 30 THz, to a traveling wave nanostrip plasmonic transmission line leads to more collection of the infrared energy than that achieved using the conventional nano dipoles.

In addition, frequency scaling law states that the antenna can be designed to operate at any frequency range by adjusting the antenna dimensions. Hence, a scaled down version of the proposed rectenna can be designed to directly harvest the visible solar energy. However, the nantenna features at such high frequency becomes significantly smaller than those of the nantenna operating in the infrared frequency. This requires advanced fabrication technology and may increase the cost of fabrication. Two nantennas with different sized can be integrated together in order to harvest both the Earth infrared and the Sun visible solar radiation. Moreover, it is well known that the solar radiation has random polarization. Consequently, to enhance the solar energy harvesting capability a nantenna with dual or circular polarization can be used. Otherwise, two orthogonal linearly polarized nantennas can be used to receive the incident random polarization.

Also, the effect of the substrate carrying the rectenna system should be studied in detail. The rectennas used in this research were simulated assuming free-space everywhere. Unlike high efficiency solar cells which must be fabricated on expensive silicon wafers, the rectenna system can be build on any low cost substrate, such as flexible polymers.

Moreover, rectenna systems are usually build in the form of large arrays. Studying the impact of the mutual coupling between the adjacent rectennas in this array, on their performance requires future study.

Finally, though the traveling wave structure for the rectifier enhances the overall system efficiency by the superior matching of nantenna impedance to the rectifier characteristic impedance, the efficiency is still very low. Carbon nanotubes and Graphene diodes need to be analyzed to examine whether they can rectify that very fast THz waves with higher quantum efficiency. Moreover, state-of-the-art transistors with nano-dimensions should be considered as an efficient alternatives of the low-efficiency MIM rectifier.

# REFERNCES